\documentclass[aps,prd,floats,floatfix,preprintnumbers,superscriptaddress,nofootinbib]{revtex4-2}

\usepackage{graphicx} 
\usepackage{dcolumn}
\usepackage{amssymb}
\usepackage{amsmath,bm}
\usepackage[hidelinks]{hyperref}
\usepackage{color}
\usepackage[utf8]{inputenc}
\usepackage{slashed}

\newcommand{\nn}{\nonumber \\}
\newcommand{\Slash}[1]{{\ooalign{\hfil/\hfil\crcr$#1$}}}

\newcommand{\be}{\begin{equation}}
\newcommand{\ee}{\end{equation}}

\newcommand{\beq}{\begin{eqnarray}}
\newcommand{\eeq}{\end{eqnarray}}

\newcommand{\calA}{\mathcal{A}}
\newcommand{\calF}{\mathcal{F}}

\newcommand{\calV}{\mathcal{V}}

\newcommand{\calS}{\mathcal{S}}

\newcommand{\calH}{\mathcal{H}}

\newcommand{\ltwp}{\boldsymbol{l}_{2T}}
\newcommand{\pqp}{\boldsymbol{p}_{qT}}

\newcommand{\tils}{\tilde{s}}
\newcommand{\tilt}{\tilde{t}}
\newcommand{\hats}{\hat{s}}
\newcommand{\hatu}{\hat{u}}
\newcommand{\hatt}{\hat{t}}
\newcommand{\hatx}{\hat{x}}
\newcommand{\hatz}{\hat{z}}
\newcommand{\pd}{\partial}

\begin{document}
\date{\today}
\preprint{ZTF-EP-24-01}

\title{Perturbative QCD contribution to transverse  single spin asymmetries  \\in Drell-Yan and  SIDIS }
\author{Sanjin Beni\' c}
\affiliation{Department of Physics, Faculty of Science, University of Zagreb, Bijenička c. 32, 10000 Zagreb, Croatia}

\author{Yoshitaka Hatta}
\affiliation{Physics Department, Brookhaven National Laboratory, Upton NY, 11973, USA}

\affiliation{RIKEN BNL Research Center, Brookhaven National Laboratory, Upton NY, 11973, USA}

\author{Abhiram Kaushik}
\affiliation{Centre for Informatics and Computing, Rudjer Bo\v skovi\' c Institute, HR-10002 Zagreb, Croatia}

\author{Hsiang-nan Li}
\affiliation{Institute of Physics, Academia Sinica, Taipei, Taiwan  11529, Republic of China} 

\begin{abstract} 

In a previous publication [Beni\' c et al., Phys. Rev. D104 (2021) 094027], we have computed the perturbative QCD   contribution to   transverse single spin asymmetries (SSAs) in semi-inclusive Deep Inelastic Scattering (SIDIS) involving the $g_T(x)$ distribution. In this paper, we first present a more efficient derivation of the asymmetries which is applicable to both transverse and longitudinal SSAs, and correct some inconsistencies  in our previous calculation. We then adapt the method to compute transverse  SSAs in  Drell-Yan proportional to $g_T(x)$ and its gluonic counterpart, and discuss the crossing symmetry between the results for SIDIS and Drell-Yan.
Finally, we  present numerical results for various asymmetries measurable at the EIC, RHIC, COMPASS and Fermilab (SpinQuest), including also part of the genuine twist-three corrections to $g_T(x)$ from a recent global analysis.  We find that the asymmetries can reach percent-level magnitude, if the kinematics predominantly probes the large-$x$ (valence) region of the polarized proton, but remain at sub-percent levels otherwise.
 
\end{abstract}


\maketitle

\section{Introduction}

In recent years, there has been renewed interest in the perturbative QCD contribution to  single spin asymmetries (SSAs) \cite{Benic:2019zvg,Benic:2021gya,Abele:2022spu,Boughezal:2023ooo} motivated by the ongoing and future experiments capable of the precision measurement of spin asymmetries up to high transverse momenta. By perturbative, we mean that the imaginary phase necessary to generate asymmetries comes from $2\to 2$ hard (partonic) scattering kernels.  For  transverse SSAs $A_{UT}$, such a contribution was studied long ago \cite{Kane:1978nd,Dharmaratna:1996xd} and had been largely neglected by the community owing to the purported proportionality to a current quark mass $A_{UT}\propto \alpha_s m_q$.  On the other hand,  longitudinal SSAs $A_{UL}$ are insensitive to the mass effect, and have been calculated within the standard collinear factorization framework in both  Drell-Yan \cite{Carlitz:1992fv,Yokoya:2007xe} and semi-inclusive DIS (SIDIS) \cite{Abele:2022spu}.

In our previous publications \cite{Benic:2019zvg,Benic:2021gya}, we have systematically extended the investigation of transverse SSAs in SIDIS
to a subleading level. 
In the $k_T$-factorization framework, we have identified the 
complete set of SSA sources up to two loops 
in hard kernels and to twist three in two-parton transverse-momentum-dependent (TMD) 
 parton distribution functions (PDFs). We then focused  on the asymmetries originating from the 
polarized quark distribution function $g_T(x)$, as well as its gluonic counterpart ${\cal G}_{3T}(x)$, 
which survives in the collinear factorization framework. 
The corresponding quark- and gluon-initiated 
${\cal O}(\alpha_s^2)$ hard kernels for SIDIS were 
computed, and the resultant asymmetries $A_{UT}$ associated with, 
e.g., the $\sin (\phi_h-\phi_S)$, $\sin \phi_S$ and $\sin (2\phi_h-\phi_S)$ 
harmonics, were predicted for measurements at the future Electron-Ion Collider (EIC). We have shown  that in SIDIS, the naive relation $A_{UT}\propto m_q$ is incorrect and should be replaced by  $A_{UT}\propto M_N$, where $M_N$ is the proton mass. We have further performed a numerical analysis and found that some of the asymmetries  could reach a percent level, manifesting the importance of high-accuracy studies in the considered 
kinematic regions. 

In this paper, we first present, in Sections II and III, a more efficient derivation of the
asymmetries in SIDIS, which allows for a unified treatment of transverse and longitudinal SSAs. We take this opportunity to correct 
some inconsistencies in our previous results \cite{Benic:2021gya}. 
 In Section IV, the formalism is  
extended to compute, for the first time, the contributions to transverse SSAs in  Drell-Yan  
proportional to $g_T$ and ${\cal G}_{3T}$ in the $q\bar{q}$ annihilation and Compton scattering channels.
Here, too, the approach is  applicable to longitudinal SSAs, and we reproduce the known results in the literature \cite{Carlitz:1992fv,Yokoya:2007xe}  for completeness. 
We then discuss how the formulas for SIDIS and Drell-Yan may be related by crossing symmetry.  
Finally in Section V, we numerically evaluate the various asymmetries. We update our predictions for the EIC and provide new predictions for the kinematics of the RHIC, COMPASS and the Fermilab SpinQuest experiments. The  possible impact of the  genuine twist-three effects in light of the recent extraction of $g_T$ \cite{Bhattacharya:2021twu,Bauer:2022mvl}  is also explored.

\section{SSA in SIDIS, quark-initiated channel}
\label{sec:sidisq}

In this and the next sections, we revisit our calculation of SSAs in SIDIS in \cite{Benic:2021gya}. After reviewing the basic kinematics of the reaction and our new approach  developed in  \cite{Benic:2019zvg,Benic:2021gya}, we propose a simpler derivation   which helps reveal the inconsistencies in the previous analysis. It also facilitates the comparison with relevant results in the literature, and allows a straightforward   extension to  the Drell-Yan case to be discussed in a later section.

\subsection{Setup}

Consider SIDIS in the so-called hadron frame where the virtual  photon with the virtuality $q^2=-Q^2$ and the transversely polarized proton with the momentum $P^\mu$ are collinear.  We neglect the proton mass $M_N$ whenever possible, so $P^\mu\approx \delta^\mu_+P^+$ is effectively a lightlike vector. We also need another lightlike vector $n^\mu = \delta^\mu_-/P^+$, such that $P\cdot n=1$. 
The spin-dependent part of the cross section can be written as 
\beq
\frac{d\Delta \sigma}{dx_B dQ^2 dz_fdq_T^2 d\phi d\chi}=\frac{\alpha_{em}^2z_f}{128\pi^4x_B^2 S_{ep}^2Q^2}\sum_{a}\int \frac{dz}{z^2}D_1^a(z)L^{\mu\nu}w_{\mu\nu}^a, \label{base}
\eeq
where $x_B=Q^2/(2P\cdot q)$ is the Bjorken variable and $S_{ep}=(P+l)^2$ is the $ep$ center-of-mass energy. $\mu$ and $\nu$ are the polarization indices of the virtual photon in the complex-conjugate amplitude and in the amplitude, respectively. The outgoing lepton has the momentum $l'^\mu = l^\mu -q^\mu$ and the azimuthal angle  $\phi$.   The incoming parton has the momentum $p^\mu= xP^\mu$.   The outgoing hadron $h$ has the azimuthal angle $\chi$ and the longitudinal momentum fraction 
\beq
z_f = \frac{P_h\cdot P}{q\cdot P} = z \frac{p_a\cdot P}{q\cdot P},
\eeq
of the virtual photon momentum. It comes from the fragmentation of a parton $a=q,\bar{q},g$ in the final state, and   $D_1^a(z)$ is the corresponding  fragmentation function. Finally, the variable
\beq
q_T^2=\frac{P^2_{h\perp}}{z_f^2} =\frac{\hat{s}\hat{u}}{\hat{t}},
\eeq
 is  the characteristic transverse momentum of the process, where the last equality has been expressed in terms the partonic  Mandelstam variables
\be
\hat{s} = (p+q)^2\,, \qquad \hat{t} = (p - p_q)^2 \,, \qquad \hat{u} = (q - p_q)^2\,.
\label{eq:collman}
\ee

The leptonic tensor is given by  the standard expression
\beq
L^{\mu\nu} = 2(l^\mu l'^\nu + l^\nu l'^\mu)-g^{\mu\nu}Q^2. \label{lepto}
\eeq
Following the literature  \cite{Meng:1991da}, we perform a tensor decomposition 
\beq
L^{\mu\nu}w_{\mu\nu} = Q^2 \sum_k^{1,2,3,4,8,9} {\cal A}_k w_{\mu\nu}\tilde{\mathcal V}_k^{\mu\nu},
\label{lepdec}
\eeq
where 
\be
\begin{split}
&\tilde{\mathcal V}^{\mu\nu}_1 =\frac{1}{2}(2T^\mu T^\nu +X^\mu X^\nu +Y^\mu Y^\nu), \\
&\tilde{\mathcal V}^{\mu\nu}_2 =T^\mu T^\nu, \\
&\tilde{\mathcal V}^{\mu\nu}_3 =-\frac{1}{2}(T^\mu X^\nu + X^\mu T^\nu), \\
&\tilde{\mathcal V}^{\mu\nu}_4 =\frac{1}{2}(T^\mu X^\nu -Y^\mu X^\nu), \\
&\tilde{\mathcal V}^{\mu\nu}_8 =-\frac{1}{2}(T^\mu Y^\nu +Y^\mu T^\nu), \\
&\tilde{\mathcal V}^{\mu\nu}_9 =\frac{1}{2}(X^\mu Y^\nu + Y^\mu X^\nu). \\
\end{split}
\ee
are defined in terms of the orthogonal four-vectors normalized as $T^2=-X^2=-Y^2=-Z^2=1$. For a later purpose, it is convenient to express them in terms of the partonic Mandelstam variables\footnote{Our convention for $\gamma_5$ and the antisymmetric tensor $\epsilon^{\mu\nu\rho\lambda}$ are the same as in \cite{Benic:2019zvg,Benic:2021gya}: $\gamma_5=+i\gamma^0\gamma^1\gamma^2\gamma^3$ and $\epsilon_{0123}=1=-\epsilon^{0123}$ so that ${\rm Tr}[\gamma_5\gamma^0\gamma^1\gamma^2\gamma^3]=-4i$. } 
\be
\begin{split}
& T^\mu = \frac{1}{Q} \left(q^\mu + \frac{2Q^2}{\hats + Q^2} p^\mu\right)\,,\\
& X^\mu = \sqrt{\frac{\hatt}{\hats\hatu}}\left[-\frac{\hats + Q^2}{\hatt}p_q^\mu - q^\mu - \frac{1}{\hats + Q^2}\left(Q^2 + \frac{\hats \hatu}{\hatt}\right)p^\mu\right]\,,\\
& Z^\mu = - \frac{q^\mu}{Q}\,,\\
& Y^\mu = \epsilon^{\mu\nu\rho\sigma} Z_\nu X_\rho T_\sigma\,.
\end{split}
\label{eq:txyz}
\ee
The coefficients are given by  
\be
\begin{split}
&{\cal A}_1= 1+\cosh^2\psi 
=\frac{2}{1-\varepsilon}, \\
&{\cal A}_2=-2, \\
&{\cal A}_3=- \sinh 2\psi \cos \phi_h=
-\frac{2\sqrt{2\varepsilon(1+\varepsilon)}}{1-\varepsilon}\cos \phi_h
, \\
&{\cal A}_4 =\sinh^2\psi \cos 2\phi_h=  
\frac{2\varepsilon}{1-\varepsilon}\cos 2\phi_h, \\
&{\cal A}_8= -\sinh 2\psi \sin \phi_h=- 
\frac{2\sqrt{2\varepsilon(1+\varepsilon)}}{1-\varepsilon}\sin \phi_h , \\
&{\cal A}_9= \sinh^2\psi \sin 2\phi_h=
\frac{2\varepsilon}{1-\varepsilon}\sin 2\phi_h,
\end{split}
\label{a1234}
\ee
where the `electron rapidity' $\psi$ is defined by $\cosh \psi \equiv 2x_BS_{eq}/Q^2-1$ and
$\phi_h\equiv \phi-\chi$ is the hadron angle relative to the lepton angle (lepton plane). We have introduced in  (\ref{a1234}) the  ratio of the longitudinal/transverse photon fluxes 
\be
\varepsilon = \frac{1-y}{1-y + \frac{y^2}{2}},
\ee
with the standard variable $y= P\cdot q/P\cdot l$ in DIS. 
Our task is to derive the hard factor $w_{\mu\nu}^a$. 
The transversely polarized proton can emit either a quark (or an antiquark) or a gluon that initiates hard scattering. We consider the quark-initiated channel in the remainder of this  section. The gluon-initiated channel will be studied in the next section.

\subsection{Quark-initiated, quark-fragmenting  channel}

The basic hard scattering processes in the quark-initiated channel include $q\gamma^*\to qg$ and $\bar{q}\gamma^*\to \bar{q}g$, where either the (anti)quark  or the gluon in the final state fragments into the detected hadron. We first consider the quark fragmenting channel, namely, $a=q$ (or $a=\bar{q}$) in (\ref{base}). 
The new approach to SSAs that we shall pursue was originally developed by Ratcliffe  \cite{Ratcliffe:1985mp}, and rediscovered and completed by us \cite{Benic:2019zvg,Benic:2021gya}. It features the following  hadronic tensor 
\be
w^q_{\mu\nu} = \frac{e_q^2}{2} \int dx g^q_T(x)  S_\perp^\lambda\frac{\pd}{\pd k_\perp^\lambda} \left(\delta\left((k+q-p_q)^2\right){\rm Tr}\left[\gamma_5 \slashed{k}S^{(q)}_{\mu\nu}(k)\right]\right)_{k = p}+\cdots \,,
\label{eq:w0}
\ee
where  $S^\mu_\perp=(0,\vec{S}_\perp,0)=M_N(0,\cos \Phi_S, \sin \Phi_S,0)$ is the spin vector of the proton with $S^2_\perp=-M_N^2$.  (In \cite{Benic:2021gya}, we adopted the normalization $S_\perp^2=-1$.)   The $\delta$-function arises from the on-shell condition for the unobserved gluon with the momentum $k+q-p_q$. 
The distribution $g_T(x)$ is defined through the matrix element
\beq
\int \frac{d\lambda}{4\pi}e^{i\lambda x}\langle PS_\perp|\bar{q}(0)\gamma^\mu_\perp \gamma_5Wq(\lambda n)|PS_\perp\rangle = S_\perp^\mu g^q_T(x),\label{pmass}
\eeq
where  $W$ denotes the Wilson line connecting $[0,\lambda n]$. Equation~(\ref{pmass}) already shows that a transverse SSA is proportional to the proton mass $|S_\perp|=M_N$, but not the current quark mass $m_q$ as often assumed in perturbative calculations \cite{Kane:1978nd}.  We set $m_q=0$ throughout this paper. 

In (\ref{eq:w0}), we have employed the Wandzura-Wilczek (WW) approximation and  omitted the contribution from the genuine twist-three $q\bar{q}g$ correlation functions. The full expression can be found in \cite{Benic:2019zvg,Benic:2021gya}. While the truncated formula (\ref{eq:w0}) has attractive features such as gauge invariance and infrared finiteness, strictly speaking, the WW approximation is not self-consistent. Our rationale is that, since the hard kernel $S^{(q)}_{\mu\nu}$ is evaluated to ${\cal O}(\alpha_s^2)$, the omitted terms in (\ref{eq:w0}) are formally of higher order in $\alpha_s$ compared to the genuine twist-three contributions previously investigated in the literature \cite{Efremov:1984ip,Qiu:1991pp,Eguchi:2006mc,Koike:2011nx,Yoshida:2016tfh}. The WW part, on the other hand, has no analogue in the usual twist-three approach to SSAs. Moreover, it can be evaluated solely from the knowledge of twist-two PDFs. It thus deserves a focused study. 

The imaginary phase necessary for a SSA arises from internal loops  in the hard kernel $S^{(q)}_{\mu\nu}(k)$. We assume that it is symmetric in $\mu\nu$, because the leptonic tensor (\ref{lepto}) is. 
Thanks to the factorization property discussed in \cite{Benic:2019zvg}, $S^{(q)}_{\mu\nu}(k)$  is calculable in perturbation theory and infrared finite.\footnote{The $k_\perp$-derivative in (\ref{eq:w0}) gives 
\beq
\left. S_\perp^\lambda \frac{\partial}{\partial k_\perp^\lambda} {\rm Tr}\left[\gamma_5\Slash k S_{\mu\nu}^{(q)}(k)\right]\right|_{k=p}= {\rm Tr}\left[\gamma_5 \Slash S_\perp S_{\mu\nu}^{(q)}(p)\right] + S_\perp^\lambda {\rm Tr}\left[\gamma_5\Slash p \frac{\partial S^{(q)}_{\mu\nu}(k)}{\partial k_\perp^\lambda}\right]_{k=p}. \label{twoterm}
\eeq
The first term alone is infrared divergent \cite{Metz:2006pe,Kovchegov:2012ga}. The divergence is cured by the second term, which takes into account a quark transverse momentum \cite{Schlegel:2012ve,Benic:2019zvg,Benic:2022qzv}.}  
It starts at ${\cal O}(\alpha_s^2)$ and involves six topologically different loop diagrams in the  $\gamma^*q\to qg$ amplitude. The tree-level amplitude consists of two diagrams, so there are twelve diagrams at the squared amplitude level (see  \cite{Benic:2021gya} for the explicit expression).  These diagrams have to be computed for a non-colinear incoming parton momentum $k^\mu = (xP^+,k_\perp,0)$, and the limit $k_\perp\to 0$ can be taken only after the differentiation  with respect to $k_\perp$.\footnote{More precisely, we required  the on-shell condition $k^2=0$ in \cite{Benic:2019zvg} for the incoming parton in the proof of gauge invariance, which implies $k^-=k_\perp^2/(2k^+)\neq 0$. However, $k^-$ can be neglected for the present purpose, since we only keep terms linear in $k_\perp$.} 
In \cite{Benic:2021gya}, we derived  \eqref{eq:w0} by brute force, taking care of the cancellation of infrared divergences from various diagrams. The calculation was quite cumbersome and the result has not been checked by other means.  In this paper we present a more efficient approach, which can be straightforwardly generalized to the Drell-Yan case. 

Our basic observation is that the symmetric tensor 
\beq
{\cal H}_{\mu\nu}(k) \equiv {\rm Tr}\left[\gamma_5 \slashed{k}S^{(q)}_{\mu\nu}(k)\right] ,
\label{eq:Hmunu}
\eeq 
consists only of the three four-vectors $k,q$ and $p_q$, and  involves the totally antisymmetric tensor because of the presence of $\gamma_5$. Furthermore, it has to satisfy the QED Ward identity $q_\mu {\cal H}^{\mu\nu}=0$. There are only two independent tensor structures that satisfy these criteria, 
\be
\begin{split}
T_1^{\mu\nu}(k) & = \left(k^\mu + \frac{k\cdot q}{Q^2}q^\mu\right)\epsilon^{\nu k q p_q} + \left(k^\nu + \frac{k\cdot q}{Q^2}q^\nu\right)\epsilon^{\mu k q p_q}\,,\\
T_2^{\mu\nu}(k) & = \left(p_q^\mu + \frac{p_q\cdot q}{Q^2}q^\mu\right)\epsilon^{\nu k q p_q} + \left(p_q^\nu + \frac{p_q\cdot q}{Q^2}q^\nu\right)\epsilon^{\mu k q p_q}\,.\\
\end{split}
\label{eq:T12}
\ee
We can thus write 
\be
{\cal H}^{\mu\nu}(k)  = T_1^{\mu\nu}(k) S_1(\tilde{s},\tilde{t},\hat{u},Q^2) + T_2^{\mu\nu}(k) S_2(\tilde{s},\tilde{t},\hat{u},Q^2)\,.
\label{eq:decompos}
\ee
Thanks to the Lorentz covariance, the scalar functions $S_{1,2}$ depend only on $Q^2$ and the Mandelstam variables
\be
\tilde{s} = (k+q)^2\,, \qquad \tilde{t} = (k - p_q)^2 \,, \qquad \tilde{u} = (q - p_q)^2=\hat{u} \,, \label{tildeman}
\ee
 which are the non-collinear generalizations of (\ref{eq:collman}).  
In terms of these variables, we have
\be
\delta\left((k+q-p_q)^2\right) = \delta(\tilde{s} + \tilde{t} + \hat{u} - k^2 + Q^2)\,. \label{deltal}
\ee
Since we only the terms linear in $k_\perp$ are needed (see (\ref{eq:w0})),  we can ignore any $k^2$ dependence in the $\delta$-function as well as in ${\cal H}^{\mu\nu}(k)$.

Given ${\cal H}^{\mu\nu}$, we invert the relation (\ref{eq:decompos}) to obtain 
$S_{1,2}$ as
\be
\begin{split}
& S_1(\tilde{s},\tilde{t},\hat{u},Q^2) = \frac{({\cal H}\cdot T_1)(T_2\cdot T_2) -  ({\cal H}\cdot T_2)(T_1\cdot T_2)}{(T_1\cdot T_1) (T_2\cdot T_2) - (T_1\cdot T_2)^2}\,,\\
& S_2(\tilde{s},\tilde{t},\hat{u},Q^2) = \frac{({\cal H}\cdot T_2)(T_1\cdot T_1) -  ({\cal H}\cdot T_1)(T_1\cdot T_2)}{(T_1\cdot T_1) (T_2\cdot T_2) - (T_1\cdot T_2)^2}\,,
\end{split}
\label{eq:S012}
\ee
where a shorthand notation $T_1 \cdot T_2 \equiv T_1^{\mu\nu} T_{2\mu\nu}$, etc., have been introduced. The hadronic tensor is then written as  
\be
\begin{split}
w^q_{\mu\nu} &= \frac{ e_q^2}{2}\int dx\delta(\hat{s} + \hat{t} + \hat{u} + Q^2)  \Biggl[x\frac{d g^q_T(x)}{d x}\frac{S_\perp \cdot p_q}{p \cdot (q - p_q)}{\cal H}_{\mu\nu}(p) \\
& \qquad \qquad + g^q_T(x) \frac{S_\perp \cdot p_q}{p \cdot (q - p_q)} x\frac{\pd {\cal H}_{\mu\nu}(p)}{\pd x} +  g^q_T(x) S_\perp^\lambda\left(\frac{\pd}{\pd k_\perp^\lambda} {\cal H}_{\mu\nu}(k)\right)_{k = p} \Biggr]\,,
\label{eq:wmain}
\end{split}
\ee
which follows from (40) in \cite{Benic:2021gya} with the partial integration.

We now come to the actual computation of $S_{1,2}$. Because $S_{1,2}$ are functions only of the Mandelstam variables $\tilde{s},\tilde{t},\tilde{u}$, it is enough to calculate them for the special case $k = p$, express the result in terms of $\hat{s},\hat{t},\hat{u}$, and then make the replacements $\hat{s},\hat{t},\hatu \to \tilde{s},\tilde{t},\tilde{u}$.  
This is a tremendous simplification compared to the brute-force calculation in \cite{Benic:2021gya}, especially because ${\cal H}^{\mu\nu}$ involves loop integrals. With this new method, there are only two loop integrals to perform (corresponding to $T^{\mu\nu}_{i=1,2}$) relative to six in  \cite{Benic:2021gya} (corresponding to the basis tensors $\tilde{\mathcal V}^{\mu\nu}_{k=1,2,3,4,8,9}$). Moreover, the integrals themselves are simpler owing to the simpler kinematics $k\to p$. The details of the calculation is relegated to Appendix A. Here we show the final expressions,  
\be
\begin{split}
& \frac{1}{g^4}S_1(\tils,\tilt,\tilde{u},Q^2) = 2C_F\frac{Q^2(\tils+\tilt)}{\tils \tilde{u}^2}\left[\frac{C_A}{\tils + Q^2} - C_F\frac{3(Q^2+\tils)+\tilt}{(\tils + Q^2)^2} - (C_A - 2C_F)\frac{1}{\tilde{u}}\ln\left(-\frac{\tilt}{\tils + Q^2}\right)\right]\,,\\
& \frac{1}{g^4} S_2(\tils,\tilt,\tilde{u},Q^2) = 2C_F\frac{Q^2}{\tils \tilde{u}^2}\left[C_A + C_F\frac{Q^2 + \tils + 3\tilt}{\tils + Q^2} + (C_A -2C_F)\frac{\tilt - \tilde{u}}{\tilde{u}}\ln\left(-\frac{\tilt}{\tils + Q^2}\right)\right]\,,\\
\end{split}
\label{eq:S012final}
\ee
where $N_c=3$ is the number of colors, $C_F=(N_c^2-1)/(2N_c)$ and $C_A = N_c$. 
Importantly, $S_{1,2}$ vanish in the photoproduction limit $Q^2\to 0$. The vanishing is necessary in order to avoid a  kinematic pole in the hard factor ${\cal H}^{\mu\nu}$ at $Q^2=0$  (see \eqref{eq:T12} and  \eqref{eq:decompos}). It then immediately follows that,  since the terms proportional to $q^\mu$ and $q^\nu$ in $T^{\mu\nu}_{1,2}$ vanish when contracted with the lepton tensor, asymmetries from the present mechanism (neglecting the genuine twist-three effects) are proportional to $Q^2$, and hence vanish at $Q^2=0$. Unfortunately, part of our previous result in \cite{Benic:2021gya} does not pass this consistency check.  


Inserting \eqref{eq:wmain} into (\ref{base}), we find 
\be
\begin{split}
\frac{d^6 \Delta\sigma}{d x_B dQ^2 d z_f d q_T^2 d\phi d\chi} &= \frac{\alpha^2_{em}\alpha_s^2 }{16 \pi^2 x_B^2 S_{ep}^2 Q^2}\frac{M_N}{q_T}\sum_q e_q^2\sum_k \calA_k {\cal S}_k\int\frac{dx}{x}\int\frac{dz}{z}D^q_1(z)\delta\left(\frac{q_T^2}{Q^2} - \left(1 - \frac{1}{\hatx}\right)\left(1 - \frac{1}{\hatz}\right)\right)\\
&\qquad \times \left(x^2\frac{d g^q_T(x)}{d x}\Delta \hat{\sigma}_{Dk}^{qq}  +x g^q_T(x) \Delta \hat{\sigma}_k^{qq}\right)\,,
\end{split}
\label{eq:sig1}
\ee
with the variables
\be
\hatx \equiv \frac{x_B}{x}= \frac{Q^2}{\hats + Q^2}\,,\qquad \hatz \equiv \frac{z_f}{z} = \frac{-\hatt}{\hats + Q^2}\,.
\label{eq:sidisvar}
\ee
Following \cite{Benic:2021gya},  we have defined  the partonic cross sections in (\ref{eq:sig1})
\be
\begin{split}
& \frac{g^4 M_N}{q_T}\calS_k \Delta \hat{\sigma}^{qq}_{D k} = \frac{S_\perp \cdot p_q}{p\cdot(q - p_q)}{\cal H}_{\mu\nu}(p)\tilde{\calV}_k^{\mu\nu}\,,\\
& \frac{g^4 M_N}{q_T}\calS_k \Delta \hat{\sigma}^{qq}_k = \left(\frac{S_\perp \cdot p_q}{p \cdot (q - p_q)} x\frac{\pd {\cal H}_{\mu\nu}(p)}{\pd x} + \left[S_\perp^\lambda\frac{\pd  {\cal H}_{\mu\nu}(k)}{\pd k_\perp^\lambda}\right]_{k = p}\right) \tilde{\calV}_k^{\mu\nu}\,,
\end{split}
\label{hard}
\ee
where ${\cal S}_k=\sin (\Phi_S-\chi)$ for $k=1,2,3,4$ and ${\cal S}_k=\cos(\Phi_S-\chi)$ for $k=8,9$. This structure arises  from the property 
\beq
T_{1,2}^{\mu\nu}\tilde{\calV}_{\mu\nu}^k=0\,, \qquad k=1,2,3,4\,, \label{t12v}
\eeq
 so that we have $\Delta\hat{\sigma}^{qq}_{Dk=1,2,3,4}=0$, and the explicit derivatives of the $T_{1,2}^{\mu\nu}$ tensors
\be
\begin{split}
x \frac{\pd {\cal H}^{\mu\nu}(p)}{\pd x} & = 
T_1^{\mu\nu}(p)\left(
2S_1(\hat{s},\hat{t},\hat{u},Q^2) +  x \frac{\pd }{\pd x}S_1(\hat{s},\hat{t},\hat{u},Q^2)\right)\\
& \quad + 
T_2^{\mu\nu}(p)\left(S_2(\hat{s},\hat{t},\hat{u},Q^2) +  x \frac{\pd}{\pd x}S_2(\hat{s},\hat{t},\hat{u},Q^2)\right)\,.\\
\end{split}
\label{eq:dx}
\ee
The $k_\perp$-derivative is evaluated as  
\be
\begin{split}
\left[S_\perp^\lambda \frac{\pd {\cal H}^{\mu\nu}(k)}{\pd k_\perp^\lambda}\right]_{k = p} & = \left[S_\perp^\lambda\frac{\pd T_1^{\mu\nu}(k)}{\pd k_\perp^\lambda}\right]_{k = p} S_1(\hat{s},\hat{t},\hat{u},Q^2) + T_1^{\mu\nu}(p)  \left[S_\perp^\lambda\frac{\pd}{\pd k_\perp^\lambda}S_1(\tilde{s},\tilde{t},\tilde{u},Q^2)\right]_{k = p}\\
&\qquad  + \left[S_\perp^\lambda\frac{\pd T_2^{\mu\nu}(k)}{\pd k_\perp^\lambda}\right]_{k = p} S_2(\hat{s},\hat{t},\hat{u},Q^2) + T_2^{\mu\nu}(p)  \left[S_\perp^\lambda\frac{\pd}{\pd k_\perp^\lambda}S_2(\tilde{s},\tilde{t},\tilde{u},Q^2)\right]_{k = p}\,, 
\end{split}
\label{eq:dkt}
\ee
with 
\be
\begin{split}
&\left[S_\perp^\lambda\frac{\pd T_1^{\mu\nu}(k)}{\pd k_\perp^\lambda}\right]_{k = p} = S_\perp^\mu \epsilon^{\nu p q p_q} + \left(p^\mu + \frac{p\cdot q}{Q^2}q^\mu\right)\epsilon^{\nu S_\perp q p_q}+(\mu\leftrightarrow \nu)
\,,\\
&\left[S_\perp^\lambda\frac{\pd T_2^{\mu\nu}(k)}{\pd k_\perp^\lambda}\right]_{k = p} = \left(p_q^\mu + \frac{p_q\cdot q}{Q^2}q^\mu\right) \epsilon^{\nu S_\perp q p_q} + \left(p_q^\nu + \frac{p_q\cdot q}{Q^2}q^\nu\right)\epsilon^{\mu S_\perp q p_q} = T_2^{\mu\nu}(S_\perp)\,.
\end{split}
\label{t2deri1}
\ee

For the derivatives on $S_{1,2}$, we exploit the fact that  $S_{1,2}$ are functions of the Mandelstam variables as in \cite{Xing:2019ovj}, 
\be
 x \frac{\pd}{\pd x} = (\hat{s} +Q^2)\frac{\pd}{\pd \hat{s}} + \hat{t} \frac{\pd}{\pd \hat{t}}\,,\qquad
 S_\perp^\lambda\frac{\pd}{\pd k_\perp^\lambda} = -2(S_\perp \cdot p_q) \frac{\pd}{\pd \tilde{t}}\,.
\ee
Some care is needed to implement these derivatives in practice.  According to the definition \eqref{tildeman}, the $k_\perp$-dependence of the hard factor is solely encoded  in $\tilde{t}$. However, if one uses the $\delta$-function constraint $\tils+\tilt+\tilde{u}=-Q^2$ to eliminate $\tilde{u}$ or $\tilt$ in $S_{1,2}(\tils,\tilt,\tilde{u})$, it may seem that one could artificially modify the $k_\perp$-dependence. In the original prescription in \cite{Xing:2019ovj}, all the Mandelstam variables are treated as being independent when the hard factor is differentiated in order to avoid this ambiguity.  However, such an approach is not practical in the present case owing to the complexity of the one-loop diagrams. In fact, we have inserted the relation $\tils+\tilt+\tilde{u}=-Q^2$ in  \eqref{eq:S012final} to simplify the calculation. This is nevertheless justified because 
\beq 
\left.\left(\frac{S_\perp\cdot p_q}{p\cdot (q-p_q)}x\frac{\partial}{\partial x} + S_\perp^\lambda\frac{\partial}{\partial k_\perp^\lambda }\right) (\tils+\tilt+\tilde{u}+Q^2)\right|_{k=p}=0,
\label{delcon}
\eeq
and the hard factor (\ref{hard}) is unaffected by this ambiguity. In other words, one can freely switch between $\tilde{u}\leftrightarrow -\tils-\tilt-Q^2$  even before taking the derivatives without changing the final result. In practice, this leads to an enormous simplification especially in Drell-Yan, where there are more diagrams to calculate.     

The hard kernels thus obtained are given by 
\be
\begin{split}
\Delta \hat{\sigma}^{qq}_{D8} & = 2 C_F(1-\hatx)\left(C_A(1-\hatx) + C_F(\hatx-1 - \hatz + 3\hatx\hatz) + (C_A - 2 C_F)(1-2\hatx)\frac{\hatz \ln\hatz}{1-\hatz}\right)\frac{Q}{q_T}\,,\\
\Delta \hat{\sigma}^{qq}_{D9} & = 2C_F (1-\hatx)^2\left(C_A + C_F(1-3\hatz)-(C_A - 2 C_F)(1-2\hatz)\frac{\ln\hatz}{1-\hatz}\right)\frac{Q^2}{q_T^2}\,,\\
\Delta \hat{\sigma}^{qq}_1 & = -3\frac{1-\hatz}{\hatz}\frac{Q}{q_T}\Delta \hat{\sigma}^{qq}_{D8}\,,\\
\Delta \hat{\sigma}^{qq}_2 & = -2\frac{1-\hatz}{\hatz}\frac{Q}{q_T}\Delta \hat{\sigma}^{qq}_{D8}\,,\\
\Delta \hat{\sigma}^{qq}_3 & = -\frac{\hatx\hatz + (1-\hatx)(1-\hatz)}{2(1-\hatx)\hatz}\Delta\hat{\sigma}^{qq}_{D8} - \frac{1-\hatz}{\hatz}\frac{Q}{q_T}\Delta\hat{\sigma}^{qq}_{D9}\,,\\
\Delta \hat{\sigma}^{qq}_4 & = \frac{1-\hatz}{\hatz}\frac{Q}{q_T}\Delta\hat{\sigma}^{qq}_{D8} - \frac{\hatx\hatz + (1-\hatx)(1-\hatz)}{(1-\hatx)\hatz}\Delta\hat{\sigma}^{qq}_{D9}\,,\\
\Delta \hat{\sigma}^{qq}_8 & = C_F \frac{1}{\hatz}\bigg(C_A (1-\hatx)(1-\hatx + \hatz - 4\hatx\hatz) + C_F(-(1 - \hatx)^2 - \hatz(6 - 13 \hatx) (1 - \hatx)  + \hatz^2(3 + \hatx (-11 + 6 \hatx)))\\
&+(C_A - 2 C_F)(3-\hatz + \hatx (-7 + 4 \hatx + 2 \hatz))\frac{\hatz\ln\hatz}{1-\hatz}\bigg)\frac{Q}{q_T}\,,\\
\Delta \hat{\sigma}^{qq}_9 & = -2C_F \frac{1-\hatx}{\hatz}\bigg(C_A(1-\hatx)(1-2\hatz) + C_F(-4 + \hatz(11 - 5 \hatz)  + \hatx (4 - 3\hatz (3 - \hatz)))\\
& + (C_A - 2 C_F)(1 -\hatz (3 - \hatz)  - \hatx (1 - 2 \hatz))\frac{\ln\hatz}{1-\hatz}\bigg)\frac{Q^2}{q_T^2}\,.
\end{split}
\label{final}
\ee
 The same expressions hold for the antiquark initiated channels $\Delta \hat{\sigma}^{\bar{q}\bar{q}}=\Delta\hat{\sigma}^{qq}$. 
 The coefficients $\Delta\hat{\sigma}_{1,\dots,4}$ have been conveniently written as the linear combinations of $\Delta \hat{\sigma}^{qq}_{D8}$ and $\Delta \hat{\sigma}^{qq}_{D9}$. This is thanks to the fact that the relations in \eqref{hard} for $k=1,2,3,4$ take a special form 
\be
\begin{split}
\frac{g^4 M_N}{q_T}\calS_k \Delta \hat{\sigma}^{qq}_k 
& = \left\{\left[S_\perp^\lambda \frac{\pd T_{1,\mu\nu}(k)}{\pd k_\perp^\lambda}\right]_{k = p}S_1(\hats,\hatt,\hatu,Q^2) + \left[S_\perp^\lambda \frac{\pd T_{2,\mu\nu}(k)}{\pd k_\perp^\lambda}\right]_{k = p}S_2(\hats,\hatt,\hatu,Q^2) \right\}\tilde{\calV}_k^{\mu\nu}\,,
\end{split}
\label{eq:hardsimple}
\ee
so the hard coefficients $\Delta \hat{\sigma}^{qq}_k$, $k = 1,\dots, 4$ are the linear combinations of $S_{1,2}$ just like $\Delta \hat{\sigma}^{qq}_{D8,9}$ (see Eq.~\eqref{hard}). 
We also find $\Delta \hat{\sigma}^{qq}_2 = 2\Delta \hat{\sigma}^{qq}_1/3$. It follows by noticing $\tilde{\calV}_2^{\mu\nu} = \frac{2}{3} \tilde{\calV}_1^{\mu\nu}+Z^\mu Z^\nu + g^{\mu\nu}$, where the extra terms, $Z^\mu Z^\nu$ and $g^{\mu\nu}$, decouple owing to the QED Ward identity and to $g_{\mu\nu} T_{1,2}^{\mu\nu}(k) = 0$, respectively.

Compared with our previous result in \cite{Benic:2021gya}, an exact agreement   for $\Delta \hat{\sigma}^{qq}_{2}$ and $\Delta \hat{\sigma}^{qq}_{Dk}$ ($k=8,9$) has been confirmed. 
However, for $\Delta \hat{\sigma}^{qq}_{1,3,4,8,9}$, only the logarithmic terms agree and the remainders are different. In particular, $\Delta \hat{\sigma}_1^{qq}$ in \cite{Benic:2021gya}  does not vanish in the photoproduction limit $Q=0$, in contradiction with the general observation made below (\ref{eq:S012final}). 


As a nontrivial check of our revised formulas in (\ref{final}), it is useful and instructive to   make a connection to the {\it longitudinal} single spin asymmetry recently discussed in \cite{Abele:2022spu}, which is associated with a longitudinally polarized proton with the spin vector $S^\mu \approx \delta^\mu_+S^+$. In this case, the hadronic tensor is given by 
\be
w^q_{\mu\nu} = 
\frac{1}{2}\frac{S^+}{P^+} e_q^2\int\frac{dx}{x} \Delta q(x) \delta((p+q-p_q)^2){\rm Tr}[\gamma_5\Slash p S^{(q)}_{\mu\nu}(p)] \,,
\label{eq:longw}
\ee
where $\Delta q(x)$ is the polarized quark PDF. The hard factor $S^{(q)}_{\mu\nu}$ is in fact exactly the same as for a transverse SSA \eqref{eq:w0}. However, there is no  parton transverse momentum $k_\perp$ involved in the longitudinal case; namely, the observable is purely twist-two.  Inserting (\ref{eq:longw}) into (\ref{base}) and using \eqref{eq:Hmunu} and \eqref{eq:decompos}, we find that only the $k=8,9$ harmonics survive, 
\be
\begin{split}
\frac{d^6 \Delta\sigma}{d x_B dQ^2 d z_f d q_T^2 d\phi d\chi} & = \frac{\alpha_{em}^2 \alpha_s^2}{16 \pi^2 x_B^2 S_{ep}^2 Q^2} \sum_qe_q^2\int \frac{dz}{z}\frac{dx}{x}D^q_1(z)  \Delta q(x) \delta\left(\frac{q_T^2}{Q^2} - \left(1 - \frac{1}{\hatx}\right)\left(1 - \frac{1}{\hatz}\right)\right)\\
&  \qquad \times \left(-\calA_8 \Delta\hat{\sigma}^{qq}_{L8} + \calA_9 \Delta\hat{\sigma}^{qq}_{L9} \right),
\end{split}
\ee
where 
\be
\begin{split}
& g^4\Delta\hat{\sigma}^{qq}_{L8} \equiv -{\cal H}_{\mu\nu}(p)\tilde{\calV}_8^{\mu\nu}\,,\\
& g^4\Delta\hat{\sigma}^{qq}_{L9} \equiv {\cal H}_{\mu\nu}(p)\tilde{\calV}_9^{\mu\nu}\,.\\
\end{split}
\label{eq:hardl}
\ee
Switching the variables as  $ d q_T^2 = d P_{hT}^2/z_f^2$, $d Q^2 = x_B S_{ep} d y$ and integrating over $\chi$,
we arrive at 
\be
\begin{split}
\frac{d^5 \Delta\sigma}{d x_B dy d z_f d P_{hT}^2 d\phi_h} & = \frac{\pi\alpha_{em}^2}{x_Bz_f Q^4}\frac{y}{1-\varepsilon} \sum_qe_q^2 \int \frac{d\hatx}{\hatx}\Delta q\left(\frac{x_B}{\hatx}\right) \frac{d\hatz}{\hatz}D^q_1\left(\frac{z_f}{\hatz}\right)\delta\left(\frac{q_T^2}{Q^2} - \left(1 - \frac{1}{\hatx}\right)\left(1 - \frac{1}{\hatz}\right)\right)\\
&\times\left(\frac{\alpha_s}{2\pi}\right)^2 \left[\sqrt{2\varepsilon(1+\varepsilon)} \Delta\hat{\sigma}^{qq}_{L8}\sin\phi_h + \varepsilon \Delta\hat{\sigma}^{qq}_{L9} \sin(2\phi_h)\right]\,.
\end{split}
\label{eq:longquark}
\ee
Comparing the above expression to (2) and (4) in \cite{Abele:2022spu}, we identify  $e_q^2\Delta\hat{\sigma}^{qq}_{L8} = C_{UL}^{\sin\phi_h,q\to q}$ and $e_q^2\Delta\hat{\sigma}^{qq}_{L9} = C_{UL}^{\sin 2\phi_h, q\to q}$, with $C_{UL}$'s being given in (18) and (19) in \cite{Abele:2022spu}. 
We find a complete agreement with our result  including the sign. 
Indeed, it follows from \eqref{hard} and  \eqref{eq:hardl} that 
\be
\Delta\hat{\sigma}^{qq}_{D8,9} 
=\pm 2(1-\hatx)\Delta\hat{\sigma}^{qq}_{L8,9} \,, \label{L89}
\ee
namely, the hard coefficients in the longitudinal SSA and (part of) transverse SSA are proportional to each other as already noticed in \cite{Abele:2022spu}.

\subsection{Quark-initiated, gluon-fragmenting  channel}

Next we investigate the gluon fragmentation channel $\gamma^*q\to qg$ with $g\to h$. 
In this case, we write  $l_1^\mu \equiv k^\mu+q^\mu-p^\mu_q = P_h^\mu/z$, instead of $p_q^\mu = P_h^\mu/z$. 
It is convenient to employ the same tensor decomposition for the hard part as in \eqref{eq:decompos}, because the hard factors $S_{1,2}$  can then be  simply obtained from \eqref{eq:S012final} by the crossing $\tilt \leftrightarrow \tilde{u}$. For the cross section, we apply the replacement $p_q \to l_1$ everywhere in \eqref{eq:sig1} to get
\be
\begin{split}
\frac{d^6 \Delta\sigma}{d x_B dQ^2 d z_f d q_T^2 d\phi d\chi} &= \frac{\alpha^2_{em}\alpha_s^2 }{16 \pi^2 x_B^2 S_{ep}^2 Q^2}\frac{M_N}{q_T}\sum_q e_q^2\sum_k \calA_k {\cal S}_k\int\frac{dx}{x}\int\frac{dz}{z}D^g_1(z)\delta\left(\frac{q_T^2}{Q^2} - \left(1 - \frac{1}{\hatx}\right)\left(1 - \frac{1}{\hatz}\right)\right)
\\
&\qquad \times \left(x^2\frac{d g^q_T(x)}{d x}\Delta \hat{\sigma}_{Dk}^{qg}  +x g^q_T(x) \Delta \hat{\sigma}_k^{qg}\right), 
\end{split}
\label{eq:sigg}
\ee
where the hard coefficients are defined as
\be
\begin{split}
& \frac{g^4 M_N}{q_T}\calS_k \Delta \hat{\sigma}^{qg}_{D k} = \frac{S_\perp \cdot l_1}{p\cdot(q - l_1)}{\cal H}_{\mu\nu}(p)\tilde{\calV}_k^{\mu\nu}\,,\\
& \frac{g^4M_N}{q_T} \calS_k \Delta \hat{\sigma}^{qg}_k = \left(\frac{S_\perp \cdot l_1}{p \cdot (q - l_1)} x\frac{\pd}{\pd x}{\cal H}_{\mu\nu}(p) + \left[S_\perp^\lambda\frac{\pd}{\pd k_\perp^\lambda} {\cal H}_{\mu\nu}(k)\right]_{k = p}\right) \tilde{\calV}_k^{\mu\nu}\,.
\end{split}
\label{eq:hard}
\ee
To perform the above contractions, the expressions for $\tilde{\calV}_k^{\mu\nu}$ in the partonic variables need to be adjusted accordingly by applying $p_q^\mu \to l_1^\mu$ to \eqref{eq:txyz}.
To compute the derivatives, we write $T_{1,2}^{\mu\nu}(k)$ in terms of $l_1$ by substituting  $p_q = k + q - l_1$, 
\be
\begin{split}
& T_1^{\mu\nu}(k) = -\left(k^\mu +\frac{k\cdot q}{Q^2}q^\mu\right)\epsilon^{\nu kq l_1} - \left(k^\nu +\frac{k\cdot q}{Q^2}q^\nu\right)\epsilon^{\mu kq l_1}\,,\\
& T_2^{\mu\nu}(k) = -\left(k^\mu - l_1^\mu +\frac{(k-l_1)\cdot q}{Q^2}q^\mu\right)\epsilon^{\nu kq l_1} - \left(k^\nu - l_1^\nu +\frac{(k-l_1)\cdot q}{Q^2}q^\nu\right)\epsilon^{\mu kq l_1}\,,\\
\end{split}
\ee
which lead to 
\be
\begin{split}
& x\frac{\pd T_1^{\mu\nu}(p)}{\pd x} = p^\lambda \frac{\pd T_1^{\mu\nu}(p)}{\pd p^\lambda} = 2 T_1^{\mu\nu}(p)\,,\\
& x\frac{\pd T_2^{\mu\nu}(p)}{\pd x} = p^\lambda \frac{\pd T_2^{\mu\nu}(p)}{\pd p^\lambda} = T_1^{\mu\nu}(p) + T_2^{\mu\nu}(p)\,,
\end{split}
\label{eq:xdt12}
\ee
and
\be
\begin{split}
\left[S_\perp^\lambda\frac{\pd T_1^{\mu\nu}(k)}{\pd k_\perp^\lambda}\right]_{k = p} & = - S_\perp^\mu \epsilon^{\nu p q l_1} 
- \left(p^\mu + \frac{p\cdot q}{Q^2}q^\mu\right)\epsilon^{\nu S_\perp q l_1}  +(\mu\leftrightarrow \nu)
\,,\\
\left[S_\perp^\lambda\frac{\pd T_2^{\mu\nu}(k)}{\pd k_\perp^\lambda}\right]_{k = p} &= - S_\perp^\mu \epsilon^{\nu p q l_1} 
 - \left(p^\mu - l_1^\mu + \frac{(p-l_1)\cdot q}{Q^2}q^\mu\right)\epsilon^{\nu S_\perp q l_1} +(\mu\leftrightarrow \nu)
 \,. \label{t2deri2}
\end{split}
\ee
The explicit expressions for the hard coefficients are found to be
\be
\begin{split}
\Delta\hat{\sigma}_{D8}^{qg} & = -2C_F\frac{(1-\hatx)(1-\hatz)}{\hatz}\left[C_A(1-\hatx) + C_F(-3 \hatx\hatz + 4\hatx + \hatz - 2) + (C_A-2C_F)(1-2\hatx)\frac{(1-\hatz)\ln(1-\hatz)}{\hatz}\right] \frac{Q}{q_T}\,,\\
\Delta\hat{\sigma}_{D9}^{qg} &= 2C_F \frac{(1-\hatx)^2(1-\hatz)^2}{\hatz^2}\left(C_A - C_F(2-3\hatz) + (C_A - 2 C_F)(1-2\hatz)\frac{\ln(1-\hatz)}{\hatz}\right)\frac{Q^2}{q_T^2}\,,\\
\Delta\hat{\sigma}_1^{qg} &= -3\frac{1-\hatz}{\hatz}\frac{Q}{q_T}\Delta\hat{\sigma}_{D8}^{qg}\,,\\
\Delta\hat{\sigma}_2^{qg} &=-2\frac{1-\hatz}{\hatz}\frac{Q}{q_T}\Delta\hat{\sigma}_{D8}^{qg}\,, \\
\Delta\hat{\sigma}_3^{qg} &= -\frac{\hatx\hatz + (1-\hatx)(1-\hatz)}{2(1-\hatx)\hatz}\Delta\hat{\sigma}^{qg}_{D8} - \frac{1-\hatz}{\hatz}\frac{Q}{q_T}\Delta\hat{\sigma}^{qg}_{D9}\,,\\
\Delta\hat{\sigma}_4^{qg} & = \frac{1-\hatz}{\hatz}\frac{Q}{q_T}\Delta\hat{\sigma}^{qg}_{D8} - \frac{\hatx\hatz + (1-\hatx)(1-\hatz)}{(1-\hatx)\hatz}\Delta\hat{\sigma}^{qg}_{D9}\,,\\
\Delta\hat{\sigma}_8^{qg} &= C_F \frac{1-\hatz}{\hatz^2}\bigg(C_A(1-\hatx)(4 \hatx \hatz-7 \hatx-\hatz+3) + C_F((\hatx (6 \hatx-11)+3) \hatz^2+\hatx (7 \hatx-6) \hatz+2 (11-8 \hatx) \hatx+\hatz-6)\\
& + (C_A - 2C_F)(\hatx (8 \hatx-2 \hatz-11)+\hatz+3)\frac{(1-\hatz)\ln(1-\hatz)}{\hatz}\bigg)\frac{Q}{q_T}\,,\\
\Delta \hat{\sigma}^{qg}_9 & = 2C_F \frac{(1-\hatx)(1-\hatz)^2}{\hatz^3}\bigg(2 C_A(-\hatx \hatz+\hatx+\hatz-1) + C_F(\hatx (3 \hatz (\hatz+2)-4)-\hatz (5 \hatz+4)+4)\\
& + (C_A - 2C_F)(\hatx (2-4 \hatz)+\hatz (\hatz+3)-2)\frac{\ln(1-\hatz)}{\hatz}\bigg)\frac{Q^2}{q_T^2}\,.\\
\end{split}
\label{q->g}
\ee
It is seen that the coefficients $\Delta\hat{\sigma}^{qg}_2$ and $\Delta\hat{\sigma}^{qg}_{Dk}$ with $k=8,9$ match exactly (56) in \cite{Benic:2021gya}. As to $\Delta\hat{\sigma}^{qg}_{1,3,4,8,9}$, only the logarithmic terms agree, with the remainder being different. 
We have explicitly cross-checked that starting from our results for $\calH_{\mu\nu}$ and calculating  the longitudinal SSAs, the resultant hard coefficients (proportional to $\Delta \hat{\sigma}_{D8,9}^{qg}$, cf., (\ref{L89})) agree exactly with those obtained in \cite{Abele:2022spu}. Note that $\Delta\hat{\sigma}_{D8,9}$ in (\ref{final}) and (\ref{q->g})  are related by the crossing symmetry  (up to a minus sign for $\Delta \hat{\sigma}_{D8}$) \cite{Abele:2022spu}
\beq
\hat{t}\to \hat{u}, \qquad \hat{z}\to 1-\hat{z}, \qquad q_T\propto \sqrt{\frac{1-\hat{z}}{\hat{z}}} \to \frac{\hat{z}}{1-\hat{z}}q_T. 
\label{eq:cross}
\eeq
Interestingly, this symmetry does not hold for $\Delta\hat{\sigma}_{1,2,3,4,8,9}$, since a transverse SSA involves the derivative of a hard factor. More specifically, the derivatives of $T_2$ for $k=1,2,3,4$ in  
(\ref{t2deri1}) and (\ref{t2deri2}) cannot be exactly mapped onto each other by the replacement $p_q\to p+q-l_1$. As for $k=8,9$,  $\Delta\hat{\sigma}_{8,9}$ contain the derivatives of $S_{1,2}$.

\section{SSA in SIDIS, Gluon-initiated channel}
\label{sec:sidisg}

In the previous section, we have assumed that the transversely polarized proton emits a quark (or an antiquark). As it emits a gluon, there is a new contribution to a SSA proportional to the ${\cal G}_{3T}(x)$ distribution \cite{Ji:1992eu,Hatta:2012jm}\footnote{Note the absence of the factor $1/2$ in our definition  of ${\cal G}_{3T}$, because of which the cross section formulas obtained below have different overall coefficients compared to those in other channels. Nevertheless, we stick to the original normalization in our previous works.  }  
\beq
\int \frac{d\lambda}{2\pi}e^{ix\lambda}\langle PS_\perp|F^{n\alpha}(0)WF^{n\beta}(\lambda n)|PS_\perp\rangle =-\frac{i}{2}\frac{S^+}{P^+}x\Delta G(x)\epsilon^{\alpha\beta Pn}- ix{\cal G}_{3T}(x)\epsilon^{\alpha\beta S_\perp n}+\cdots,
\eeq
which is the gluonic analog of the $g_T(x)$ distribution. The relevant diagrams that provide an imaginary phase have been identified and computed in \cite{Benic:2021gya}. In this section we revisit the calculation following the new approach introduced in the previous section.  

\subsection{Gluon-initiated quark-fragmenting channel}

Our starting point is  (32) in \cite{Benic:2021gya}
\beq
w^g_{\mu\nu} &=& i e_q^2  \int \frac{dx}{x}\mathcal{G}_{3T}(x)\delta((p+q-p_q)^2)\epsilon^{n\alpha\beta S_\perp} S^{(g)\alpha'\beta'}_{\mu\nu}(p)\omega_{\alpha'\alpha}\omega_{\beta'\beta}\nn
&& - i e_q^2 \int dx \mathcal{G}_{3T}(x)\left(g_\perp^{\beta\lambda} \epsilon^{\alpha P n S_\perp} -g_\perp^{\alpha\lambda} \epsilon^{\beta P n S_\perp} \right)\frac{\partial}{\partial k_\perp^\lambda} \left(\delta((k+q-p_q)^2)S^{(g)}_{\mu\nu\alpha\beta}(k)\right)_{k = p}\,,
\label{eq:Wglue}
\eeq
where $g_\perp^{\mu\nu}=g^{\mu\nu}-P^\mu n^\nu - n^\mu P^\nu$ and $\omega^{\mu\nu}=g^{\mu\nu}-P^\mu n^\nu$, and $\alpha$ and $\beta$ are the polarization indices of the gluon in the complex-conjugate amplitude and in the amplitude, respectively. 
The explicit expression of the hard factor  $S^{(g)}_{\mu\nu\alpha\beta}$ is referred to (58)-(61) of  \cite{Benic:2021gya}. 
We rewrite (\ref{eq:Wglue}) in a way similar to (\ref{eq:w0}). 
For the first term, we use the identity 
\be
\epsilon^{n\alpha\beta S_\perp} S^{(g)\alpha'\beta'}_{\mu\nu} \omega_{\alpha'\alpha}\omega_{\beta'\beta} = \epsilon^{n P \beta S_\perp} S^{(g)}_{\mu\nu n \beta} - \epsilon^{n P \alpha S_\perp} S^{(g)}_{\mu\nu \alpha n} = -\epsilon^{\alpha\beta S_\perp n}S^{(g)}_{\mu\nu\alpha\beta}\,,
\ee
where the second equality follows from the Schouten identity 
\be
(P\cdot n) \epsilon^{\alpha\beta S_\perp n} = n^\alpha \epsilon^{P\beta S_\perp n} + n^\beta \epsilon^{\alpha P S_\perp n} .
\ee
Applying also the Schouten identity  to the second term of \eqref{eq:Wglue}, we obtain a compact formula 
\be
w^g_{\mu\nu} = -i e_q^2  \int \frac{dx}{x}\mathcal{G}_{3T}(x)S_\perp^\lambda\frac{\pd}{\pd k_\perp^\lambda}  \left(\delta((k+q-p_q)^2)\epsilon^{\alpha\beta k n}S^{(g)}_{\mu\nu\alpha\beta}(k)\right)_{k = p}\,.
\label{eq:Wglue2}
\ee
We then notice that in the present approximation $k^2\approx 0$, it is permissible to write 
\beq
(k\cdot q) \epsilon^{\alpha\beta k n}S^{(g)}_{\mu\nu\alpha\beta}(k)
= (k\cdot n) \epsilon^{\alpha\beta k q}S_{\mu\nu\alpha\beta}^{(g)}(k), \label{schou}
\eeq
where we have employed the Schouten identity and the Ward identity $k^\alpha S_{\mu\nu\alpha\beta}^{(g)}=k^\beta S_{\mu\nu\alpha\beta}^{(g)}=0$ again. This allows us to write 
\be
w^g_{\mu\nu} =
 e_q^2  \int dx\mathcal{G}_{3T}(x)S_\perp^\lambda\frac{\pd}{\pd k_\perp^\lambda}  \left(\delta((k+q-p_q)^2)\frac{-2i}{\tilde{s}+Q^2}\epsilon^{\alpha\beta kq}S^{(g)}_{\mu\nu\alpha\beta}(k)\right)_{k=p} \,,
\label{eq:Wglue3}
\ee
for $k\cdot n/k\cdot q=2x/(\tilde{s}+Q^2)$ commutes with the  $k_\perp$-derivative. The form (\ref{eq:Wglue3}) is convenient with the hard part in the brackets consisting only of the vectors $k,q$ and $p_q$, so the same argument leading to \eqref{eq:decompos} holds; we employ the trick (\ref{schou}) to eliminate the vector $n$ from the antisymmetric tensor. 
We then have the parametrization  
\beq 
 {\cal H}^{(g)}_{\mu\nu}(k) &\equiv & \frac{-2i}{\tilde{s}+Q^2}\epsilon^{\alpha\beta kq}S^{(g)}_{\mu\nu\alpha\beta}(k) 
 \nn
 &=& T_{1\mu\nu}(k) S^{(g)}_1(\tilde{s},\tilde{t},\hat{u},Q^2) + T_{2\mu\nu}(k) S^{(g)}_2(\tilde{s},\tilde{t},\hat{u},Q^2)\,,
\eeq
with the same tensors $T_{1,2}^{\mu\nu}$ as in (\ref{eq:T12}). 

Analogous to the quark-initiated case, \eqref{eq:Wglue3} yields the contribution to the cross section 
\be
\begin{split}
\frac{d^6 \Delta\sigma}{d x_B dQ^2 d z_f d q_T^2 d\phi d\chi} &= \frac{\alpha^2_{em}\alpha_s^2}{8 \pi^2 x_B^2 S_{ep}^2 Q^2}\frac{M_N}{q_T}\sum_qe_q^2\sum_k \calA_k {\cal S}_k\int\frac{dx}{x}\int\frac{dz}{z}D^q_1(z)\delta\left(\frac{q_T^2}{Q^2} - \left(1 - \frac{1}{\hatx}\right)\left(1 - \frac{1}{\hatz}\right)\right)
\\
&\qquad \times \left(x^2\frac{d \mathcal{G}_{3T}(x)}{d x} \Delta\hat{\sigma}_{Dk}^{gq} + x \mathcal{G}_{3T}(x)\Delta \hat{\sigma}_k^{gq}\right), 
\end{split}
\label{eq:sigglue}
\ee
where the summation is over the quark and antiquark flavors $q$. 
The  hard coefficients are defined as
\be
\begin{split}
& \frac{g^4 M_N}{q_T}\calS_k \Delta \hat{\sigma}^{gq}_{D k} = \frac{S_\perp \cdot p_q}{p\cdot(q - p_q)}{\cal H}^{(g)}_{\mu\nu}(p)\tilde{\calV}_k^{\mu\nu}\,,\\
& \frac{g^4 M_N}{q_T}\calS_k \Delta \hat{\sigma}^{gq}_k = \left(\frac{S_\perp \cdot p_q}{p \cdot (q - p_q)} x\frac{\pd {\cal H}^{(g)}_{\mu\nu}(p)}{\pd x} + \left[S_\perp^\lambda\frac{\pd  {\cal H}^{(g)}_{\mu\nu}(k)}{\pd k_\perp^\lambda}\right]_{k = p}\right) \tilde{\calV}_k^{\mu\nu}\,.
\end{split}
\label{eff}
\ee

As in \cite{Benic:2021gya}, we evaluate $S^{(g)}_{1,2}$ analytically. Again thanks to the new method, the calculation is much simpler. We omit the detailed derivation, since it is similar to the quark case as explained in Appendix A. The result is given by
\be
\begin{split}
\frac{1}{g^4}S_1^{(g)}(\tils,\tilt,\hatu,Q^2) &= 2T_R (C_A-2C_F)\frac{Q^2}{\tilt^3 \hatu^3}\Bigg[\frac{\tilt \hatu}{(\tils + Q^2)^2} \left((\tilt^2 - \hatu^2)(\tils + \tilt) + 2\tilt^2 \hatu\right) \\
& + \tilt^2 (\tilt + \tils)\ln\left(-\frac{\tilt}{Q^2 + \tils}\right) + \hatu^2 (\tilt - \tils)\ln\left(-\frac{\hatu}{Q^2 + \tils}\right)\Bigg] \,,\\
\frac{1}{g^4}S_2^{(g)}(\tils,\tilt,\hatu,Q^2) &= 2T_R(C_A-2C_F)\frac{Q^2(\tils + Q^2)}{\tilt^3 \hatu^3}\Bigg[\frac{\tilt\hatu}{(\tils + Q^2)^2}(\tilt^2 + \hatu^2)\\
& +\tilt^2\ln\left(-\frac{\tilt}{Q^2 + \tils}\right) + \hatu^2\ln\left(-\frac{\hatu}{Q^2 + \tils}\right)\Bigg]\,,
\end{split}
\label{eq:S012g}
\ee
from which we arrive at
\be
\begin{split}
\Delta \hat{\sigma}^{gq}_{D8} &= 2T_R(C_A - 2C_F)\frac{(1-\hatx)^2}{\hatz^2}\Bigg[\hatx \hatz (1 - 2 \hatz) - (1 - \hatx)\ln(1 - \hatz) + (1-\hatx)\frac{\hatz\ln\hatz}{1-\hatz} \Bigg]\frac{Q}{q_T}\,,\\
\Delta \hat{\sigma}^{gq}_{D9} &= -2T_R(C_A - 2C_F)\frac{(1-\hatx)^3}{\hatz^2}\Bigg[2\hatz(1-\hatz)-1 - \frac{(1-\hatz)\ln(1-\hatz)}{\hatz} - \frac{\hatz\ln\hatz}{1-\hatz}\Bigg]\frac{Q^2}{q_T^2}\,,\\
\Delta \hat{\sigma}^{gq}_1 &= -3 \frac{1-\hatz}{\hatz}\frac{Q}{q_T}\Delta \hat{\sigma}^{gq}_{D8}\,,\\
\Delta \hat{\sigma}^{gq}_2 &= -2 \frac{1-\hatz}{\hatz}\frac{Q}{q_T}\Delta \hat{\sigma}^{gq}_{D8}\,,\\
\Delta\hat{\sigma}_3^{gq} &= -\frac{\hatx\hatz + (1-\hatx)(1-\hatz)}{2(1-\hatx)\hatz}\Delta\hat{\sigma}^{gq}_{D8} - \frac{1-\hatz}{\hatz}\frac{Q}{q_T}\Delta\hat{\sigma}^{gq}_{D9}\,,\\
\Delta\hat{\sigma}_4^{gq} & = \frac{1-\hatz}{\hatz}\frac{Q}{q_T}\Delta\hat{\sigma}^{gq}_{D8} - \frac{\hatx\hatz + (1-\hatx)(1-\hatz)}{(1-\hatx)\hatz}\Delta\hat{\sigma}^{gq}_{D9}\,,\\
\Delta \hat{\sigma}^{gq}_8 &= T_R(C_A - 2C_F)\frac{1-\hatx}{\hatz^3}\Bigg[\hatz \bigl(4 + \hatx (-11 + \hatz + 2 \hatz^2 + \hatx (7 - 4 \hatz^2))\bigr)\\
&\quad  + (1 - \hatx) (3 - 3 \hatz - \hatx (5 - 2 \hatz)) \ln(1-\hatz)+(1 - \hatx) (-1 + \hatx + 3 \hatz - 2 \hatx \hatz)\frac{\hatz\ln\hatz}{1-\hatz}\Bigg]\frac{Q}{q_T}\,,\\
\Delta \hat{\sigma}^{gq}_9 &= 2T_R(C_A - 2C_F)\frac{(1-\hatx)^2}{\hatz^3}\Bigg[-2 + 3 \hatz - (1 - \hatz) (-2 \hatx + \hatx \hatz - 2\hatz^2 (1 - \hatx))\\
& \quad -(2 - \hatx (2 - \hatz) - 2 \hatz)\frac{(1-\hatz)\ln(1-\hatz)}{\hatz}-(1 - \hatx - \hatz(2 - \hatx)) \frac{\hatz\ln\hatz}{1-\hatz}\Bigg]\frac{Q^2}{q_T^2}\,,
\end{split}
\label{eq:coefsgq}
\ee
with $T_R=1/2$. 
It is noticed that the hard coefficients $\Delta\hat{\sigma}_{2,9}^{gq}$ and $\Delta\hat{\sigma}_{Dk}^{gq}$, $k=8,9$, agree exactly with (66) and (67) in \cite{Benic:2021gya} (after the adjustment of the conventional factor of 2). The logarithmic terms of the coefficients $\Delta\hat{\sigma}_{1,3,4,8}^{gq}$ agree with (67) in \cite{Benic:2021gya}, but the remainder is different.

\subsection{Gluon-initiated antiquark-fragmenting channel}

The hard coefficients in the antiquark fragmenting channel $g\to \bar{q} \to h$ are written as (cf. (\ref{eff}))
\be
\begin{split}
& \frac{g^4 M_N}{q_T}\calS_k \Delta \hat{\sigma}^{g\bar{q}}_{D k} = \frac{S_\perp \cdot l_1}{p\cdot(q - l_1)}{\cal H}^{(g)}_{\mu\nu}(p)\tilde{\calV}_k^{\mu\nu}\,,\\
& \frac{g^4M_N}{q_T} \calS_k \Delta \hat{\sigma}^{g\bar{q}}_k = \left(\frac{S_\perp \cdot l_1}{p \cdot (q - l_1)} x\frac{\pd}{\pd x}{\cal H}^{(g)}_{\mu\nu}(p) + \left[S_\perp^\lambda\frac{\pd}{\pd k_\perp^\lambda} {\cal H}^{(g)}_{\mu\nu}(k)\right]_{k = p}\right) \tilde{\calV}_k^{\mu\nu}\,,
\end{split}
\label{eq:hardgqbar}
\ee
where $l^\mu_1=P_h^\mu/z$ and the crossing $\tilde{t}\leftrightarrow \tilde{u}$ is implied in ${\cal H}_{\mu\nu}^{(g)}$. 
Performing the derivatives as in \eqref{eq:xdt12} and \eqref{t2deri2}, we obtain the hard coefficients
\be
\begin{split}
 \Delta\hat{\sigma}_{Dk}^{g\bar{q}}  =  \Delta\hat{\sigma}_{Dk}^{gq}\,, \qquad 
 \Delta\hat{\sigma}_k^{g\bar{q}} =  \Delta\hat{\sigma}_k^{gq}\,.\\
\end{split}
\label{eq:coefsgqbar}
\ee
The result for $\Delta\hat{\sigma}_{D8,9}^{g\bar{q}}$ can be inferred in a way which manifests the crossing symmetry under {\eqref{eq:cross},   
\beq
\left. \Delta\hat{\sigma}_{D8,9}^{g\bar{q}}(\hat{x},\hat{z}) = \mp\Delta\hat{\sigma}_{D8,9}^{gq}(\hat{x},1-\hat{z})\right|_{q_T\to \hat{z}q_T/(1-\hat{z})} .
\eeq

\subsection{Comparison to the longitudinal polarization case}

It is again illustrative to make a comparison to the case with a longitudinally polarized proton, where the hadronic tensor takes the form (cf. \eqref{eq:Wglue2})
\beq
w^g_{\mu\nu} =-i\frac{e_q^2}{2}\frac{S^+}{P^+}\int\frac{dx}{x}\Delta G(x)\epsilon^{\alpha\beta P n}S^{(g)}_{\mu\nu\alpha\beta}(p) 
&=& \frac{e_q^2}{2}\frac{S^+}{P^+}\int \frac{dx}{x}\Delta G(x){\cal H}^{(g)}_{\mu\nu}(p) 
.
\eeq
 Inserting ${\cal H}^{(g)}_{\mu\nu}(p)$ derived above, we find a formula  analogous  to  \eqref{eq:longquark} with the replacement $\Delta q(x)\to \Delta G(x)$. It is seen  from (\ref{eff}) that the hard coefficients $\Delta\hat{\sigma}_{L8,9}^{gq}$ are related to $\Delta\hat{\sigma}^{gq}_{D8,9}$ via  
 \be
 \begin{split}
 \Delta \hat{\sigma}_{D8,9}^{gq}= \pm 2(1-\hat{x})\Delta\hat{\sigma}^{gq}_{L8,9}.
 \label{eq:crossg}
 \end{split}
 \ee
  We observe a perfect agreement $\Delta\hat{\sigma}^{gq}_{L8} = C_{UL}^{\sin\phi_h}$ and $\Delta\hat{\sigma}^{gq}_{L9} = C_{UL}^{\sin(2\phi_h)}$, where $C_{UL}$'s   are given in (18) and (19) (with $g\to q$)  in \cite{Abele:2022spu}. 
In the antiquark fragmentation channel, 
\beq
\left.  \Delta\hat{\sigma}^{g\bar{q}}_{L8,9}(\hat{x},\hat{z})=\Delta\hat{\sigma}^{gq}_{L8,9}(\hat{x},\hat{z}) = \mp \Delta\hat{\sigma}^{gq}_{L8,9}(\hat{x},1-\hat{z})\right|_{q_T\to \frac{\hat{z}}{1-\hat{z}}q_T}  , 
\eeq
also confirms the consistency with \cite{Abele:2022spu}.

\section{SSA in Drell-Yan}

The new mechanism to generate transverse SSAs in SIDIS described in the previous sections has a direct counterpart in Drell-Yan, as the two processes are related by crossing symmetry.   
In this section, we show that the $g_T(x)$ and ${\cal G}_{3T}(x)$ distributions generate SSAs in the polarized Drell-Yan $pp^\uparrow \to l^+l^-X$ or $\bar{p}p^\uparrow \to l^+l^-X$. The theoretical framework is entirely similar to that for SIDIS, but there are practical differences owing to the timelike/spacelike nature of a virtual photon.

\subsection{Setup}

We first review the basics of the unpolarized Drell-Yan $pp\to \gamma^* X\to l^+l^-X$ and set up our notations. 
The important subprocesses include  the $q\bar{q}$ annihilation  $q(p)+\bar{q}(p')\to \gamma^*(q)g(p+p'-q)$ and the Compton scattering  
$q(p)+g(p')\to \gamma^*(q) q(p+p'-q)$, followed by the subsequent decay of the timelike  virtual photon $\gamma^*(q)\to l^+(l_1)l^-(l_2)$ with $q^2=Q^2>0$.  Here we assign $p^\mu=xP^\mu$, $p'^\mu=x'P'^\mu$, $p^2=p'^2=0$, $l_1^2=l_2^2=0$, and the center-of-mass energy $s=(P+P')^2\approx 2P\cdot P'$.  We assume that the photon transverse momentum is large, $q_\perp\gg \Lambda_{\rm QCD}$, so that collinear factorization is applicable, but do not assume a particular ordering between $q_\perp$ and $Q$. The partonic Mandelstam variables are defined as 
\beq
\hat{s}=(p+p')^2=xx's, \quad \hat{t}=(p-q)^2, \quad \hat{u}=(p'-q)^2. \label{dyman}
\eeq
We parameterize the virtual photon momentum in the proton-proton center-of-mass frame as 
\beq
q^\mu=(q^+,\vec{q}_\perp,q^-), \qquad q^\pm = \frac{m_\perp}{\sqrt{2}}e^{\pm y}, 
\eeq
where $m_\perp=\sqrt{Q^2+q_\perp^2}$ is the transverse mass and $y$ is the rapidity. The following relations are useful
\beq
\hat{s}+\hat{t}+\hat{u}=Q^2,\qquad q_\perp^2=\frac{\hat{t}\hat{u}}{\hat{s}}, \qquad m_\perp^2= \frac{(\hat{s}+\hat{t})(\hat{s}+\hat{u})}{\hat{s}}. 
\eeq
The differential cross section is given by 
\beq
d\sigma = \frac{e^4}{2s Q^4} \frac{d^3l_1 d^3l_2}{(2\pi)^32l_1^0 (2\pi)^32l_2^0}L^{\mu\nu}W_{\mu\nu},
\label{sigmady}
\eeq
where the hadronic and leptonic tensors take the form 
\beq
W_{\mu\nu}&=&\int d^4 y e^{-iq\cdot y}\langle PP'|J_\mu(y) J_\nu(0)|PP'\rangle, \nn
L^{\mu\nu}&=& 4(l^\mu_1 l^\nu_2+l_1^\nu l^\mu_2-g^{\mu\nu}l_1\cdot l_2) = 2(q^\mu q^\nu -l^\mu l^\nu -q^2g^{\mu\nu}), \label{declep}
\eeq
with  $l\equiv l_1-l_2$. Inserting $1=\int d^4q \delta^{(4)}(q-l_1-l_2)$ and integrating over the phase space of the dilepton, we get 
\beq
\frac{d\sigma}{d^4q} &=&
\frac{4\pi \alpha_{em}^2}{3sQ^4(2\pi)^4}(q^\mu q^\nu -q^2g^{\mu\nu})W_{\mu\nu} =  
\frac{4\pi \alpha_{em}^2}{3sQ^2(2\pi)^4}W^{\mu\nu}(-g_{\mu\nu}) \nn
 &=&\frac{4\pi\alpha_{em}^2}{ 3 s Q^2N_c} \frac{\alpha_s}{2\pi^2}\int \frac{dx}{x}\frac{dx'}{x'}\sum_{q} e_q^2 \bigl(\sigma_{q\bar{q}} q(x)\bar{q}(x') + \sigma_{qg} q(x)G(x')+\sigma_{gq}G(x)q(x')\bigr)\delta(\hat{s}+\hat{t}+\hat{u}-Q^2) .  
 \label{eq:dyunpol}
\eeq 
The second line shows the QCD factorization formula for the unpolarized cross section with the summation over $q=u,d,\bar{u},\bar{d},...$. The lowest-order cross sections for the $q\bar{q}\to \gamma^*g$ and $qg\to \gamma^*q$ subprocesses are 
\beq
\sigma_{q\bar{q}} = 2C_F\left(\frac{\hat{u}}{\hat{t}} + \frac{\hat{t}}{\hat{u}} + \frac{2Q^2\hat{s}}{\hat{u}\hat{t}}\right),
\qquad 
\sigma_{qg}= 2T_R \left(\frac{\hat{s}}{-\hat{t}} + \frac{-\hat{t}}{\hat{s}} -\frac{2Q^2 \hat{u}}{\hat{s}\hat{t}}\right), \qquad \sigma_{gq}= 2T_R \left(\frac{\hat{s}}{-\hat{u}} + \frac{-\hat{u}}{\hat{s}} -\frac{2Q^2 \hat{t}}{\hat{s}\hat{u}}\right).   \label{feng}
\eeq

As we shall see in the following subsections,  it is necessary to keep the dependence on lepton angles in order to have nonvanishing SSAs from the present mechanism.  We thus return to (\ref{sigmady}) and work in the so-called Collins-Soper (CS) frame \cite{Collins:1977iv}, in which experimental data are commonly presented. In the CS frame, $q^\mu=(Q,0,0,0)$ and the lepton momenta take the form 
\beq
l_1^\mu = \frac{Q}{2}(1,\sin\theta_{cs} \cos\phi_{cs},\sin\theta_{cs} \sin \phi_{cs},\cos \theta_{cs}), \quad l_2^\mu =  \frac{Q}{2}(1,-\sin\theta_{cs} \cos\phi_{cs},-\sin\theta_{cs} \sin \phi_{cs},-\cos \theta_{cs}). 
\eeq
We thus have 
\beq
\frac{d^3l_1d^3l_2}{l_1^0l_2^0} = \frac{d^3q d^3\left(\frac{l_1-l_2}{2}\right)}{(l^0_1)^2} 
= \frac{1}{2}d^4qd\Omega,
\eeq
and 
\beq
\frac{d\sigma}{d^4qd\Omega}= \frac{\alpha_{em}^2}{64\pi^4s Q^4}L^{\mu\nu}W_{\mu\nu},
\eeq
with $d\Omega= d\cos\theta_{cs}d\phi_{cs}$. In complete analogy to the SIDIS case (\ref{lepdec})--(\ref{a1234}), we decompose 
the lepton tensor (\ref{declep}) in this frame  as \cite{Tangerman:1994eh,Arnold:2008kf,Boglione:2011zw} 
\be
L^{\mu\nu} =
2Q^2 \sum_k \calA_k \tilde{\calV}_k^{\mu\nu}\,, \label{lepdy}
\ee
where 
\be
\begin{split}
&\calA_1 = 1 + \cos^2 \theta_{cs}\,,\\
&\calA_2 = -2\,,\\
&\calA_3 = \sin 2\theta_{cs} \cos\phi_{cs}\,,\\
&\calA_4 = \sin^2\theta_{cs} \cos 2\phi_{cs}\,,\\
&\calA_8 = -\sin2\theta_{cs} \sin\phi_{cs}\,,\\
&\calA_9 = \sin^2\theta_{cs} \sin 2\phi_{cs}\,.
\end{split}
\label{dyangle}
\ee
and 
\be
\begin{split}
&\tilde{\calV}_1^{\mu\nu} = \frac{1}{2}\left(X^\mu X^\nu + Y^\mu Y^\nu - 2 Z^\mu Z^\nu\right)\,,\\
&\tilde{\calV}_2^{\mu\nu} = -Z^\mu Z^\nu \,,\\
&\tilde{\calV}_3^{\mu\nu} = -\frac{1}{2}\left(Z^\mu X^\nu + X^\mu Z^\nu\right)\,,\\
&\tilde{\calV}_4^{\mu\nu} = \frac{1}{2}\left(- X^\mu X^\nu + Y^\mu Y^\nu\right)\,,\\
& \tilde{\calV}_8^{\mu\nu} = \frac{1}{2}(Z^\mu Y^\nu + Y^\mu Z^\nu)\,,\\
& \tilde{\calV}_9^{\mu\nu} =  -\frac{1}{2}\left(X^\mu Y^\nu + Y^\mu X^\nu\right)\,.
\end{split}
\ee
In the above expressions the orthonormal vectors $T,X,Y,Z$ coincide with $T^\mu=(1,0,0,0)$, $X^\mu=(0,1,0,0)$, $Y^\mu=(0,0,1,0)$ and $Z^\mu=(0,0,0,1)$ in the CS frame. In  generic frames, we may write  
\be
\begin{split}
& T^\mu = \frac{q^\mu}{Q}\,,\\
& Z^\mu = \frac{2}{s m_\perp}\left((q\cdot P')\tilde{P}^\mu - (q\cdot P) \tilde{P}'^\mu\right)=\frac{2}{\hats m_\perp}\left((q\cdot p')\tilde{p}^\mu - (q\cdot p) \tilde{p}'^\mu\right)\,,\\
& X^\mu = - \frac{2Q}{s q_\perp m_\perp}\left((q\cdot P')\tilde{P}^\mu + (q\cdot P) \tilde{P}'^\mu\right)= - \frac{2Q}{\hats q_\perp m_\perp}\left((q\cdot p')\tilde{p}^\mu + (q\cdot p) \tilde{p}'^\mu\right)\,,\\
& Y^\mu = -\epsilon^{\mu\nu \rho\sigma} X_\nu Z_\rho T_\sigma\,,
\end{split}
\ee
with $\tilde{P}^\mu = P^\mu - P\cdot qq^\mu/q^2$, etc.

\subsection{Quark-initiated channel: $q\bar{q}$ annihilation }

We now turn to SSAs, assuming that the proton with momentum $P^\mu\approx \delta^\mu_+P^+$ is transversely polarized. It emits either a quark or a gluon, which participates in hard scattering. In the collinear kinematics with $q_\perp\gg \Lambda_{\rm QCD}$, it has been known that SSAs can arise from the genuine twist-three $q\bar{q}g$ and $ggg$ correlation functions  \cite{Qiu:1991pp,Ji:2006vf,Zhou:2010ui,Koike:2011nx}. We first demonstrate that the $g_T(x)$ distribution gives rise to a novel source of SSAs, considering the case in which the polarized proton emits a quark. The unpolarized proton emits either an antiquark to proceed with the $q\bar{q}$ annihilation or a gluon to proceed with the Compton scattering. These two cases are studied separately in this and the next subsections.  

The derivation of SSAs is completely parallel with that in SIDIS  \cite{Benic:2019zvg}. In the annihilation channel $q\bar{q}\to \gamma^*g$, by following the same steps as in \cite{Benic:2019zvg} with trivial modifications,   we can immediately derive the hadronic tensor (cf. \eqref{eq:w0})
\beq
W^{q\bar{q}}_{\mu\nu}&=& 
 \sum_q \frac{ e_q^2}{4N_c}  \int dx g^q_T(x) \int \frac{dx'}{x'}  \bar{q}(x')  S_\perp^\lambda \frac{\partial}{\partial k_\perp^\lambda} \Bigl( \delta((k+p'-q)^2)   {\rm Tr}\left[\gamma_5\Slash k S^{q\bar{q}}_{\mu\nu}(k)\right] \Bigr)_{k= p}  +\cdots,
\label{k}
\eeq
where again the genuine twist-three correlators have been neglected.  The diagrams contributing to the hard factors $S^{q\bar{q}}$ have been identified in \cite{Pire:1983tv} in the context of longitudinal SSAs, and are displayed in  Fig.~\ref{fig:qqbardy}. For transverse SSAs, the same set of diagrams are relevant, except that  we have to compute them for the non-collinear incoming parton momentum $k^\mu=(xP^+,k_\perp,0)$.   

\begin{figure}
\includegraphics[scale = 0.3]{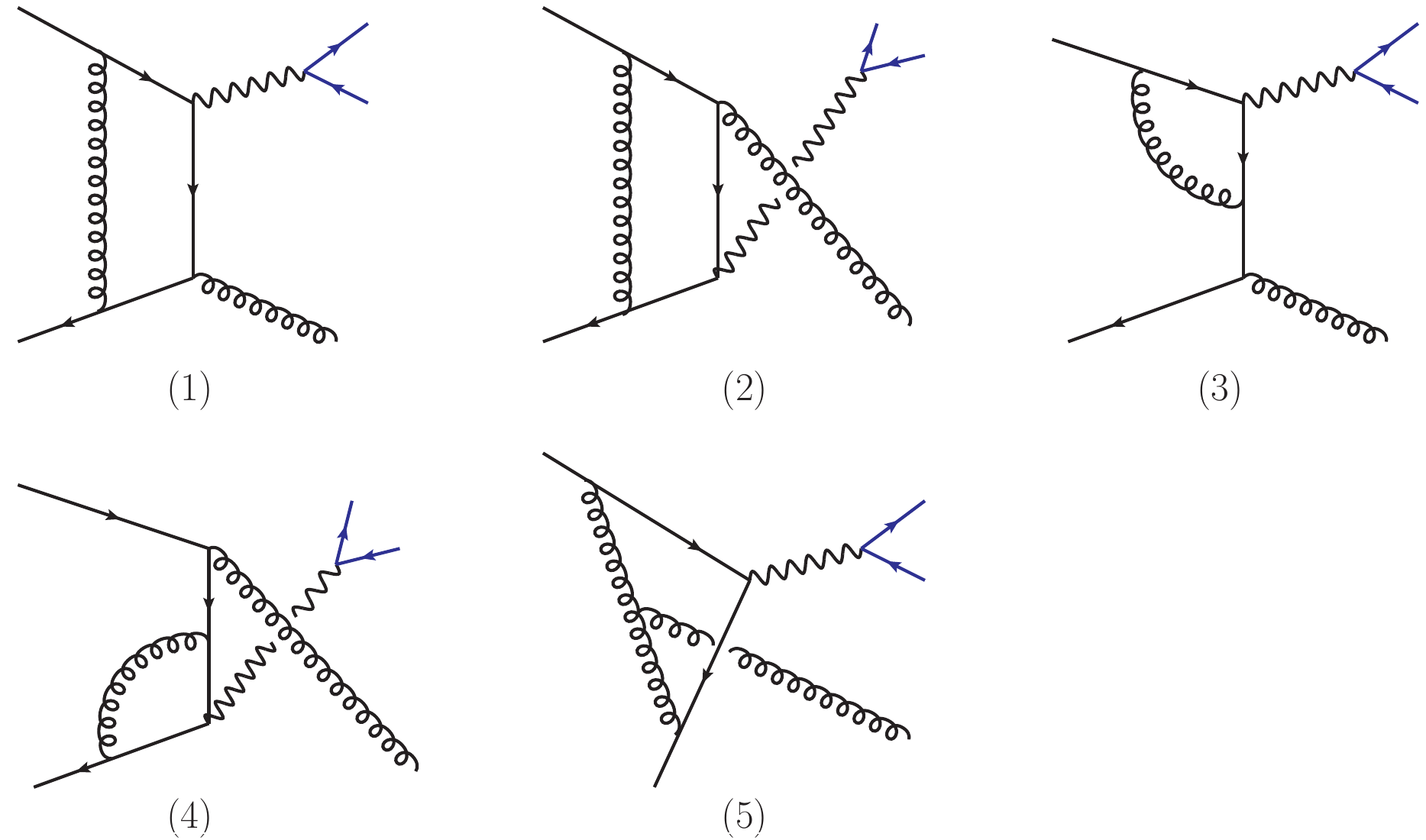}
\caption{A class of one-loop diagrams in the $q\bar{q} \to g \gamma^*$ channel that contain imaginary parts. The short blue fermion lines denote the lepton pair from the virtual photon via $\gamma^* \to l^+ l^-$.}
\label{fig:qqbardy}
\end{figure}

We adopt the same strategy as in SIDIS; the available vectors to construct $S^{q\bar{q}}_{\mu\nu}$ are $k,p'$ and $q$, and $S^{q\bar{q}}_{\mu\nu}$ has to satisfy the Ward identity  $q^\mu S_{\mu\nu}^{q\bar{q}}=0$. After tracing with a  $\gamma_5$,  the only allowed tensor structures are 
\beq
{\cal H}_{\mu\nu}^{q\bar{q}}\equiv {\rm Tr}[\gamma_5\Slash k S^{q\bar{q}}_{\mu\nu}(k)] &=&  A^{q\bar{q}}(\tilde{s},\tilde{t},\tilde{u})\left[\left(k_\mu-\frac{q\cdot k}{q^2}q_\mu\right)\epsilon_\nu^{\ k  qp'}+\left(k_\nu-\frac{q\cdot k}{q^2}q_\nu\right)\epsilon_\mu^{\ k  qp'} \right] \nn
 && +B^{q\bar{q}}(\tilde{s},\tilde{t},\tilde{u})\left[\left(p'_\mu -\frac{q\cdot p'}{q^2}q_\mu\right)\epsilon_\nu^{\ k  qp'} +\left(p'_\nu -\frac{q\cdot p'}{q^2}q_\nu\right)\epsilon_\mu^{\ k  qp'}\right] \nn 
 &\equiv& A^{q\bar{q}}{\cal T}^{1}_{\mu\nu}(k)+B^{q\bar{q}}{\cal T}^2_{\mu\nu}(k), \label{sab}
\eeq
 where $\tilde{s},\tilde{t},\tilde{u}$ are the non-collinear ($p\to k$) versions of the Mandelstam variables in \eqref{dyman}.
An immediate consequence of (\ref{sab}) is that 
\beq
g^{\mu\nu}W_{\mu\nu}^{q\bar{q}}(k)=0,
\eeq
namely, the spin-dependent contribution to the cross section \eqref{eq:dyunpol} integrated over the lepton phase space vanishes. This explains why we need to keep the lepton angles.  In contrast, SSAs  from the genuine twist-three correlation functions  \cite{Qiu:1991pp,Ji:2006vf,Zhou:2010ui,Koike:2011nx} do not vanish after the integration over the lepton angles, which take the form $d\Delta\sigma \sim \vec{S}_\perp \times \vec{q}_\perp$ in the  laboratory frame. 

Concerning the coefficients $A^{q\bar{q}}$ and $B^{q\bar{q}}$, 
again it is enough to compute them for the collinear kinematics in (\ref{dyman}) and then implement the replacement $\hat{s},\hat{t},\hat{u}\to \tilde{s},\tilde{t},\tilde{u}$. 
In the case of SIDIS discussed in the previous section, we have calculated the hard factors   analytically and demonstrated the efficiency of our new approach. For Drell-Yan, in order to save even more efforts and reduce the risk of algebraic errors, we opt to evaluate them using the Mathematica package `Package-X' \cite{Patel:2015tea}. This allows us to directly obtain the imaginary part of each diagram as a discontinuity in the kinematical variables. The diagrams 1 and 2 in Fig.~\ref{fig:qqbardy} have  cuts  in both $\hat{s}>0$ and $Q^2>0$ that need to be added. The diagrams 3, 4 and 5 have cuts only in $Q^2>0$. By default, individual diagrams are computed in $4-2\epsilon$ dimensions in Package-X, and typically contain $1/\epsilon$ poles even in the imaginary parts. These poles must cancel in the end, serving as  a consistency check of the result. 
We find 
\beq
A^{q\bar{q}}(\tils,\tilt,\tilde{u})&=&g^4C_F\Biggl[C_F\frac{2Q^2(3Q^4-\tils^2-6Q^2\tilt-2\tils\tilt+2\tilt^2)}{(\tils+\tilt)(\tils+\tilde{u})^2\tilt\tilde{u}} \nn
&&-(C_A-2C_F)\frac{2Q^2}{\tilt^2\tilde{u}^2}\left\{\frac{Q^2(\tilt-\tilde{u})-2\tilt(\tils+\tilt)}{\tils+\tilde{u}}  - 
  \frac{(\tils+\tilt)\tilt}{\tilde{u}}\ln \frac{\tils}{\tils+\tilde{u}}-\frac{(\tils-\tilt)\tilde{u}}{\tilt}\ln \frac{\tils}{\tils+\tilt}
    \right\} \Biggr], 
\label{aqqbar}
\eeq
 and $B^{q\bar{q}}(\tils,\tilt,\tilde{u})=
   A^{q\bar{q}}(\tils,\tilde{u},\tilt)$. Note that $A^{q\bar{q}}$ and $B^{q\bar{q}}$ are proportional to $Q^2$, so transverse SSAs vanish in real photon (`direct' photon)  production with $Q^2=0$,  similar to the SIDIS case.\footnote{Strictly speaking, the CS frame is not defined for $Q=0$. However, since the hard coefficients are Lorentz invariant,  one can still  conclude that transverse SSAs vanish in the laboratory frame. } The same observation holds for the other channels discussed below.  This is not the case for SSAs from the genuine twist-three distributions \cite{Qiu:1991pp,Ji:2006vf,Zhou:2010ui,Koike:2011nx}.

\subsection{Quark-initiated channel: Compton scattering}
\label{qg}

\begin{figure}
\includegraphics[scale = 0.3]{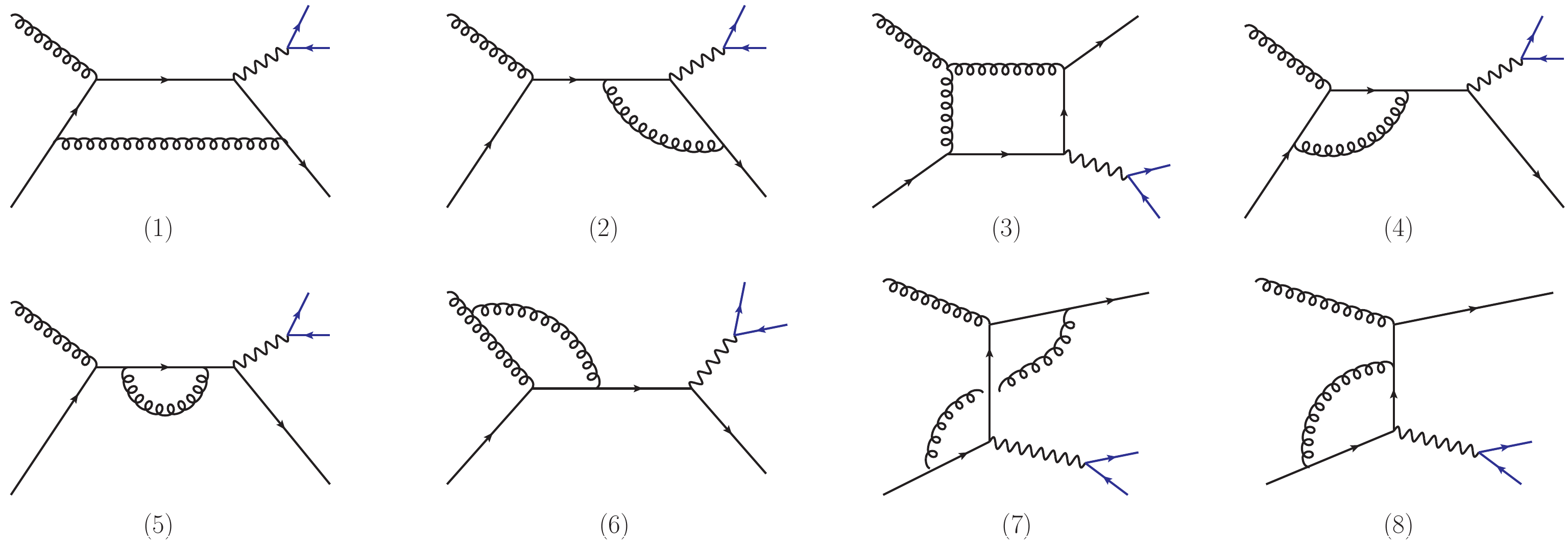}
\caption{A class of one-loop diagrams in the $qg \to q \gamma^*$ channel that contain imaginary parts. The short blue fermion lines denote the lepton pair from the virtual photon via $\gamma^* \to l^+ l^-$.}
\label{fig:qgdy}
\end{figure}

Another important subprocess in Drell-Yan is the Compton scattering $qg\to \gamma^*q$, where the gluon comes from the unpolarized proton. The corresponding hadronic tensor takes the form  
\beq
W^{\rm C}_{\mu\nu}= \sum_q \frac{e_q^2}{4N_c}  \int dx \frac{dx'}{x'} g^q_T(x) G(x')  S_\perp^\lambda \frac{\partial}{\partial k_\perp^\lambda} \Bigl( \delta(\tilde{s}+\tilde{t}+\tilde{u}-Q^2)   {\rm Tr}\left[\gamma_5\Slash kS^{\rm C}_{\mu\nu}(k)\right] \Bigr)_{k= p}, 
\label{kk}
\eeq
with the unpolarized gluon PDF $G(x)$. As in (\ref{sab}), we adopt the parametrization  
\beq
{\cal H}_{\mu\nu}^{\rm C}\equiv {\rm Tr}[\gamma_5\Slash kS^{\rm C}_{\mu\nu}(k)]&=& A^{\rm C}(\tilde{s},\tilde{t},\tilde{u}){\cal T}^1_{\mu\nu}(k)
  +B^{\rm C}(\tilde{s},\tilde{t},\tilde{u}){\cal T}_{\mu\nu}^2(k). 
\eeq
In this case there are eight diagrams that contain imaginary parts as listed in Fig.\ref{fig:qgdy}. Diagrams 1, 2 and 3 have discontinuities in both $s>0$ and $Q^2>0$. Diagrams 4, 5 and 6 have discontinuities in $s>0$, and diagrams 7 and 8 have discontinuities in $Q^2>0$. 
After confirming the cancellation of  $1/\epsilon$ poles, we obtain 
\be
\begin{split}
&A^{\rm C}(\tils,\tilt,\tilde{u})=g^4T_R\Biggl[C_F\frac{2Q^2(\tils+3\tilt-4Q^2)}{\tils\tilt(\tils+\tilde{u})^2}\\
& \qquad \qquad \qquad  -(C_A-2C_F)\frac{2 Q^2}{\tils^3(\tils+\tilde{u})\tilt^3} \left\{ \tils\tilt Q^2(\tils-\tilt)+(\tils+\tilde{u})(\tils+\tilt)\left(s^2 \ln \frac{\tils}{\tils+\tilt}+\tilt^2 \ln \frac{(\tils+\tilde{u})(\tils+\tilt)}{\tilt\tilde{u}}\right)\right\} \Biggr],\\
&B^{\rm C}(\tils,\tilt,\tilde{u})=
g^4T_R\Biggl[-C_F \frac{6Q^2}{\tils\tilt(\tils+\tilde{u})} -(C_A-2C_F)\frac{Q^2}{\tils^3\tilt^3} \Biggl\{\frac{\tils\tilt}{(\tils+\tilt)^2}\left(Q^2 (\tils + \tilt)^2 + 2\tils\tilt\tilde{u}\right) \\
& \qquad \qquad \qquad + \tils^2(\tils+\tilde{u})\ln \frac{\tils}{\tils+\tilt}+\tilt^2(\tils-\tilde{u})\ln \frac{(\tils+\tilt)(\tils+\tilde{u})}{\tilt\tilde{u}} 
\Biggr\}\Biggr],
\end{split}
\label{comptonab}
\ee
which vanish for real photon production at $Q=0$ as mentioned earlier.

\subsection{Crossing symmetry}

One might expect that the results for SSAs in SIDIS and in Drell-Yan would be related by crossing symmetry. A quick inspection reveals that this is not the case. For example, \eqref{eq:S012final} and \eqref{aqqbar} are not exactly related by $\tils\leftrightarrow \tilt$, although some parts are. Even the numbers of diagrams involved are different in the two cases. The reason is that we are focusing on the imaginary parts of loop diagrams. Because of different kinematics ($q^2<0$ or $q^2>0$),  topologically the same diagrams provide imaginary parts in one case but not in the other.  Nevertheless, crossing symmetry should work for the total (real plus imaginary) amplitudes as illustrated in  \cite{Korner:2000zr}, where the authors extracted the imaginary  parts of the one-loop amplitudes in SIDIS and Drell-Yan from the known  one-loop result for  $e^+e^-\to \gamma^*\to q\bar{q}g$ \cite{Korner:1984xd}, even though the latter has no imaginary part. This was achieved  by analytically continuing logarithms and dilogarithms, which are the sources of the imaginary parts, into appropriate kinematical regions. As a matter of fact, our direct evaluation of the aforementioned diagrams agrees perfectly with their result inferred by crossing symmetry. Specifically, $A^{q\bar{q}}, B^{q\bar{q}}$ match $H_8^a,H_9^a$ in (4.3) of  \cite{Korner:2000zr}, and $A^{qg},B^{qg}$ match $H_8^{C_q},H_9^{C_q}$ in  (4.4) of  \cite{Korner:2000zr} up to overall constants.  
This might seem surprising at first sight, given that the results in  \cite{Korner:2000zr}  were meant for different  observables, i.e., the so-called time-reversal odd (or `T-odd') asymmetries in {\it unpolarized} Drell-Yan via $W$-boson production, which had been originally analyzed in  \cite{Hagiwara:1984hi,Hagiwara:1982cq}.  The equivalence to the present calculation can be understood as follows (see also \cite{Abele:2022spu}). The $W$-boson coupling is proportional to $1-\gamma_5$. As one picks the `1' term in the amplitude and the $\gamma_5$ term in the complex-conjugate amplitude in unpolarized Drell-Yan (or DIS), the configuration is equivalent to  longitudinally polarized Drell-Yan (or DIS) with  photon production (or photon exchange); one can move the $\gamma_5$ term from the $W$-boson vertex to the proton matrix element by repeatedly anticommuting it with $\gamma$ matrices along a fermion flow. In the case of transverse SSAs,  there is an additional complication having to do with the non-collinear kinematics. However, the hard factors $S_{1,2}$ are essentially the same and can be derived via a careful implementation of crossing symmetry \cite{Korner:1984xd}.

\subsection{Gluon-initiated channel}

So far  we have assumed that the transversely polarized proton  emits a quark. In parallel with our discussion in  SIDIS (see Sec.~III), the polarized proton can also emit a gluon leading to another contribution to SSAs proportional to the ${\cal G}_{3T}(x)$ distribution in Drell-Yan. The relevant subprocess is the Compton scattering $qg\to \gamma^*q$ with the hard factor 
\beq
W^{\rm C(g)}_{\mu\nu}&=& -i
\sum_q \frac{e_q^2}{2N_c}  \int \frac{dx}{x} \frac{dx'}{x'} {\cal G}_{3T}(x) q(x')  S_\perp^\lambda \frac{\partial}{\partial k_\perp^\lambda} \left( \delta(\tilde{s}+\tilde{t}+\tilde{u}-Q^2)\epsilon^{\alpha\beta k n} S^{{\rm C}(g)}_{\mu\nu\alpha\beta}(k)\right)_{k= p}
\nn
&=&\sum_q \frac{e_q^2}{2N_c}  \int dx \frac{dx'}{x'} {\cal G}_{3T}(x) q(x')  S_\perp^\lambda \frac{\partial}{\partial k_\perp^\lambda} \left( \delta(\tilde{s}+\tilde{t}+\tilde{u}-Q^2)\frac{-2i}{\tilde{s}} \epsilon^{\alpha\beta k p'} S^{{\rm C}(g)}_{\mu\nu\alpha\beta}(k)\right)_{k= p},
\eeq
which can be compared to \eqref{eq:Wglue3} and \eqref{kk}.  This time we use the relation 
\beq
k\cdot p' \epsilon^{\alpha\beta kn}S^{(g)}_{\mu\nu\alpha\beta}(k)=k\cdot n \epsilon^{\alpha\beta kp'}S^{(g)}_{\mu\nu\alpha\beta}(k),
\eeq  
following from the Schouten and Ward identities, to eliminate the vector $n$. Note that $
k\cdot n/k\cdot p'=2x/\tilde{s}$ commutes with the $k_\perp$-derivative. 

As before, we parameterize the hard part with two independent tensors,  
\beq
 {\cal H}^{{\rm C}(g)}_{\mu\nu}\equiv \epsilon^{\alpha\beta kp'}\frac{-2i}{\tilde{s}}S^{{\rm C}(g)}_{\mu\nu\alpha\beta}(k) &=&
 A^{{\rm C}(g)}{\cal T}_{\mu\nu}^1(k) 
 +B^{{\rm C}(g)}{\cal T}_{\mu\nu}^2(k). 
 \eeq
 The relevant diagrams are the same as in Fig.~\ref{fig:qgdy}, but we need to employ the replacement $p\leftrightarrow p'$ (or $\hat{t}\leftrightarrow \hat{u}$) and contract the gluon indices $\alpha,\beta$ with the antisymmetric tensor $\epsilon^{\alpha\beta p p'}$. The results are
\beq
&&A^{{\rm C}(g)}(\tils,\tilt,\tilde{u})=-B^{\rm C}(\tils,\tilde{u},\tilt) -g^4T_R(C_A-2C_F)\frac{4Q^2(\tils+\tilt)}{\tils \tilde{u}^3} \left(\frac{(\tils+2\tilde{u})\tilde{u}}{(\tils+\tilde{u})^2}+\ln \frac{\tils}{\tils+\tilde{u}}\right),\nn
&& B^{{\rm C}(g)}(\tils,\tilt,\tilde{u})=-A^{\rm C}(\tils,\tilde{u},\tilt)-g^4T_R(C_A-2C_F)\frac{4Q^2}{\tils\tilde{u}^3(\tils+\tilde{u})}\left(\tils\tilde{u}+(\tils+\tilde{u})^2\ln \frac{\tils}{\tils+\tilde{u}}\right). 
\eeq
It is seen that the cross section is for the most part related to the one in the Compton scattering channel (\ref{comptonab})  via the crossing $\hat{t}\leftrightarrow \hat{u}$, as naively expected. The minus sign is due to $\epsilon^{\mu p q p'}=-\epsilon^{\mu p' q p}$.  However, there are extra ${\cal O}(1/N_c)$ terms from different contractions of Lorentz indices. Again the above expressions vanish at $Q=0$.

\subsection{Summary of the results}
\label{summary}

After computing the hard factors in all the subprocesses, we can write down the formulas for transverse SSAs. 
Inserting (\ref{lepdy}) into (\ref{sigmady}) and using the formula
\be
S_\perp^\lambda \left(\frac{\pd}{\pd k_\perp^\lambda}\delta\left(\tils + \tilt + \hatu - k^2 - Q^2\right)\right)_{k = p} = -\frac{S_\perp \cdot q}{p\cdot(p' - q)}x\frac{\pd}{\pd x}\delta\left(\hats + \hatt + \hatu - Q^2\right)\,,
\ee
we find 
\beq
&&\frac{d^6\sigma^{q\bar{q}}}{d^4 q d\Omega} = 
\frac{\alpha_{em}^2 \alpha_s^2M_N }{8\pi^2  N_cs Q^3}\sum_k \calA_k {\cal S}_k\sum_q e_q^2 \int \frac{dx}{x}\int\frac{dx'}{x'} \bar{q}(x')\delta(\hats + \hatt + \hatu - Q^2) \left(x^2 \frac{d g_T(x)}{d x} \Delta \hat{\sigma}^{q\bar{q}}_{Dk} + x g^q_T(x)\Delta \hat{\sigma}^{q\bar{q}}_k\right),
\label{eq:csdynew}
\\
&&\frac{d^6\sigma^{\rm C}}{d^4 q d\Omega} = 
\frac{\alpha_{em}^2 \alpha_s^2M_N }{8\pi^2N_c s Q^3 }\sum_k \calA_k {\cal S}_k\sum_q e_q^2 \int \frac{dx}{x}\int\frac{dx'}{x'}  G(x')\delta(\hats + \hatt + \hatu - Q^2) \left(x^2 \frac{d g^q_T(x)}{d x} \Delta \hat{\sigma}^{\rm C}_{Dk} + x g_T(x)\Delta \hat{\sigma}^{\rm C}_k\right), \\
&& \frac{d^6\sigma^{{\rm C}(g)}}{d^4 q d\Omega} = 
\frac{\alpha_{em}^2 \alpha_s^2M_N }{ 4\pi^2N_c s Q^3 }\sum_k \calA_k {\cal S}_k\sum_q e_q^2 \int \frac{dx}{x}\int\frac{dx'}{x'}  q(x')\delta(\hats + \hatt + \hatu - Q^2) \left(x^2 \frac{d {\cal G}_{3T}(x)}{d x} \Delta \hat{\sigma}^{{\rm C}(g)}_{Dk} + x {\cal G}_{3T}(x)\Delta \hat{\sigma}^{{\rm C}(g)}_k\right),\nn
\eeq
where the partonic cross sections are given by\footnote{Similar to (\ref{delcon}), we have the relation
\beq
\left.\left(\frac{S_\perp\cdot q}{p\cdot (p'-q)}x\frac{\partial }{\partial x} + S_\perp^\lambda \frac{\partial}{\partial k_\perp^\lambda}\right)(\tils+\tilt+\tilde{u}-Q^2)\right|_{k=p}=0,
\label{eq:dysighat}
\eeq
which allows us to apply the constraint $\tils+\tilt+\tilde{u}=Q^2$ to the hard factors $A^i,B^i$  before taking the derivatives in (\ref{dyhard}), as  having done already in the previous subsection.} 
\be
\begin{split}
& \frac{g^4M_N}{Q}{\cal S}_k \Delta\hat{\sigma}_{Dk}^{i} =\frac{S_\perp\cdot q}{p\cdot(p'-q)}{\cal H}^{i}_{\mu\nu}(p)\tilde{\calV}_k^{\mu\nu} , \\
& \frac{g^4M_N}{Q}{\cal S}_k \Delta\hat{\sigma}_k^{i}=\left(\frac{S_\perp\cdot q}{p\cdot(p'-q)}x\frac{\pd {\cal H}^{i}_{\mu\nu}(p)}{\pd x} + \left[S_\perp^\lambda \frac{\pd {\cal H}^{i}_{\mu\nu}(k)}{\pd k_\perp^\lambda} \right]_{k = p}\right)\tilde{\calV}_k^{\mu\nu},
\end{split}
\label{dyhard}
\ee
with $i=q\bar{q},{\rm C}, {\rm C}(g)$. 
We define  $\calS_k = \sin(\Phi_S)$ for $k = 1,2,3,4$ and $\calS_k = \cos(\Phi_S)$ for $k = 8,9$.\footnote{More precisely, $\Phi_S\to \Phi_S-\phi_{q_\perp}$ where $\phi_{q_\perp}$ is the azimuthal angle of the virtual photon in the center-of-mass frame. In  the CS frame, one sets $\phi_{q_\perp}=0$ as  part of the frame definition.}

As in the SIDIS case, we present the partonic cross sections in terms of the new variables
\beq
x_a=\frac{Q^2}{2p\cdot q}=\frac{Q^2}{Q^2-\hat{t}}, \qquad x_b=\frac{Q^2}{2p'\cdot q}=\frac{Q^2}{Q^2-\hat{u}}, 
\qquad c= x_a+x_b-x_ax_b =\frac{Q^2\hat{s}}{(\hat{s}+\hat{t})(\hats+\hatu)} =\frac{Q^2}{m_\perp^2},
\label{eq:abc}
\eeq
which can be often found in the literature. 
For the $q\bar{q}$ annihilation channel, we have
\be
\begin{split}
\Delta\hat{\sigma}_{D8}^{q\bar{q}} &= \frac{2 C_F\sqrt{c}}{ x_a(1-x_b)} \bigg(2C_F(x_a^2 - x_b^2) + (C_A - 2 C_F)x_a x_b\left(-\frac{\ln\left(c/x_a\right)}{1-x_a} + \frac{\ln\left(c/x_b\right)}{1-x_b}\right)\bigg)\,,\\
\Delta\hat{\sigma}_{D9}^{q\bar{q}} &= \frac{2C_F}{\sqrt{1-c} x_a(1-x_b)} \bigg(C_A x_a x_b(2-x_a-x_b) + C_F ( x_a(x_a-2)c+ x_b (x_b-2)c+x_a^2+x_b^2)\\
& - (C_A - 2 C_F)c\left(x_a (1-x_b)\frac{\ln\left(c/x_a\right)}{1-x_a} + (1-x_a)x_b\frac{\ln\left(c/x_b\right)}{1-x_b}\right)\bigg)\,,
\end{split}
\ee
\be
\begin{split}
\Delta\hat{\sigma}_1^{q\bar{q}} &= \frac{3}{2}\frac{x_a (1-x_b)}{x_b\sqrt{1-c}}\Delta\hat{\sigma}_{D8}^{q\bar{q}}\,,\\
\Delta\hat{\sigma}_2^{q\bar{q}} &= \frac{x_a (1-x_b)}{x_b\sqrt{1-c}}\Delta\hat{\sigma}_{D8}^{q\bar{q}}\,,\\
\Delta\hat{\sigma}_3^{q\bar{q}} &= -\frac{1}{2}\frac{x_a}{(1-x_a)x_b}\left(\sqrt{c} \Delta\hat{\sigma}_{D8}^{q\bar{q}} + \sqrt{1-c}\Delta\hat{\sigma}_{D9}^{q\bar{q}}\right)\,,\\
\Delta\hat{\sigma}_4^{q\bar{q}} &= \frac{1}{2}\frac{x_a}{(1-x_a)x_b}\left(\sqrt{1-c}\Delta\hat{\sigma}_{D8}^{q\bar{q}} + 2\sqrt{c}\Delta\hat{\sigma}_{D9}^{q\bar{q}}\right)\,,\\
\Delta\hat{\sigma}_8^{q\bar{q}} &= \frac{C_F\sqrt{c}}{x_b(1-c)}\bigg[-C_A x_a x_b(c + 3 x_a (1-x_a)+x_b) + C_F\Bigl(-c^2 (x_b+3)+c (x_a (x_a (5 x_a+2)+3)-x_b (3 x_b+4)-7)
\\
&-9 x_a^3+7 x_a+7 x_b (x_b+1)\Bigr) + (C_A - 2 C_F)\left(x_a (2x_b + c)\frac{\ln(c/x_a)}{1-x_a} - x_b(2x_a -c)\frac{\ln(c/x_b)}{1-x_b}\right)\bigg]\,,\\
\Delta\hat{\sigma}_9^{q\bar{q}} &= \frac{C_F}{ x_a x_b(1-c)^{3/2}}\bigg[-C_A x_a x_b\bigl(c^2-c x_a(x_a-3) + c(-3 x_b + 1) + x_a(1-x_a)-x_b\bigr)+ C_F\Bigl(-c^3 (x_b+3)\\
& + c^2 (x_a(x_a^2-1)+x_b (3 x_b+11)+2) - c (2 x_a (2 x_a (x_a-2) + 1)+10 x_b^2+x_b-1) - x_a(x_a^2 + 1) - x_b(x_b + 1)\Bigr)\\
& + (C_A - 2 C_F)c\left(x_a^2 (x_a + 3)(1-x_b)\frac{\ln\left(c/x_a\right)}{1-x_a} + (1-x_a)x_b(2x_a - x_b - c)\frac{\ln\left(c/x_b\right)}{1-x_b}\right)\bigg]\,.
\end{split}
\ee

For the quark-initiated Compton channel, we get 
\be
\begin{split}
\Delta\hat{\sigma}_{D8}^{\rm C} &= \frac{2 T_R x_a}{c^{3/2}}\bigg(-C_A x_b(c (2-x_b)-x_a-x_b) + C_F(c+x_a)(c - 2x_b^2) - (C_A - 2C_F) x_a^2 x_b(1-x_b)\frac{\ln(1-c)}{c}\bigg)\,,\\
\Delta\hat{\sigma}_{D9}^{\rm C} &= \frac{2 T_R\sqrt{1-c} x_a}{ c (1-x_a)}\bigg(C_A x_b(- c + x_a + 1) + C_F(c (-x_a + 2 x_b + 1) - 2 x_a (1-x_a) - 4 x_b)\\
& + (C_A - 2 C_F)x_a\left(-\frac{\ln(c/x_a)}{1-x_a} + (1-x_a)x_b\frac{\ln(1-c)}{c}\right)\Bigg)\,,
\end{split}
\ee
\be
\begin{split}
\Delta\hat{\sigma}_1^{\rm C} &= \frac{3}{2}\frac{x_a (1-x_b)}{x_b\sqrt{1-c}}\Delta\hat{\sigma}_{D8}^{\rm C}\,,\\
\Delta\hat{\sigma}_2^{\rm C} &= \frac{x_a (1-x_b)}{x_b\sqrt{1-c}}\Delta\hat{\sigma}_{D8}^{\rm C}\,,\\
\Delta\hat{\sigma}_3^{\rm C} &= -\frac{1}{2}\frac{x_a}{(1-x_a)x_b}\left(\sqrt{c} \Delta\hat{\sigma}_{D8}^{\rm C} + \sqrt{1-c}\Delta\hat{\sigma}_{D9}^{\rm C}\right)\,,\\
\Delta\hat{\sigma}_4^{\rm C} &= \frac{1}{2}\frac{x_a}{(1-x_a)x_b}\left(\sqrt{1-c}\Delta\hat{\sigma}_{D8}^{\rm C} + 2\sqrt{c}\Delta\hat{\sigma}_{D9}^{\rm C}\right)\,,\\
\Delta\hat{\sigma}_8^{\rm C} & = \frac{T_R x_a}{(1-x_a) x_b c^{3/2}}\bigg[-C_A x_b\bigl(c^2 (2 x_a+2 x_b-3)+c (x_a (x_a+6)+x_b+4)-4 (x_a^2+x_a+x_b)\bigr)\\
& + C_F\Bigl(c^3 (4 x_a-2)+c^2 ((x_a-9) x_a+4 x_b^2-2 x_b-16)+2 c (2 x_a (x_a+6)+x_b (x_b+12)+8)\\
& -8 (x_a (x_a+2)+x_b (x_b+2))\Bigr) + (C_A - 2 C_F)x_a^2(1-x_b)\left(c\frac{\ln(c/x_a)}{1-x_a} - x_b(-5 c + 4 x_a)\frac{\ln(1-c)}{c}\right)\bigg]\,,\\
\Delta\hat{\sigma}_9^{\rm C} & = \frac{T_R x_a\sqrt{1-c}}{(1-x_a)^2 x_b c}\bigg[C_A x_b\bigl(2 c^2+c (x_a (2 x_a - 5) - 2) + x_a (-3 (x_a-1) x_a-1)\bigr)\\
& + C_F\Bigl(-2 c^2 (2 x_a^2-3 x_a+ 2 x_b+2)+c (x_a (x_a(5 x_a - 18)+7)+10 x_b-2)+2 (4 x_a^3+x_a+x_b)\Bigr)\\
& + (C_A - 2 C_F)x_a\left((c(x_a + 4) - x_a)\frac{\ln(c/x_a)}{1-x_a} + (1-x_a)x_b(-5 c + 4 x_a + x_b)\frac{\ln(1-c)}{c}\right)\bigg]\,.
\end{split}
\ee

For the gluon initiated Compton channel, we have 
\be
\begin{split}
\Delta\hat{\sigma}_{D8}^{{\rm C}(g)} & = \frac{2 T_R (1-x_a)x_b}{x_a(1-x_b) c^{3/2}}\bigg(-C_A\bigl(x_a x_b(c(2-x_a)-x_a-x_b)+2cx_a^2 \bigr) + C_F\bigl(x_b(c+x_b)(c-2x_a^2)+4cx_a^2 \bigr)\\
&\quad -(C_A - 2 C_F)x_a(1-x_a)x_b^3\frac{\ln(1-c)}{c}\bigg)\,,\\
\Delta\hat{\sigma}_{D9}^{{\rm C}(g)} & = \frac{2 T_R (1 - x_a) x_b^2 \sqrt{1-c}}{x_a (1-x_b)^2 c }\bigg(C_A x_a(c - x_b + 1) + C_F\bigl(c (-2 x_a + x_b - 1)+2 x_b (1 - x_b)\bigr)\\
&\quad - (C_A - 2 C_F) x_b\left(\frac{\ln(c/x_b)}{1-x_b} + x_a(1-x_b)\frac{\ln(1-c)}{c}\right)\bigg)\,,
\end{split}
\ee
\be
\begin{split}
\Delta\hat{\sigma}_1^{{\rm C}(g)} &= \frac{3}{2}\frac{x_a (1-x_b)}{x_b\sqrt{1-c}}\Delta\hat{\sigma}_{D8}^{{\rm C}(g)}\,,\\
\Delta\hat{\sigma}_2^{{\rm C}(g)} &= \frac{x_a (1-x_b)}{x_b\sqrt{1-c}}\Delta\hat{\sigma}_{D8}^{{\rm C}(g)}\,,\\
\Delta\hat{\sigma}_3^{{\rm C} (g)} &= -\frac{1}{2}\frac{x_a}{(1-x_a)x_b}\left(\sqrt{c} \Delta\hat{\sigma}_{D8}^{{\rm C} (g)} + \sqrt{1-c}\Delta\hat{\sigma}_{D9}^{{\rm C} (g)}\right)\,,\\
\Delta\hat{\sigma}_4^{{\rm C} (g)} &= \frac{1}{2}\frac{x_a}{(1-x_a)x_b}\left(\sqrt{1-c}\Delta\hat{\sigma}_{D8}^{{\rm C} (g)} + 2\sqrt{c}\Delta\hat{\sigma}_{D9}^{{\rm C} (g)}\right)\,,\\
\Delta\hat{\sigma}_8^{{\rm C}(g)} & = \frac{T_R}{c^{3/2} x_a (1-x_b)}\bigg[C_A x_a^2 \bigl(-4 c^3+c^2 (3-2 x_a)+c (x_a-3 x_b^2-4 x_b-4)+4 (x_a+x_b)+4 x_b^2\bigr)\\
&\quad + C_F\Bigl(c^3 (8 x_a^2+x_b+6)+c^2 (2 (x_a-3) x_a (2 x_a+3)+x_b (3 x_b-10)-24)\\
&\quad +c (-2 x_a^3+20 x_a^2+40 x_a+4 x_b (x_b+8)+24) -8 (x_a (x_a (x_a+2)+3)+x_b (x_b+3))\Bigr)\\
&\quad + (C_A - 2 C_F)x_a (1 - x_a) x_b^2\left(c\frac{\ln(c/x_b)}{1-x_b} + x_b(c - 4 x_b)\frac{\ln(1-c)}{c}\right)\bigg]\,,\\
\Delta\hat{\sigma}_9^{{\rm C}(g)} & = \frac{T_R \sqrt{1-c}}{x_a (1-x_b)^2 c}\bigg[C_A x_a\bigl(-4 c^3+c^2 (2 x_a+4 x_b-1)+c (x_a-x_b^2+5 x_b+3)-3 (x_a+x_b)+x_b^3-2 x_b^2\bigr)\\
&\quad + C_F\Bigl(8 c^3 (x_a+1) + c^2 (-2 x_a (2 x_a+3)+(x_b-11) x_b+6)\\
&\quad +c (-2 x_a (x_a+6)-3 (x_b-2) (x_b-1) x_b-6)+6 x_a (x_a+1)-4 x_b^3+6 x_b\Bigr)\\
&\quad + (C_A - 2 C_F)x_b^2\left((c(x_a - 2) - x_a)\frac{\ln(c/x_b)}{1-x_b} + x_a^2(x_a - 3)(1-x_b)\frac{\ln(1-c)}{c}\right)\bigg]\,.\\
\end{split}
\ee

\subsection{Longitudinal SSA in Drell-Yan} 

Finally,  we examine the longitudinal polarization case for a consistency check of the previous results in the literature   \cite{Pire:1983tv,Carlitz:1992fv,Yokoya:2007xe}. The hard factor in the annihilation channel is written as 
\beq
W^{q\bar{q}}_{\mu\nu}&=& 
 \sum_q \frac{ e_q^2}{4N_c} \frac{S^+}{P^+} \int \frac{dx}{x} \Delta q(x) \int \frac{dx'}{x'}  \bar{q}(x')\delta(\hat{s}+\hat{t}+\hat{u}-Q^2)   {\rm Tr}[\gamma_5\Slash p S^{q\bar{q}}_{\mu\nu}(p)].
\label{pire}
\eeq 
Contracting (\ref{sab}) with the lepton tensor in (\ref{lepdy}) and noticing $T_{\mu\nu}^{1,2}\tilde{\calV}_k^{\mu\nu}=0$ for $k=1,2,3,4$,  we derive, including the other channels, 
\beq
\frac{d\Delta \sigma^{q\bar{q}}}{d^4qd\Omega}&=& \frac{\alpha_{em}^2\alpha_s^2}{8\pi^2 s Q^2N_c}\sum_qe_q^2\frac{S^+}{P^+} \int \frac{dx}{x} \Delta q(x) \int \frac{dx'}{x'}  \bar{q}(x')  \delta(\hat{s}+\hat{t}+\hat{u}-Q^2) \nn && \qquad \times \bigl(\sin2\theta_{cs}\sin \phi_{cs}\Delta\hat{\sigma}_{L8}^{q\bar{q}}  +\sin^2\theta_{cs}\sin  2\phi_{cs}\Delta\hat{\sigma}_{L9}^{q\bar{q}}\bigr),
\eeq
\beq
\frac{d\Delta \sigma^{\rm C}}{d^4qd\Omega}&=& \frac{\alpha_{em}^2\alpha_s^2}{8\pi^2 s Q^2N_c}\sum_qe_q^2\frac{S^+}{P^+} \int \frac{dx}{x} \Delta q(x) \int \frac{dx'}{x'}  G(x')  \delta(\hat{s}+\hat{t}+\hat{u}-Q^2) \nn && \qquad \times \bigl(\sin2\theta_{cs}\sin \phi_{cs}\Delta\hat{\sigma}_{L8}^{\rm C}  +\sin^2\theta_{cs}\sin  2\phi_{cs}\Delta\hat{\sigma}_{L9}^{\rm C}\bigr),
\eeq
\beq
\frac{d\Delta \sigma^{{\rm C}(g)}}{d^4qd\Omega}&=& \frac{\alpha_{em}^2\alpha_s^2}{8\pi^2 s Q^2N_c}\sum_qe_q^2\frac{S^+}{P^+} \int \frac{dx}{x} \Delta G(x) \int \frac{dx'}{x'}  q(x')  \delta(\hat{s}+\hat{t}+\hat{u}-Q^2) \nn && \qquad \times \bigl(\sin2\theta_{cs}\sin \phi_{cs}\Delta\hat{\sigma}_{L8}^{{\rm C}(g)}  +\sin^2\theta_{cs}\sin  2\phi_{cs}\Delta\hat{\sigma}_{L9}^{{\rm C}(g)}\bigr),
\eeq
where
\be
\begin{split}
&g^4\Delta\hat{\sigma}^{i}_{L8}=-\tilde{\calV}_8^{\mu\nu}(A^{i}T^1_{\mu\nu}+B^{i}T^2_{\mu\nu})=
 \frac{\hat{t}\hat{u}(-(\hat{s}+\hat{u})A^{i}+(\hat{s}+\hat{t})B^{i})}{4q_\perp m_\perp}
,\\ 
&g^4\Delta\hat{\sigma}^{i}_{L9}=\tilde{\calV}_9^{\mu\nu}(A^{i}T^1_{\mu\nu}+B^{i}T^2_{\mu\nu})=
\frac{\hat{t}\hat{u}((\hat{s}+\hat{u})A^{i}+(\hat{s}+\hat{t})B^{i})}{4Q m_\perp},
\end{split}
\label{eq:sigdyL}
\ee
with $i=q\bar{q},{\rm C}, {\rm C}(g)$. Comparing \eqref{dyhard} to \eqref{eq:sigdyL}, we deduce the relations between the hard coefficients for longitudinal SSAs and the corresponding ones for transverse SSAs
\be
\Delta\hat{\sigma}^i_{D 8,9} = \pm 2\sqrt{\frac{1-c}{c}}\frac{x_b}{1-x_b}\Delta\hat{\sigma}^i_{L8,9}\,,
\ee
which verify the agreement with the previous results \cite{Carlitz:1992fv,Yokoya:2007xe}.\footnote{An earlier work   \cite{Pire:1983tv} computed the ${\cal O}(1/N_c)$ term of $\Delta\hat{\sigma}_{L8}^{q\bar{q}}$ in the annihilation channel. The logarithmic terms agree, but the non-logarithmic terms do not.  This disagreement was already noted in \cite{Carlitz:1992fv}.} We collect the explicit formulas below for completeness,   
\begin{equation}
\begin{split}
&\Delta\hat{\sigma}^{q\bar{q}}_{L8}=
C_F\frac{-2 c}{\sqrt{1-c}}\left[C_F\left(\frac{x_b}{x_a}-\frac{x_a}{x_b}\right)+\frac{C_A-2C_F}{2}\left(\frac{\ln (c/x_a)}{1-x_a}-\frac{\ln (c/x_b)}{1-x_b}\right)\right],
\\
&\Delta\hat{\sigma}^{q\bar{q}}_{L9}
=C_F\sqrt{c}\left[C_F\left(\frac{x_a}{x_b}+\frac{x_b}{x_a}\right) -(C_A-2C_F)\left\{ \frac{1}{1-x_a}\left(1-\frac{c \ln (c/x_a)}{c-x_a}\right) + \frac{1}{1-x_b}\left(1-\frac{c\ln (c/x_b)}{c-x_b}\right)\right\}\right],\label{l2} 
\end{split}
\end{equation}
\begin{equation}
\begin{split}
&\Delta\hat{\sigma}_{L8}^{\rm C} 
= T_R\frac{2(c-x_b)}{\sqrt{1-c}}\left[C_F\left(\frac{x_a}{x_b}-\frac{1+x_a}{2}\right)+\frac{C_A-2C_F}{2}\left\{x_b-1 +\frac{x_a x_b}{c} -\frac{x_a(c-x_b)}{c^2}\ln (1-c)\right\}\right],
\\
& \Delta\hat{\sigma}_{L9}^{\rm C} 
= T_R\frac{2(c-x_b)}{\sqrt{c}}\left[C_F \left(\frac{x_a}{2x_b}+\frac{1+x_a}{2}\right)+\frac{C_A-2C_F}{2}\left\{ x_b -\frac{1}{1-x_a}-\frac{x_a\ln(1-c)}{c}+\frac{x_a\ln (c/x_a)}{x_b(1-x_a)^2}\right\}\right],
\end{split}
\end{equation}
\begin{equation}
\begin{split}
&\Delta\hat{\sigma}_{L8}^{{\rm C}(g)} 
= \left. \Delta \hat{\sigma}^{\rm C}_{L8}\right|_{\hatt\leftrightarrow \hatu} -T_R(C_A-2C_F)\frac{2x_a(1-x_a)}{\sqrt{1-c}},
\\
&\Delta\hat{\sigma}_{L9}^{{\rm C}(g)} 
= \left. -\Delta \hat{\sigma}^{\rm C}_{L9}\right|_{\hatt\leftrightarrow \hatu}
-T_R(C_A-2C_F)\frac{2x_b(1-x_a)}{\sqrt{c}(1-x_b)}\left(1-\frac{x_b\ln (c/x_b)}{c-x_b}\right).
\end{split}
\end{equation}

A numerical evaluation of the asymmetries was carried out in \cite{Yokoya:2007xe} for the RHIC and J-PARC kinematics with $q_\perp <Q$, which were found to reach up to a few percent in the forward rapidity region.

\section{Numerical results}

\begin{figure}
  \begin{center}
  \includegraphics[scale = 0.5]{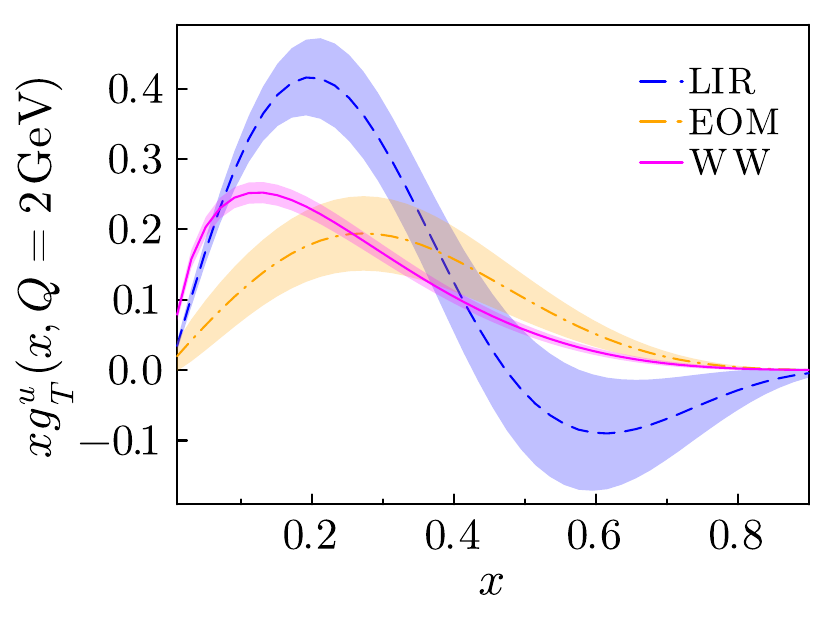}
   \includegraphics[scale = 0.5]{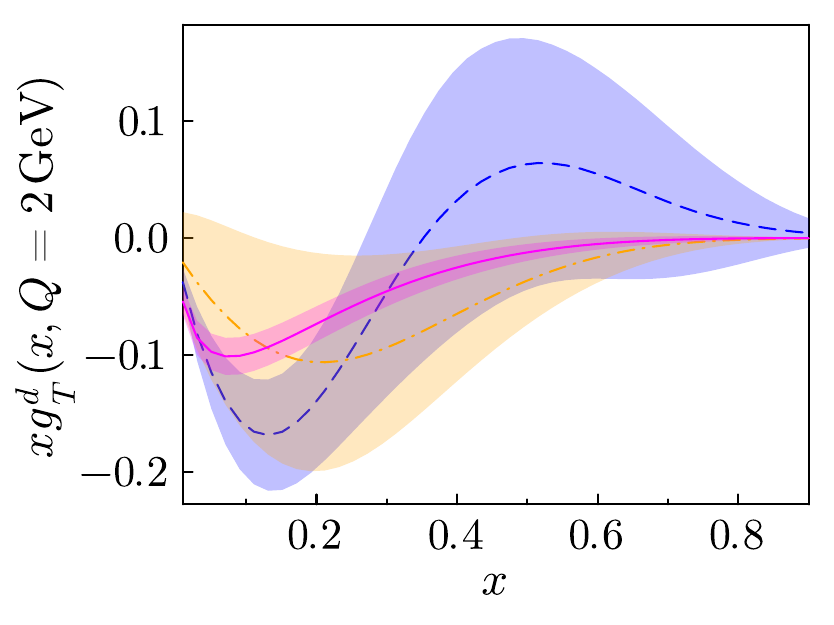}
  \end{center}
  \caption{$xg_T(x)$ for $u$ (left) and $d$ (right) quarks in the three different scenarios defined in \eqref{www} and \eqref{eq:gTscene}. The blue and magenta bands roughly match the blue and magenta bands on Fig.~2 in \cite{Bauer:2022mvl}.}
  \label{fig:gT}
\end{figure}

\begin{figure}
  \begin{center}
  \includegraphics[scale = 0.5]{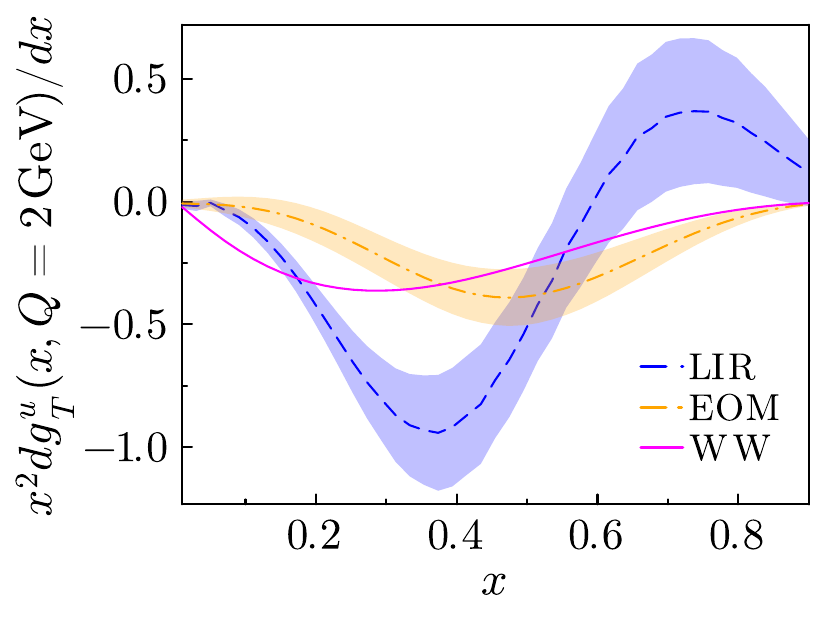}
   \includegraphics[scale = 0.5]{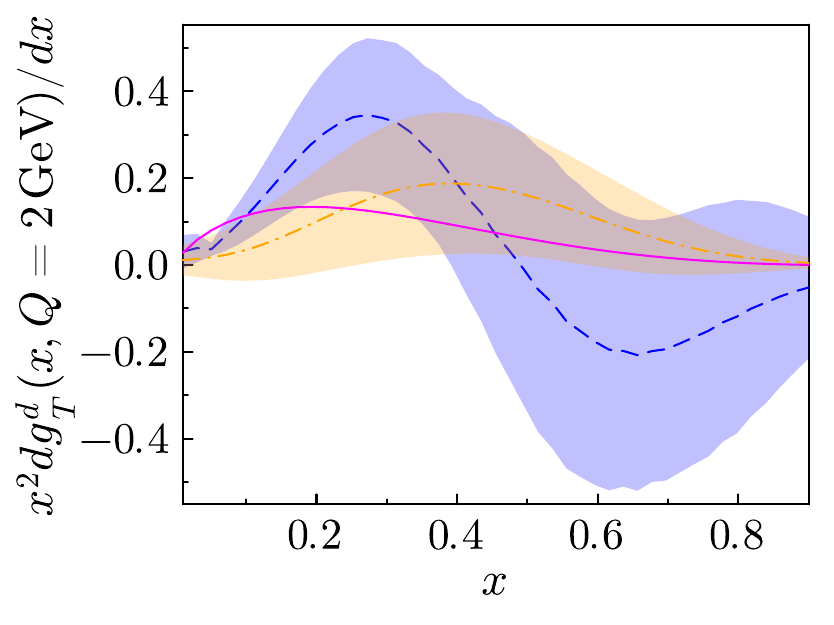}
  \end{center}
  \caption{$x^2 dg_T(x)/dx$ for $u$ and $d$ quarks in the three different scenarios defined in \eqref{www} and \eqref{eq:gTscene}.}
\label{fig:dgT}
\end{figure}

In this section we revise our predictions made in \cite{Benic:2021gya} for transverse SSAs in SIDIS in light of the  corrected formulas presented in Sections II and III. We then make new predictions for SSAs in Drell-Yan based on the results in Section  IV. 

\subsection{Parton distributions } 

Since we have neglected the genuine-twist three terms initially as indicated by  (\ref{eq:w0}) and (\ref{k}), we should, to be consistent, employ  the Wandzura-Wilczek (WW) approximation for the $g_T$ and ${\cal G}_{3T}$ distributions 
\be
\begin{split}
& xg^q_T(x) \approx x\int_x^1 dx'\frac{ \Delta q(x')}{x'}, \qquad ({\rm WW}) \\
& x{\cal G}_{3T}(x) \approx \frac{x}{2}\int_x^1 dx' \frac{\Delta G(x')}{x'}.  \qquad ({\rm WW})
\end{split}
\label{www}
\ee
However, there have been recent attempts to constrain the genuine twist-three part of the $g_T(x)$ distribution from a global analysis of the ``worm-gear" TMD $g_{1T}(x,k_\perp)$  \cite{Bhattacharya:2021twu} supplemented with QCD equations of motion \cite{Bauer:2022mvl} and small-$x$ resummation \cite{Santiago:2023rfl}. Motivated by these developments, we try, in addition to (\ref{www}), two different estimations \cite{Bauer:2022mvl} based on the equation of motion (EOM)  and the Lorentz invariant relation (LIR) 
\be
\begin{split}
& x g^q_T(x) \approx g_{1T}^{(1)q}(x), \qquad ({\rm EOM})\\ 
& x g^q_T(x) \approx x\Delta q(x) + x \frac{d g_{1T}^{(1)q}(x)}{dx} ,
\qquad ({\rm LIR}) 
\end{split}
\label{eq:gTscene}
\ee
where $g_{1T}^{(1)}$ is the second moment of the TMD PDF $g_{1T}(x,k_\perp)$ from \cite{Bhattacharya:2021twu}. 
Of course, partly including  genuine twist-three corrections this way is not entirely consistent, since we have neglected the other twist-three corrections in  (\ref{eq:w0}), (\ref{k}) from the outset. Still, it is a useful exercise to get a rough estimate of the potential impact of genuine twist-three corrections.  According to the derivation of (\ref{eq:w0}) in \cite{Benic:2019zvg,Benic:2021gya}, we consider the EOM scenario to be more natural in the present context. However, the LIR scenario cannot be excluded. 

We plot $x g^{u,d}_T(x)$ in Fig.~\ref{fig:gT}
and $x^2 d g^{u,d}_T(x)/dx$ in Fig.~\ref{fig:dgT}, which enter the computation of SSAs (see, e.g., (\ref{eq:csdynew})). The $g_T^q(x)$ PDFs of the remaining flavors are set to zero  in the EOM and LIR scenarios, while  they are computed from $\Delta q(x)$ in the WW scenario, for which the result in \cite{Ethier:2017zbq} is adopted.    
The outcomes in the EOM and LIR scenarios display the importance of the genuine twist-three corrections, although the uncertainty bands are broad.  
In particular, the PDFs in the LIR scenario are qualitatively different,   exhibiting nodes when going from moderate to large $x$ \cite{Bauer:2022mvl}, because  the (numerical)  derivatives of $g_{1T}^{(1)}(x)$ are involved. 
For $\mathcal{G}_{3T}(x)$, we use  $\Delta G(x)$ from \cite{Ethier:2017zbq}.
 These plots already make clear that SSAs from the $g_T(x)$ distribution can be sizable only when the large-$x$, or valence quark region of the polarized proton is probed (cf., Fig.~6 of \cite{Benic:2021gya}).  

For the unpolarized cross section in the denominator,  we take the PDFs and FFs in the WW scenario from \cite{Owens:2012bv} and  \cite{Ethier:2017zbq}, respectively, including all the light flavors $q=u,d,s$ and the gluon. In the EOM and LIR scenarios, we take the PDFs from \cite{Lai:2010vv} and the FFs from \cite{Cammarota:2020qcw} in order to be consistent with \cite{Bhattacharya:2021twu}.  We will consider both the proton-proton and pion-proton Drell-Yan reactions. In the latter case,  we employ the $\pi^-$ PDF from \cite{Barry:2021osv}.


Our numerical evaluations take into account the following sources of uncertainty. The uncertainties in the $g_T^q(x)$ and $\mathcal{G}_{3T}(x)$ PDFs are inherited from $\Delta q(x)$ and $\Delta G(x)$ in the WW scenario, and from $g_{1T}^{(1) q}(x)$ in the EOM and LIR scenarios. The 1-$\sigma$ uncertainty from the Monte Carlo replica method is included.  Second, we set the factorization and renormalization scales in the PDFs/FFs and the running coupling $\alpha_S$ to $\xi Q$ and estimate the  scale uncertainty by changing the parameter $\xi$ within $0.5 < \xi < 2$. 
The described uncertainties are combined in quadrature in  the numerical results presented below.

\subsection{SSA in SIDIS revised}

We first revise  our predictions for SIDIS made in \cite{Benic:2021gya}. The calculation is based on the definitions of the SIDIS asymmetries given in (79), (75), (70) and (20) of \cite{Benic:2021gya}, which are expressed in terms of the $g^q_T(x)$ and $\mathcal{G}_{3T}(x)$ PDFs (and their derivatives) and the hard coefficients having been corrected in Secs.~\ref{sec:sidisq} and \ref{sec:sidisg}.

\begin{figure}
  \begin{center}
  \includegraphics[scale = 0.8]{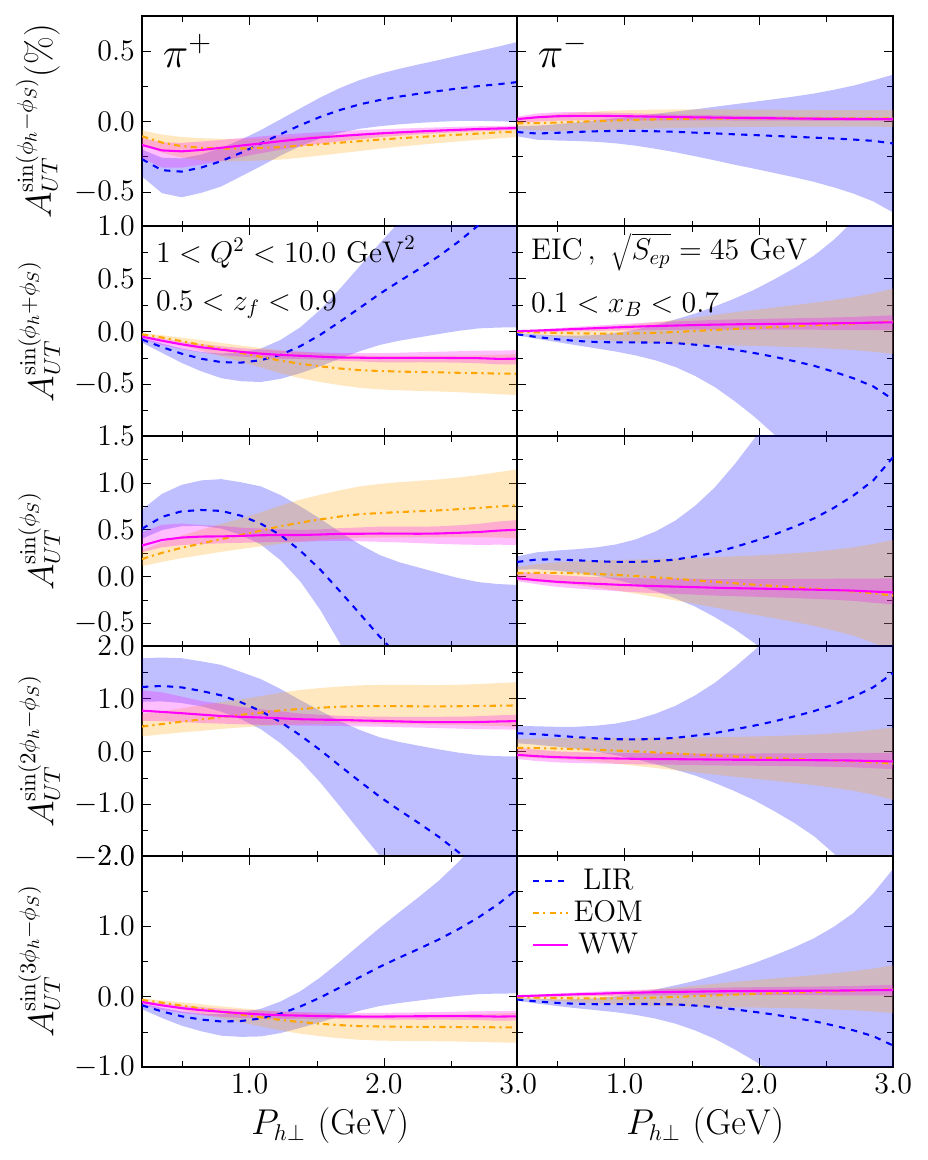}
  \end{center}
  \caption{The $P_{h\perp}$ dependencies of all the SIDIS asymmetries for the EIC kinematics at $\sqrt{S_{ep}} = 45$ GeV in the WW, EOM and LIR scenarios. Left (right) is for $\pi^+$ ($\pi^-$).}
  \label{fig:eicrevise}
\end{figure}

\begin{figure}
  \begin{center}
  \includegraphics[scale = 0.8]{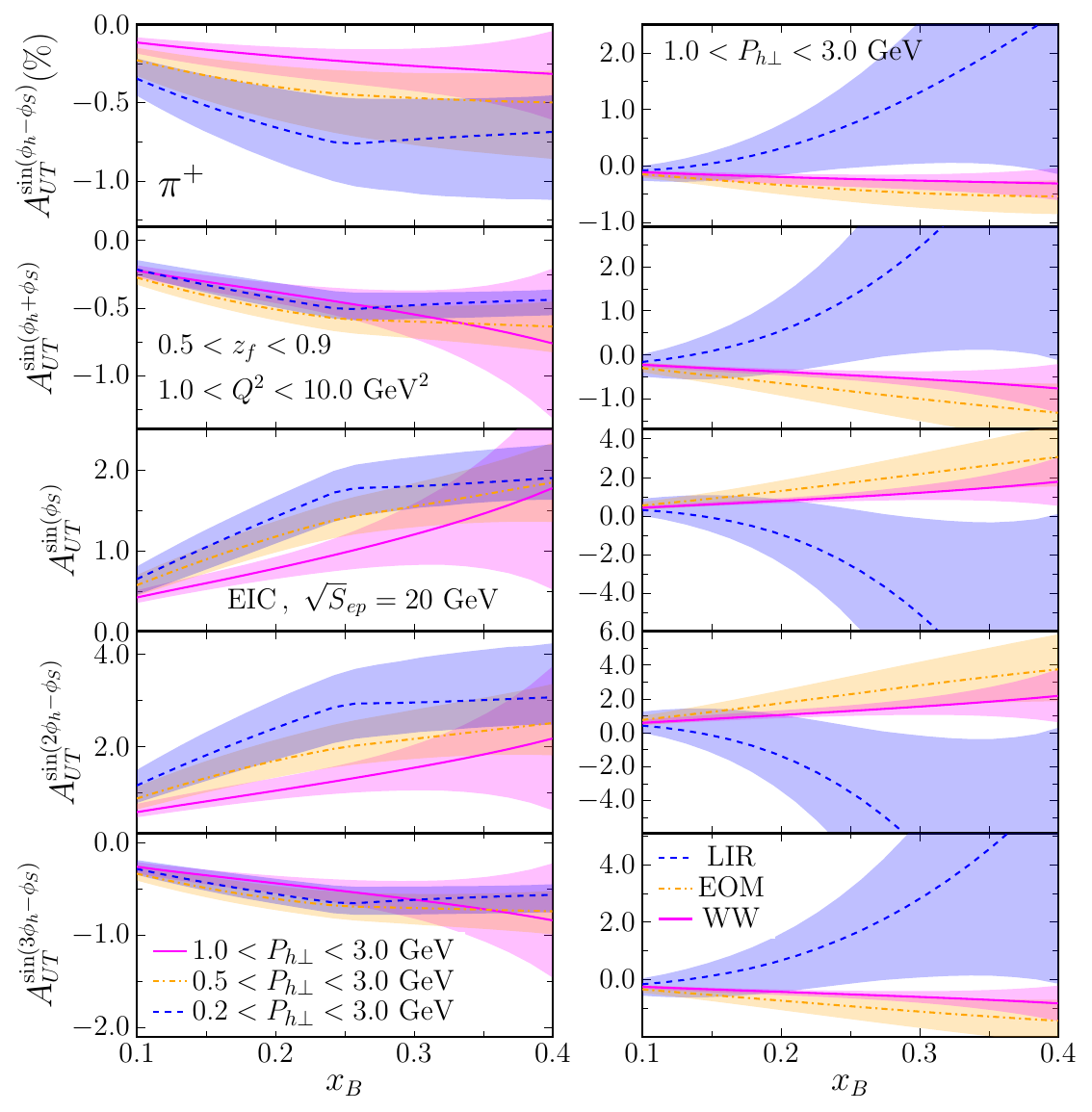}
  \end{center}
  \caption{The $x_B$ dependencies of all the SIDIS  $\pi^+$ asymmetries for the EIC kinematics at  $\sqrt{S_{ep}} = 20$ GeV. Left: different lower $P_{h\perp}$ cutoffs in the WW  scenario. Right: different scenarios.  }
  \label{fig:eicrevise2}
\end{figure}

We focus on the comparison with Fig.~8 in \cite{Benic:2021gya}, which was one of the main results there.  
In Fig.~\ref{fig:eicrevise} we show the results of all the asymmetries as a function of $P_{h\perp}$ for $\pi^+$ (left) and $\pi^-$ (right), considering the typical EIC kinematics with $\sqrt{S_{ep}} = 45 \, {\rm GeV}$, $0.5 < z_f < 0.9$ and $1 <  Q^2 < 10$ GeV$^2$. We have imposed a lower cut $W > 5$ GeV on $W^2 = (q + P)^2 = Q^2(1-x_B)/x_B$.
The cuts for Fig.~\ref{fig:eicrevise} are the same as in Fig.~8 of \cite{Benic:2021gya}, except that we extend the $P_{h\perp}$ range to below 1 GeV without rigorous justification. Comparing the WW result in the left plot of Fig.~\ref{fig:eicrevise} (for $\pi^+$) and the first column in Fig.~8 of \cite{Benic:2021gya}, we notice that the Sivers moment $A_{UT}^{\sin(\phi_h - \phi_S)}$ drops to sub-percent level, compared to the percent level in Fig.~8 of \cite{Benic:2021gya}. The difference is attributed to the correction of the hard coefficient $\Delta\hat{\sigma}^i_1$. Results for the other moments $A_{UT}^{\sin(\phi_h + \phi_S)}$ (Collins asymmetry),  $A_{UT}^{\sin(3\phi_h - \phi_S)}$,  $A_{UT}^{\sin(\phi_S)}$  and $A_{UT}^{\sin(2\phi_h - \phi_S)}$ are not significantly affected, because they are dominated by the derivative terms ($d g^q_T/dx$ and $d\mathcal{G}_{3T}/dx$), for which the respective hard coefficients $\Delta\hat{\sigma}_{Dk}$ do not change. We observe that the $\pi^-$ asymmetries are smaller than the $\pi^+$ ones in the WW scenario partly owing to $|g_T^u|>|g_T^d|$.  
Figure~\ref{fig:eicrevise} also contains new predictions based on $g_T^q(x)$ from the EOM and LIR scenarios, that are not present in \cite{Benic:2021gya}.


The modifications on the other results in \cite{Benic:2021gya} are qualitatively similar and we will not repeat them here. Instead,  we explore alternative kinematics to optimize the impact of our new mechanism and find that the asymmetries tend to become larger at lower center-of-mass energies. We show in Fig.~\ref{fig:eicrevise2} new predictions for $\pi^+$ as functions of $x_B$ at the EIC with $\sqrt{S_{ep}}=20$ GeV, $0.01<y<0.95$, $0.5<z_f<0.9$ and $1<Q<10$ GeV, and without a cut on $W$. Three curves in the left panel correspond to the WW scenario with different lower $P_{h\perp}$ cuts, on which the asymmetries depend rather sensitively. The $\sin \phi_S$ and $\sin (2\phi_h-\phi_S)$ asymmetries reach a few percent now, and can get even larger after the twist-three corrections are included as shown in the right panel.

\subsection{SSA in Drell-Yan} 

We then present the results for transverse SSAs in Drell-Yan.  
The spin-dependent part of the cross section can be decomposed  as (cf. (\ref{dyangle}))
\be
\begin{split}
\frac{d\Delta\sigma}{d^4 q d\Omega} & = \sin\Phi_S \left(\calF_1 + \calF_2 \cos\phi_{cs} + \calF_3  \cos 2\phi_{cs}\right) + \cos\Phi_S \left(\calF_4 \sin\phi_{cs} + \calF_5 \sin 2\phi_{cs}\right)\\
& = F^{\sin\Phi_S} \sin\Phi_S + F^{\sin(\phi+\Phi_S)}\sin(\phi_{cs} + \Phi_S) + F^{\sin(\phi_{cs} - \Phi_S)}\sin(\phi_{cs} - \Phi_S)\\
& \qquad  + F^{\sin(2\phi_{cs}+\Phi_S)}\sin(2\phi_{cs} + \Phi_S) + F^{\sin(2\phi_{cs} - \Phi_S)}\sin(2\phi_{cs} - \Phi_S)\,,
\end{split}
\ee
where
\be
\begin{split}
&F^{\sin\Phi_S} = \calF_1\,,\\
& F^{\sin(\phi_{cs}+\Phi_S)} = \frac{1}{2}\left(\calF_2 + \calF_4\right)\,,\\
& F^{\sin(\phi_{cs}-\Phi_S)} = \frac{1}{2}\left(-\calF_2 + \calF_4\right)\,,\\
& F^{\sin(2\phi_{cs}+\Phi_S)} = \frac{1}{2}\left(\calF_3 + \calF_5\right)\,,\\
& F^{\sin(2\phi_{cs}-\Phi_S)} = \frac{1}{2}\left(-\calF_3 + \calF_5\right)\,.
\end{split}
\ee
Explicitly, we write
\be
\begin{split}
\mathcal{F}_1 &= \frac{3}{8} \alpha_s\calF_0 \int \frac{d x}{x}\int\frac{dx'}{x'}\delta(\hats + \hatt + \hatu - Q^2)\sum_q e_q^2\\
&\quad \times \left(1+\cos^2\theta_{cs}-\frac{4}{3}\right)\left[\bar{q}(x')x g_{T}^q(x)\Delta\hat{\sigma}_1^{q\bar{q}} + G(x') x g_T^q(x)\Delta\hat{\sigma}_1^{\rm C} + 2 q(x')x\mathcal{G}_{3T}(x)\Delta\hat{\sigma}_1^{{\rm C}(g)}\right]\\
\mathcal{F}_2 & = \frac{3}{8} \alpha_s\calF_0 \int \frac{d x}{x}\int\frac{dx'}{x'}\delta(\hats + \hatt + \hatu - Q^2)\sum_q e_q^2\\
& \quad\times \sin 2\theta_{cs}\left[\bar{q}(x')x g_T^q(x)\Delta\hat{\sigma}_3^{q\bar{q}} + G(x') x g^q_T(x)\Delta\hat{\sigma}_3^{\rm C} + 2q(x') x\mathcal{G}_{3T}(x)\Delta\hat{\sigma}_3^{{\rm C}(g)}\right]\,,\\
\mathcal{F}_3 & =  \frac{3}{8} \alpha_s\calF_0 \int \frac{d x}{x}\int\frac{dx'}{x'}\delta(\hats + \hatt + \hatu - Q^2)\sum_q e_q^2\\
&\quad \times \sin^2\theta_{cs}\left[\bar{q}(x')x g_T^q(x)\Delta\hat{\sigma}_4^{q\bar{q}} + G(x') x g^q_T(x)\Delta\hat{\sigma}_4^{\rm C}  + 2q(x') x\mathcal{G}_{3T}(x)\Delta\hat{\sigma}_4^{{\rm C}(g)}\right]\,,\\
\mathcal{F}_4 & =  \frac{3}{8} \alpha_s\calF_0 \int \frac{d x}{x}\int\frac{dx'}{x'}\delta(\hats + \hatt + \hatu - Q^2)\sum_q e_q^2\\
&\quad \times (-\sin 2\theta_{cs})\bigg[\bar{q}(x')\left(x^2 \frac{d g^q_T}{dx}\Delta\hat{\sigma}_{D8}^{q\bar{q}} + x g^q_T(x)\Delta\hat{\sigma}_8^{q\bar{q}}\right) + G(x') \left(x^2 \frac{d g_T^q}{dx}\Delta\hat{\sigma}_{D8}^{\rm C} + x g^q_T(x)\Delta\hat{\sigma}_8^{\rm C}\right)\\
&\quad  + 2q(x')\left(x^2\frac{d\mathcal{G}_{3T}}{dx}\Delta\hat{\sigma}_{D8}^{{\rm C}(g)} + x\mathcal{G}_{3T}(x)\Delta\hat{\sigma}_8^{{\rm C}(g)}\right)\bigg]\,,\\
\mathcal{F}_5 & =  \frac{3}{8} \alpha_s\calF_0 \int \frac{d x}{x}\int\frac{dx'}{x'}\delta(\hats + \hatt + \hatu - Q^2)\sum_q e_q^2\\
&\quad \times \sin^2\theta_{cs}\bigg[\bar{q}(x')\left(x^2 \frac{d g^q_T}{dx}\Delta\hat{\sigma}_{D9}^{q\bar{q}} + x g^q_T(x)\Delta\hat{\sigma}_9^{q\bar{q}}\right) + G(x') \left(x^2 \frac{d g_T^q}{dx}\Delta\hat{\sigma}_{D9}^{\rm C} + x g^q_T(x)\Delta\hat{\sigma}_9^{\rm C}\right)\\
&\quad  + 2q(x')\left(x^2\frac{d\mathcal{G}_{3T}}{dx}\Delta\hat{\sigma}_{D9}^{{\rm C}(g)} + x\mathcal{G}_{3T}(x)\Delta\hat{\sigma}_9^{{\rm C}(g)}\right)\bigg]\,,\\
\end{split}
\label{f12345}
\ee
with
\be
\calF_0 = \frac{\alpha_{em}^2 \alpha_sM_N}{3 s Q^3 \pi^2 N_c}\,.
\ee
In ${\cal F}_1$, we have eliminated $\Delta \hat{\sigma}_2^i$, $i=q\bar{q},{\rm C},{\rm C}(g)$, by means of the relation $\Delta \hat{\sigma}_2^i=2\Delta\hat{\sigma}_1^i/3$. As a consequence, the $\theta_{cs}$-dependence has been altered from the more familiar form $1+\cos^2\theta_{cs}$.  Note that $\int_{-1}^1 d\cos\theta_{cs}\left(\cos^2\theta_{cs}-1/3\right)=0$, namely, the asymmetries will vanish if the lepton angles $\theta_{cs},\phi_{cs}$ are integrated over as already mentioned.

The COMPASS Collaboration has reported in \cite{COMPASS:2017jbv,COMPASS:2023vqt} the first measurements of transverse SSAs in pion-induced Drell-Yan. In their work, the asymmetries in the CS frame are normalized by the unpolarized cross section which, in the COMPASS kinematics  $Q\gg q_\perp$, exhibits the angular distribution 
\beq
\frac{1}{\sigma}\frac{d\sigma}{d\Omega}=\frac{3}{16\pi} \left( \frac{Q^2+\frac{3}{2}q_\perp^2}{Q^2+q_\perp^2}+\frac{Q^2-\frac{1}{2}q_\perp^2}{Q^2+q_\perp^2}\cos^2\theta_{cs}+\cdots\right)\approx \frac{3}{16\pi}(1+\cos^2\theta_{cs}).
\eeq
Following the expectation based on TMDs \cite{Arnold:2008kf}, they assumed ${\cal F}_1\propto 1+\cos^2\theta_{cs}$, so that the $\theta_{cs}$-dependence drops out in the $\sin \Phi_S$ asymmetry whereas the $\sin (2\phi_{cs}\pm \Phi_S)$ harmonics come with the prefactor $\sin^2\theta_{cs}/(1+\cos\theta_{cs}^2)$. The $\sin(\phi_{cs}\pm \Phi_S)$ harmonics have been neglected, since they do not arise from twist-two TMDs \cite{Arnold:2008kf}. 
 In this paper we do not assume that $q_\perp$ is small compared to $Q$. Besides, in the present mechanism ${\cal F}_1$ is not proportional to $1+\cos^2\theta_{cs}$ as we have seen.  It is then more convenient to normalize the asymmetries using the angular-integrated cross section \eqref{eq:dyunpol}. 
We thus define 
\beq
\frac{3}{16\pi}d^{\alpha,\beta}(\theta_{cs})A_{UT}^{\sin(\alpha  \phi_{cs}+\beta \Phi_S)}(Q,q_\perp,y) &\equiv& \frac{2\int_0^{2\pi} d\phi_{cs} \int_{0}^{2\pi} \frac{d\Phi_S}{2\pi} \sin(\alpha\phi_{cs}+\beta \Phi_S)\left[d\sigma(\phi_{cs},\Phi_S) - d\sigma(\phi_{cs},\Phi_S + \pi)\right]}{\frac{1}{2}\int_{-1}^1d\cos\theta_{cs}\int_0^{2\pi} d\phi_{cs} \int_{0}^{2\pi} \frac{d\Phi_S}{2\pi} \left[d\sigma(\phi_{cs},\Phi_S) + d\sigma(\phi_{cs},\Phi_S + \pi)\right]} \nn
&=& \left( \frac{d\sigma}{d^4q}\right)^{-1} 2\int_0^{2\pi} d\phi_{cs} \int_{0}^{2\pi} \frac{d\Phi_S}{2\pi} \sin(\alpha\phi_{cs}+\beta \Phi_S)d\sigma(\phi_{cs},\Phi_S) \,, \label{compdef}
\eeq
where 
$d\sigma(\phi_{cs},\Phi_S) \equiv \frac{d \sigma}{d^4 q d\Omega}$. 
The `depolarization factors' are defined as $d^{0,1}(\theta_{cs})=\cos^2\theta_{cs}-1/3$, $d^{1,\pm 1}(\theta_{cs})=\sin 2\theta_{cs}$ and $d^{2,\pm 1}(\theta_{cs})=\sin^2\theta_{cs}$. 
The above relations yield 
\beq
A_{UT}^{\sin\Phi_s}&=& \frac{\alpha_s M_N} {Q}
\frac{
\sum_q e_q^2\left[\bar{q}(x')x g_{T}^q(x)\Delta\hat{\sigma}_1^{q\bar{q}} + G(x') x g_T^q(x)\Delta\hat{\sigma}_1^{\rm C} + 2 q(x')x\mathcal{G}_{3T}(x)\Delta\hat{\sigma}_1^{{\rm C}(g)}\right]}{ 
\sum_{q} e_q^2\bigl(\sigma_{q\bar{q}} q(x)\bar{q}(x') + \sigma_{qg} q(x)G(x') + \sigma_{gq}G(x)q(x')\bigr)} ,
\label{siversqq}
\eeq
\beq
A_{UT}^{\sin(\phi_{cs}\pm \Phi_s)}&=& \frac{\alpha_s M_N}{Q} \frac{\frac{1}{2}
\sum_q e_q^2\left[\pm \bar{q}(x')xg_T^q(x)\Delta\hat{\sigma}_3^{q\bar{q}}-\bar{q}(x')\left(x^2 \frac{d g^q_T}{dx}\Delta\hat{\sigma}_{D8}^{q\bar{q}} + x g^q_T(x)\Delta\hat{\sigma}_8^{q\bar{q}}\right) +\cdots \right]}{ 
\sum_{q} e_q^2\bigl(\sigma_{q\bar{q}} q(x)\bar{q}(x') + \sigma_{qg} q(x)G(x')  + \sigma_{gq}G(x)q(x')\bigr)} ,
\label{acompton}
\eeq
\beq
A_{UT}^{\sin(2\phi_{cs}\pm \Phi_s)}&=& \frac{\alpha_s M_N}{Q}
 \frac{\frac{1}{2}
 \sum_q e_q^2\left[\pm \bar{q}(x')xg_T^q(x)\Delta\hat{\sigma}_4^{q\bar{q}}+\bar{q}(x')\left(x^2 \frac{d g^q_T}{dx}\Delta\hat{\sigma}_{D9}^{q\bar{q}} + x g^q_T(x)\Delta\hat{\sigma}_9^{q\bar{q}}\right) +\cdots\right]}{ 
 \sum_{q} e_q^2\bigl(\sigma_{q\bar{q}} q(x)\bar{q}(x') + \sigma_{qg} q(x)G(x')  + \sigma_{gq}G(x)q(x')\bigr)} ,
\label{acomptong}
\eeq
where the phase space integral $\int \frac{dx}{x} \int \frac{dx'}{x'} \delta(\hat{s}+\hat{t}+\hat{u}-Q^2)$ is implied in both the numerator and denominator. The summation over $q$ includes antiquarks. 
 The dots in (\ref{acompton}) and (\ref{acomptong}) stand for  the contributions from the Compton subprocesses, which can be easily deduced from (\ref{f12345}). In the small-$q_\perp$ region, our definition of $A_{UT}^{\sin(2\phi_{cs}\pm \Phi_S)}$ in terms of the measured cross section (\ref{compdef}) agrees with that in \cite{COMPASS:2017jbv},  but $A_{UT}^{\sin \Phi_S}$ does not because of the different $\theta_{cs}$-dependence. We note that the COMPASS Collaboration  did not directly confirm  the  $1+\cos^2\theta_{cs}$ dependence of the $\sin\Phi_S$ asymmetry from the data.


The integration over $x'$ can be performed by using the delta function constraint  
\beq
\delta(\hat{s}+\hat{t}+\hat{u}-Q^2)  
&=& \delta\left( -s \left(x'-\frac{m_\perp e^{-y}}{\sqrt{s}}\right) \left(x-\frac{m_\perp e^{y}}{\sqrt{s}}\right) +q_\perp^2\right) .
\eeq
The minimum of $x$ is attained when $x'=1$, so the integration range of $x$ is given by
\beq
1>x \ge x_{\rm min} = \frac{m_\perp e^y}{\sqrt{s}} + \frac{q_\perp^2}{s \left(1-\frac{m_\perp e^{-y}}{\sqrt{s}}\right)}.
\eeq
We also need the following relations in the center-of-mass frame
\beq
x_a=\frac{e^y Q^2}{xm_\perp \sqrt{s}}, \qquad x_b= \frac{Q^2(x\sqrt{s}-e^y m_\perp)}{m_\perp(x\sqrt{s}m_\perp -e^y Q^2)}, \qquad c=\frac{Q^2}{q_\perp^2 + Q^2}.
\eeq
Alternatively introducing the `partonic rapidity' $Y$ as  $x,x'=\sqrt{\frac{\hat{s}}{s}}e^{\pm Y}$, we proceed with 
\beq
\int \frac{dx dx'}{xx'}\delta(\hat{s}+\hat{t}+\hat{u}-Q^2)f(x)g(x')\cdots = \int \frac{2dY}{\hat{s}-Q^2}f\left(\sqrt{\frac{\hat{s}}{s}}e^Y\right)g\left(\sqrt{\frac{\hat{s}}{s}}e^{-Y}\right)\cdots,
\label{eq:partY}
\eeq
where 
\beq
\sqrt{\hat{s}}= m_\perp \cosh(Y-y)+\sqrt{m_\perp^2\cosh^2(Y-y)-Q^2},
\eeq
 and
\beq
x_a= \frac{Q^2}{\sqrt{\hat{s}}m_\perp }e^{y-Y}, \qquad x_b= \frac{Q^2}{\sqrt{\hat{s}}m_\perp }e^{Y-y}.
\eeq
We have checked the consistency between the two methods.
 
\begin{figure}
  \begin{center}
  \includegraphics[scale = 0.8]{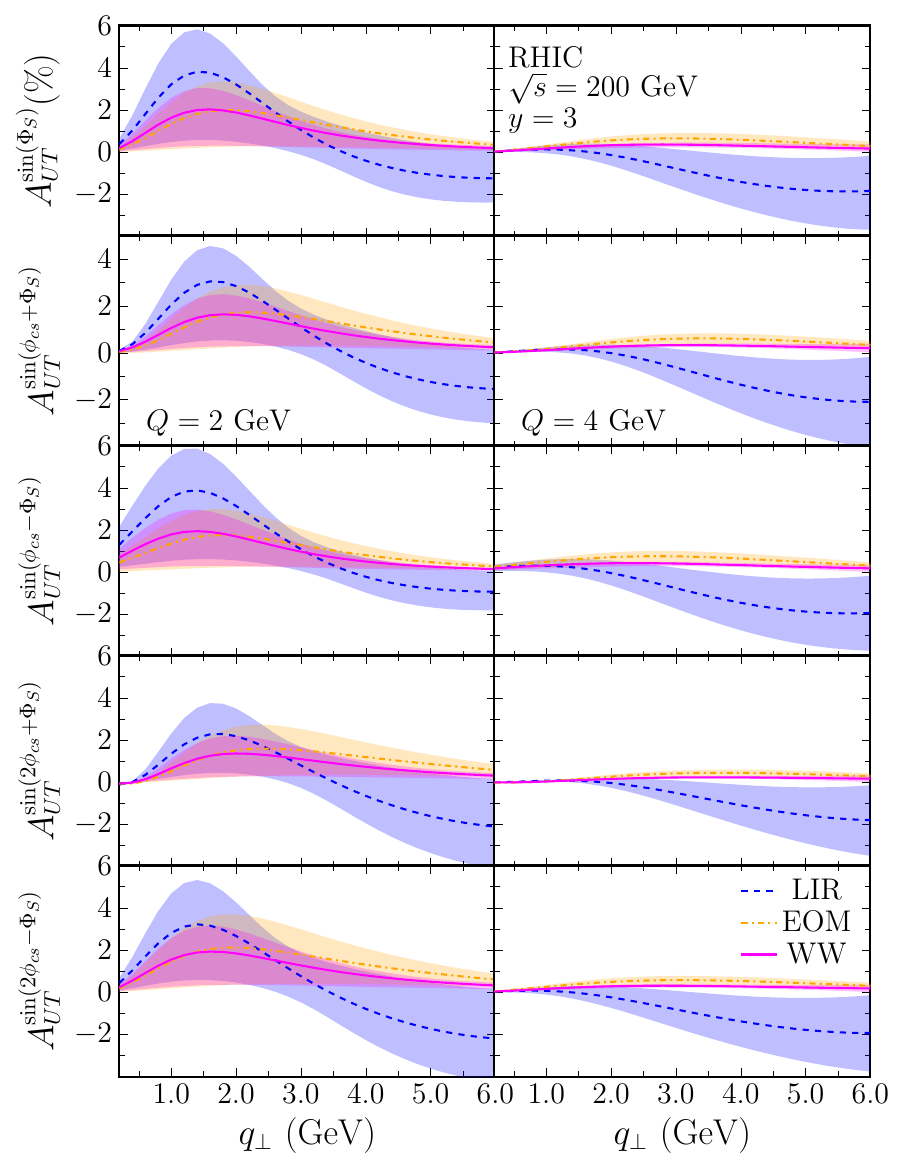}
  \end{center}
  \caption{The $q_\perp$ dependencies of all the DY asymmetries for the $\sqrt{s} = 200$ GeV RHIC kinematics. Here we have fixed $y = 3$ and $Q = 2$ GeV (left) and $Q=4$ GeV (right).}
  \label{fig:dyrhic}
\end{figure}

The numerical outcomes for the DY asymmetries defined by Eqs.~\eqref{siversqq}-\eqref{acomptong} are firstly given in Fig.~\ref{fig:dyrhic}
for the RHIC kinematics with $\sqrt{s} = 200$ GeV. All the DY azimuthal moments are displayed as functions of $q_T$ for the forward case $y = 3$ and with $Q = 2$ GeV (left) and $Q=4$ GeV (right). When $Q=2$ GeV, all the considered scenarios, i.e., WW, EOM and LIR, for $g_T^q(x)$ and $\mathcal{G}_{3T}(x)$, yield percent-level and positive asymmetries in the region $q_T \approx 1-5$ GeV. The $\sin (\phi_{cs}\pm \Phi_S)$ asymmetries, expected to be small in the TMD framework, are comparable to the other asymmetries. 
In the large $q_T$ region, some of the asymmetries tend to be negative in the LIR scenario. While the results are generally largest in magnitude in the LIR scenario, so is the overall uncertainty which is mostly due to  scale variations. Interestingly, the results exhibit a strong $Q$-dependence. Shifting to $Q = 4$ GeV, the asymmetries drop to about half of a percent (see a discussion in the concluding section).


\begin{figure}
  \begin{center}
  \includegraphics[scale = 0.8]{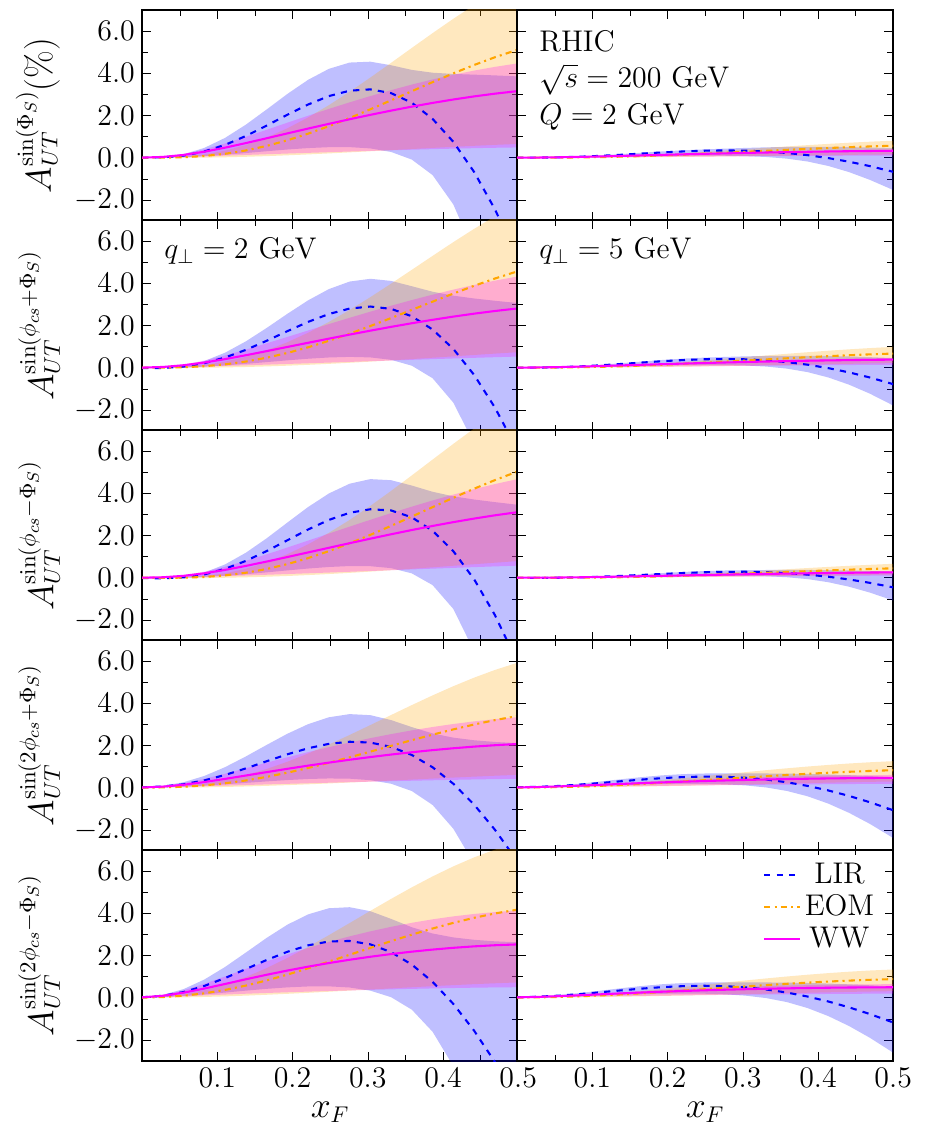}
  \end{center}
  \caption{The $x_F$ dependencies of all the DY asymmetries for the $\sqrt{s} = 200$ GeV RHIC kinematics. We have fixed $Q = 2$ GeV and $q_\perp=2 (5)$ GeV left (right).}
    \label{fig:dyrhicxF}
\end{figure}

In Fig.~\ref{fig:dyrhicxF} the RHIC results are presented as functions of the Feynman-$x$ defined by $x_F = 2 q^3/\sqrt{s} = 2 m_\perp\sinh(y)/\sqrt{s}$, so that large and positive $x_F$ corresponds to the forward region, i.e., large $x$ in the polarized proton. Here we have fixed $Q = 2$ GeV, and $q_\perp = 2$ GeV (left) and $q_\perp=5$ GeV  (right).  $x_F$ can be understood as a proxy for $y$ with the $x_F = 0.5$ corresponding to the very forward case $y \approx 4.5$. We have also explored the negative $x_F$ region, observing very small, sub-percent asymmetries. All the DY asymmetries become enhanced in the large $x_F$ region sensitive to the valence content of the polarized proton. For $q_\perp = 2$ GeV, the asymmetries are mostly positive, reaching up to about $5\%$ in their central values, with increasing uncertainties. The asymmetries change sign above $x_F \approx 0.4$ in the LIR scenario, reflecting the sign change in the $g_T^q$ PDF and its derivative from Figs.~\ref{fig:gT} and \ref{fig:dgT}.

\begin{figure}
  \begin{center}
  \includegraphics[scale = 0.8]{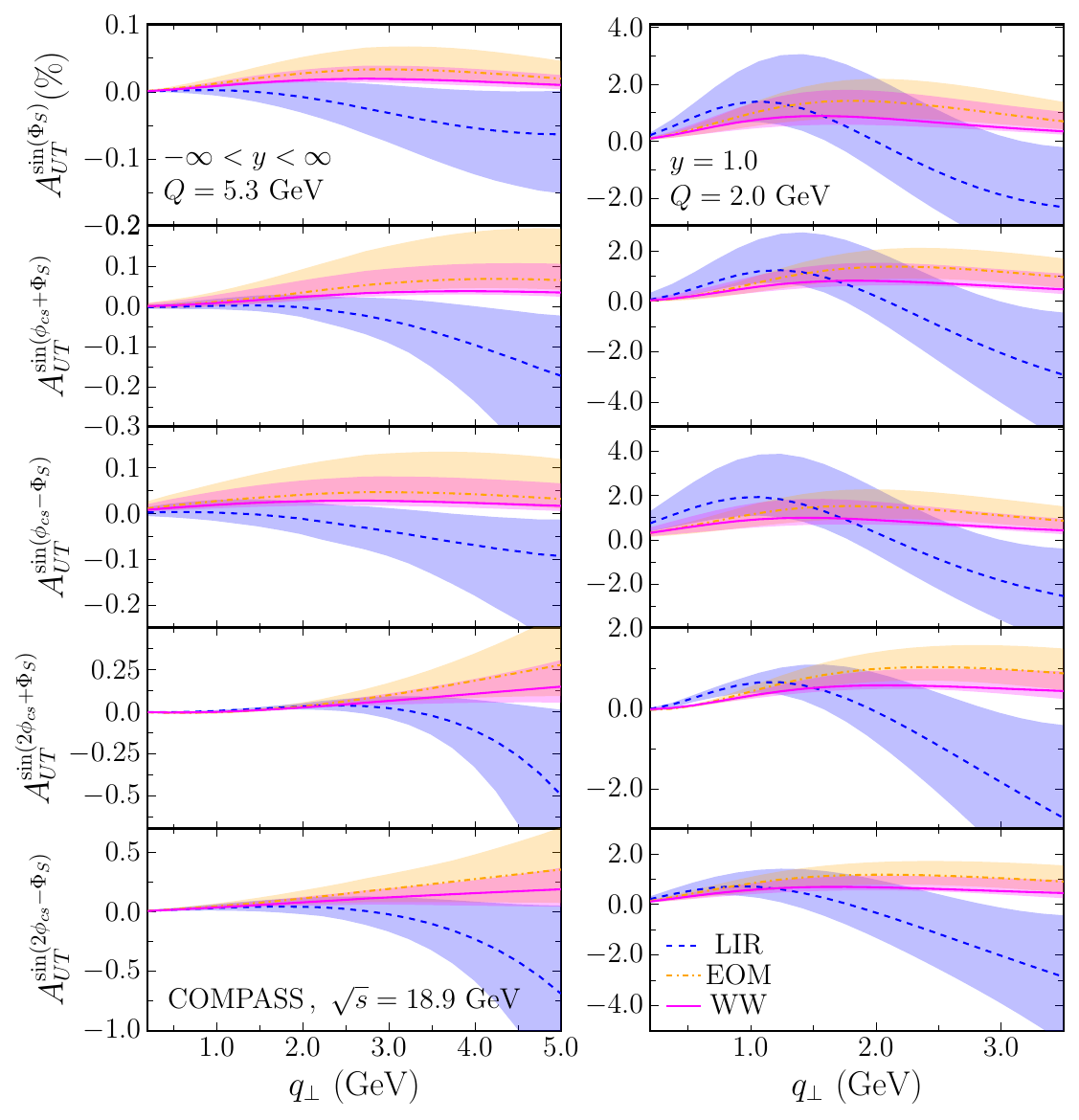}
  \end{center}
  \caption{The asymmetries $A_{UT}$'s as functions of $q_T$ for the COMPASS kinematics at$\sqrt{s} = 18.9$ GeV. On the left panel we have fixed $Q = 5.3$ GeV and integrated over the rapidity $y$ of the DY pair. On the right panel the results are obtained for $Q = 2.0$ GeV and $y = 1.0$. }
  \label{fig:dycompass}
\end{figure}

Next we turn to the analysis on the pion induced Drell-Yan that was measured at COMPASS, where $\pi^-$ with the momentum 190 GeV collides with a proton at rest \cite{COMPASS:2017jbv} corresponding to $\sqrt{s} \approx 18.9$ GeV. We take the average value of the Drell-Yan virtuality $Q=5.3$ GeV at COMPASS. Our predictions, shown in Fig.~\ref{fig:dycompass} (left), as  functions of $q_T$ are obtained by integrating over the rapidity $y$ of the Drell-Yan pair.
COMPASS is mostly sensitive to the valence region, so the $\bar{u} u \to g\gamma$ channel dominates, with the $\bar{u}$ being the valence quark in $\pi^-$. The large values of $x$ probed in the polarized proton lead to qualitatively different results for the three scenarios. Whereas in the WW and the EOM scenarios the asymmetries are positive, 
 we find mostly negative results in the LIR scenario.  COMPASS \cite{COMPASS:2017jbv,COMPASS:2023vqt} reported ${\cal O}(10\%)$ asymmetries for the $\sin \Phi_S$, $\sin (2\phi_{cs}\pm \Phi_S)$ harmonics in the region $q_\perp >1$ GeV, although there are only two data points and experimental errors are as large as the central values. It is clear that the perturbative contribution is very small in the considered kinematics compared to other TMD-based nonperturbative contributions, e.g., \cite{Cammarota:2020qcw,Echevarria:2020hpy,Bury:2021sue,Gurjar:2023uho}.  On the other hand, the asymmetries can reach percent-level magnitude for lower $Q$ values and when large $x$ in the polarized proton is probed, similar to the RHIC case investigated above. This is demonstrated by the right panel of Fig.~\ref{fig:dycompass} for $Q = 2$ GeV and $y = 1$, which correspond to $0.24<x_F<0.48$ and $0.2<q_\perp<3.5$ GeV.

\begin{figure}
  \begin{center}
  \includegraphics[scale = 0.8]{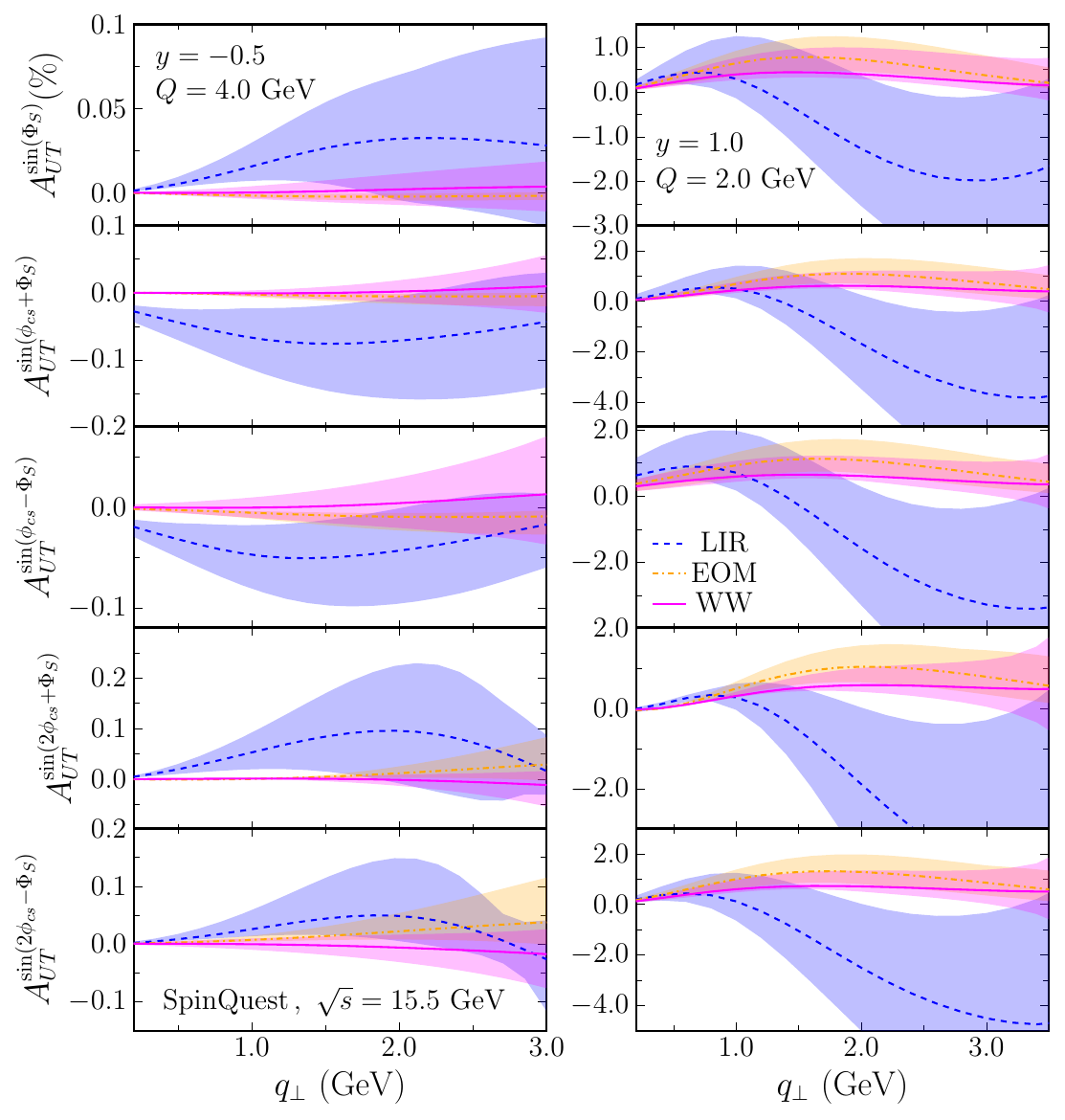}
  \end{center}
  \caption{The asymmetries $A_{UT}$'s as  functions of $q_T$ for the Fermilab SpinQuest collision energy $\sqrt{s} = 15.5$ GeV. We have taken $y = -0.5$, $Q = 4.0$ GeV (left) and $y = 1.0$, $Q = 2.0$ GeV (right).}
  \label{fig:dyfermilab}
\end{figure}

Finally, we show in Fig.~\ref{fig:dyfermilab} the predictions for the future fixed target measurement in the SpinQuest experiment at Fermilab \cite{Brown:2014sea,SeaQuest:2019hsx}, where a 120 GeV proton (instead of a $\pi^-$)  collides with a polarized target proton at rest, corresponding to $\sqrt{s} \approx 15.5$ GeV.  On the left panel we take $Q = 4$ GeV within the intended windows $3 < Q < 10$ GeV and $y = -0.5$, 
reflecting SpinQuest's preference  towards the small-$x$ (sea-quark) regions in the polarized target. With this kinematics the asymmetries are all sub-percent even after the  uncertainty bands are taken into account, which increase as $q_\perp$ is increased towards the kinematic threshold.  On the other hand, when we go to lower $Q$ regions, the perturbative contributions are at the percent level as shown on the right panel in Fig.~\ref{fig:dyfermilab} for $Q = 2$ GeV and $y = 1$. It is seen that the behaviors of various asymmetries are qualitatively similar to the COMPASS $\pi^-$ results.

\section{Discussions and conclusions}

In this paper,  we have proposed a unified formlaism to analyze SSAs in  SIDIS and Drell-Yan and for both longitudinally and transversely polarized protons. 
The new method is more efficient than our previous brute-force approach to transverse SSAs  in SIDIS   \cite{Benic:2021gya} and actually helps uncover mistakes in part of the earlier calculation \cite{Benic:2021gya}. The predictions for  transverse SSAs in Drell-Yan from the present mechanism are entirely new. We reproduced the known results for the longitudinal polarization case in  the literature on both SIDIS \cite{Abele:2022spu} and Drell-Yan \cite{Carlitz:1992fv,Yokoya:2007xe}. We emphasize that the complete QCD results for SIDIS and Drell-Yan have been presented, which should supersede the naive parton model argument $A_{UT}\sim \alpha_s m_q/\sqrt{s}$  \cite{Kane:1978nd} frequently quoted in the literature. 

 We have   observed that the asymmetries in the WW approximation vanish in the limit $Q\to 0$, i.e., the photoproduction limit of SIDIS and  direct photon production in $pp$ collisions, and by extension, also in light-hadron production in $pp$. In these processes and within the WW approximation, SSAs may be indeed proportional to a current quark mass $m_q$. An evaluation n the parton model with nonzero $m_q$ was performed in \cite{Dharmaratna:1996xd}. In QCD, the second term in (\ref{twoterm}) needs to be included, which will cancel the logarithmic singularity $\ln m_q$ in the first term. After this cancellation we expect a contribution of the form  
\beq
A_{UT}\propto \alpha_s M_N m^2_q g_T(x),
\eeq
which is even more suppressed  than the naive estimate $A_{UT}\propto \alpha_s m_q$ for light quarks.

Based on the analytical formulas, we have numerically evaluated the asymmetries in Drell-Yan  relevant to the experiments at RHIC, COMPASS, SpinQuest and EIC. The asymmetries can reach a few percent level in some preferable kinematics associated with the large-$x$ (valence) region of the polarized proton, but stay at sub-percent level elsewhere. Yet, they are one or two  orders of magnitude larger  than the common prejudice  $A_{UT}\sim {\cal O}(10^{-4})$ based on the naive estimate $A_{UT}\propto \alpha_s m_q/\sqrt{s}$. The enhancement is mostly attributed to the ratio $M_N/m_q\sim {\cal O}(10^2)$, but there is also  a suppression factor $x\Delta q(x)/q(x)<1$ from the PDFs in the WW approximation.  

The dependencies of $A_{UT}$ on $\sqrt{s}$, $q_\perp$ and $Q$ can be roughly inferred from the analytic expressions for the partonic cross sections summarized in Sec. \ref{summary}, which, however, will be significantly modified by the convolution with PDFs. In general, there exists no simple analytical formula for $A_{UT}(\sqrt{s},q_\perp,Q)$. Nevertheless, we have observed that the dependence on $\sqrt{s}$ is relatively weak in Drell-Yan, and an approximate power-law behavior in the limit $Q\gg q_\perp$,
\beq 
A_{UT}^{\sin (\alpha \phi_{cs}+\beta \Phi_S)} \propto \left(\frac{q_\perp\, {\rm or} \, M_N}{Q }\right)^{B_{\alpha,\beta}}, \qquad Q \gg q_\perp,M_N .\label{scaling}
\eeq
A numerical extraction gives the exponents 
\beq
B_{0,1} \sim 1.7, \quad B_{1,1} \sim 1.4, \quad B_{1,-1} \sim 0.7, \quad B_{2,1}\sim 0.7, \quad B_{2,-1} \sim 1.2 ,
\eeq
at $y=0$ and $q_\perp=2$. 
These exponents are sensitive to $y$ (and also to $q_\perp$ to a lesser extent), since dominant partonic channels vary with values of $y$. In practice, the logarithmic dependence $\ln Q^2/q_\perp^2$ is also expected in the hard coefficients, which is not taken into account in the above parameterization. We should mention that the scaling (\ref{scaling}) is seen only in a region where the magnitude of $A_{UT}$ is already negligibly small.   Similarly, in the opposite regime with
\beq 
A_{UT}^{\sin (\alpha \phi_{cs}+\beta \Phi_S)} \propto \left(\frac{Q\, {\rm or} \, M_N}{q_\perp }\right)^{C_{\alpha,\beta}}, \qquad q_\perp \gg Q,M_N,
\eeq
we find, for $y=2$ and $Q=2$, 
\beq
C_{0,1} \sim 2,0, \quad C_{1,1} \sim  1.2, \quad C_{1,-1} \sim 1.2 ,\quad C_{2,1}\sim 0.8 ,\quad C_{2,-1} \sim 0.8.
\eeq
These non-integer exponents indicate that the actual $q_\perp$ or $Q$ dependence is more complicated than what naively follows from the dimensional or twist counting argument $A_{UT}\sim 1/q_\perp$ or $A_{UT}\sim 1/Q$. A similar comment applies to the SIDIS case.

At last, we have also explored the possible impact of the genuine twist-three corrections using the recent extraction of $g_T(x)$ from the global analysis \cite{Bhattacharya:2021twu} and the equations of motion \cite{Bauer:2022mvl}. It was found that these corrections tend to enhance the asymmetries, and can even flip the sign of the asymmetries in the LIR scenario. However, since  our formulas involve the derivative of $g_T(x)$, uncertainties of $g_T(x)$ are amplified. A more accurate determination of $g_T(x)$ is therefore welcome for future studies.

\acknowledgements

We thank Shohini Bhattacharya, Genki Nukazuka, Marc Schlegel, Werner Vogelsang, Shinsuke Yoshida and Feng Yuan for discussions and correspondences. The work of S.~B. and A.~K. is supported by the Croatian Science Foundation (HRZZ) no. 5332 (UIP-2019-04). 
Y.~H. is supported by the U.S. Department of Energy under Contract No. DE-SC0012704, and also by  Laboratory Directed Research and Development (LDRD) funds from Brookhaven Science Associates. H.N.L. is supported by National Science and Technology Council of the Republic of 
China under Grant No. MOST-110-2112-M-001-026-MY3. S.~B. thanks for the warm hospitality at the University of T\" ubingen in 2022, where part of this work was done.

\appendix
\section{Computation of the hard kernel in SIDIS}

In this appendix we outline the main steps to arrive at (\ref{eq:S012final}). 
 Starting from (34) of \cite{Benic:2021gya}, 
we write 
\be
 {\cal H}_{\mu\nu}(p) T_r^{\mu\nu} = -\frac{2\pi i g^4}{N_c}\int\frac{d^4 l_2}{(2\pi)^4}(2\pi)\delta\left(l_2^2\right)(2\pi)\delta\left((p+q-l_2)^2\right)W_r\,,
\ee
where
\be
W_r \equiv  T_{r,\mu\nu}{\rm Tr}\left[\gamma_5 \slashed{p} A^{\alpha\mu}(p+q-p_q)\bar{M}_{\alpha\beta}(p+q-p_q,l_2)A^{\nu\beta}(l_2)\right]\,.
\ee
The building blocks $\bar{M}$ and $A$ were introduced in \cite{Benic:2021gya}, which are given by  
\be
\bar{M}_{\alpha \beta} = \sum_{l = 1}^3 c_{l} \frac{m_{l,\alpha \beta}}{d_l}\,,
\label{eq:trcm}
\ee
\be
\begin{split}
& c_1 = \frac{N_c(N_c^2 - 1)}{4} \,, \qquad m_{1,\alpha\beta} = -V_{\alpha\beta\rho}(p+q-p_q,l_2) \slashed{p}_q \gamma^\rho (\slashed{p} + \slashed{q} - \slashed{l}_2)\,, \qquad d_1 = (p + q - p_q - l_2)^2 \,,\\
& c_2 = \frac{(N_c^2 - 1)^2}{4 N_c} \,, \qquad m_{2,\alpha\beta} = \slashed{p}_q \gamma_\alpha (\slashed{p} + \slashed{q}) \gamma_\beta (\slashed{p} + \slashed{q} - \slashed{l}_2)\,, \qquad d_2 = (p+q)^2\,,\\
&c_3 = \frac{N_c^2 - 1}{4 N_c} \,, \qquad m_{3,\alpha\beta} = - \slashed{p}_q \gamma_\beta (\slashed{p}_q - \slashed{l}_2)\gamma_\alpha(\slashed{p} + \slashed{q} - \slashed{l}_2)\,, \qquad d_3 = (p_q - l_2)^2 \,,\\
\end{split}
\ee
with $V$ being the three-gluon vertex and 
\be
\begin{split}
& A^{\alpha \mu} = \sum_{i=1}^2\frac{a_{Li}^{\alpha \mu}}{d_{Lj}}\,, \qquad a_{L1}^{\alpha\mu} = \gamma^{\alpha} (\slashed{p}_q - \slashed{q})\gamma^{\mu} \,, \qquad a_{L2}^{\alpha\mu} = \gamma^{\mu}(\slashed{p} + \slashed{q})\gamma^{\alpha}\,, \qquad d_{L1} = (p_q - q)^2 \,, \qquad d_{L2} = (p+q)^2\,,\\
& A^{\nu\beta} = \sum_{j=1}^2\frac{a_{Rj}^{\nu \beta}}{d_{Rj}}\,, \qquad a_{R1}^{\nu\beta} = \gamma^{\nu} (\slashed{p} - \slashed{l}_2)\gamma^{\beta} \,, \qquad a_{R2}^{\nu\beta} = \gamma^{\beta} (\slashed{p} + \slashed{q})\gamma^{\nu}\,,\qquad d_{R1} = (p - l_2)^2 \,, \qquad d_{R2} = (p + q)^2\,.
\end{split}
\label{eq:ALR}
\ee
We perform the loop integral as  
\be
\int\frac{d^4 l_2}{(2\pi)^4}(2\pi)\delta\left(l_2^2\right)(2\pi)\delta\left((p+q-l_2)^2\right) = \frac{1}{32\pi^2}\sum_{a_2 = \pm}\int_0^1 d\Delta_2\int_0^{2\pi}d\phi_2\,,
\label{eq:loop}
\ee
where $l_{2(a_2)}^{\pm}$ denote the solutions of the $\delta$-function conditions $l_2^2 = 0$ and $(p+q - l_2)^2 = 0$, namely
\be
l_{2 (a_2)}^+ = \frac{p^+  + q^+}{2}(1 + a_2 \Delta_2)\,, \qquad l_{2(a_2)}^- = \frac{q^-}{2}(1-a_2 \Delta_2) \,, \qquad \Delta_2 = \sqrt{1 - \frac{4 {\bm l}_{2\perp}^2}{s}} \,, \qquad a_2 = \pm 1\,.
\ee
In addition, we also have the on-shell conditions for the tagged parton $p_q^2 = 0$ and $(p + q - p_q)^2 = 0$, that produce
\be
p_{q (a_q)}^+ = \frac{p^+  + q^+}{2}(1 + a_q \Delta_q)\,, \qquad p_{q(a_q)}^- = \frac{q^-}{2}(1-a_q \Delta_q) \,, \qquad \Delta_q = \sqrt{1 - \frac{4 {\bm p}_{q\perp}^2}{s}} \,, \qquad a_q = \pm 1\,.
\ee

Next we write the Dirac trace as
\be
W_r = \sum_{ij=1}^2\sum_{l=1}^3 c_l \frac{W_{r,ilj}}{d_{Li}d_l d_{Rj}}\,,
\ee
where
\be
W_{r,ilj} = T_{r,\mu\nu}{\rm Tr}\left[\gamma_5 \slashed{k} a_{Li}^{\alpha\mu} m_{l,\alpha\beta}a_{Rj}^{\nu\beta}\right]\,,
\ee
and $d_l$  in the denominator contains the $\phi_2$ angular dependence. After the computation of the Dirac traces, the angular dependence in the numerator takes the form 
\be
W_{r,ilj} = \sum_{g = 0}^3 w^{(g)}_{r,ilj}\left(\frac{{\bm p}_{q\perp} \cdot {\bm l}_{2\perp}}{s}\right)^g\,.
\ee
The $\phi_2$ integral can then be performed analytically, leading to
\be
 \int_0^{2\pi} d\phi_2 \frac{1}{d_l}\left(\frac{{\bm p}_{q\perp} \cdot {\bm l}_{2\perp}}{s}\right)^g=\frac{2\pi}{s} I_{g,l} \,,
\ee
where 
\be
\begin{split}
& I_{0,1}^{(a_q,a_2)} = - \frac{2}{|a_q \Delta_q + a_2 \Delta_2|}\,,\\
& I_{1,1}^{(a_q,a_2)} = -\frac{1}{2}\left(1 - \frac{1 + a_q a_2 \Delta_q\Delta_2}{|a_q\Delta_q + a_2 \Delta_2|}\right)\,,\\
& I_{2,1}^{(a_q,a_2)} = \frac{1}{8}(1 + a_q a_2 \Delta_q\Delta_2)\left(1 - \frac{1 + a_q a_2 \Delta_q\Delta_2}{|a_q\Delta_q + a_2 \Delta_2|}\right)\,,\\
& I_{3,1}^{(a_q,a_2)} = -\frac{1}{32}(1-\Delta_q^2)(1-\Delta_2^2)\left[\frac{1}{2} + \frac{(1 + a_q a_2 \Delta_q\Delta_2)^2}{(1-\Delta_q^2)(1-\Delta_2^2)}\left(1 - \frac{1 + a_q a_2 \Delta_q\Delta_2}{|a_q\Delta_q + a_2 \Delta_2|}\right)\right]\,,\\
& I_{0,2}^{(a_q,a_2)} = 1\,,\\
& I_{1,2}^{(a_q,a_2)} = 0\,,\\
& I_{2,2}^{(a_q,a_2)} = \frac{1}{32}(1-\Delta_q^2)(1-\Delta_2^2)\,,\\
& I_{3,2}^{(a_q,a_2)} = 0\,,\\
& I_{0,3}^{(a_q,a_2)} = - \frac{2}{|a_q \Delta_q - a_2 \Delta_2|}\,,\\
& I_{1,3}^{(a_q,a_2)} = \frac{1}{2}\left(1 - \frac{1 - a_q a_2 \Delta_q\Delta_2}{|a_q\Delta_q - a_2 \Delta_2|}\right)\,,\\
& I_{2,3}^{(a_q,a_2)} = \frac{1}{8}(1 - a_q a_2 \Delta_q\Delta_2)\left(1 - \frac{1 - a_q a_2 \Delta_q\Delta_2}{|a_q\Delta_q - a_2 \Delta_2|}\right)\,,\\
& I_{3,3}^{(a_q,a_2)} = \frac{1}{32}(1-\Delta_q^2)(1-\Delta_2^2)\left[\frac{1}{2} + \frac{(1 - a_q a_2 \Delta_q\Delta_2)^2}{(1-\Delta_q^2)(1-\Delta_2^2)}\left(1 - \frac{1 - a_q a_2 \Delta_q\Delta_2}{|a_q\Delta_q - a_2 \Delta_2|}\right)\right]\,.\\
\end{split}
\ee
The above relations allow us to get 
\be
\begin{split}
{\cal H}(p)\cdot T_r 
= -\frac{i g^4}{8 N_c s} \sum_{a_2 = \pm}\int_0^1 d\Delta_2 \sum_{ilj}\frac{1}{d_{Li}d_{Rj}}\sum_g w_{r,ilj}^{(g)(a_q,a_2)} I^{(a_q,a_2)}_{g,l}\,,
\end{split}
\label{eq:delta2int}
\ee
with 
\be
\Delta_q = a_q\left(1 + \frac{2 t}{s + Q^2}\right)\,.
\ee
Because there is no collinear divergence  in the sum over $ijlg$, the $\Delta_2$ integral can be safely done. The result is actually independent of $a_2$, so the summation over $a_2$ simply gives a factor of 2. 
It is then straightforward to derive  (\ref{eq:S012final}). 

\bibliographystyle{apsrev4-1}
\bibliography{references}

\begin{thebibliography}{49}%
\makeatletter
\providecommand \@ifxundefined [1]{%
 \@ifx{#1\undefined}
}%
\providecommand \@ifnum [1]{%
 \ifnum #1\expandafter \@firstoftwo
 \else \expandafter \@secondoftwo
 \fi
}%
\providecommand \@ifx [1]{%
 \ifx #1\expandafter \@firstoftwo
 \else \expandafter \@secondoftwo
 \fi
}%
\providecommand \natexlab [1]{#1}%
\providecommand \enquote  [1]{``#1''}%
\providecommand \bibnamefont  [1]{#1}%
\providecommand \bibfnamefont [1]{#1}%
\providecommand \citenamefont [1]{#1}%
\providecommand \href@noop [0]{\@secondoftwo}%
\providecommand \href [0]{\begingroup \@sanitize@url \@href}%
\providecommand \@href[1]{\@@startlink{#1}\@@href}%
\providecommand \@@href[1]{\endgroup#1\@@endlink}%
\providecommand \@sanitize@url [0]{\catcode `\\12\catcode `\$12\catcode
  `\&12\catcode `\#12\catcode `\^12\catcode `\_12\catcode `\%12\relax}%
\providecommand \@@startlink[1]{}%
\providecommand \@@endlink[0]{}%
\providecommand \url  [0]{\begingroup\@sanitize@url \@url }%
\providecommand \@url [1]{\endgroup\@href {#1}{\urlprefix }}%
\providecommand \urlprefix  [0]{URL }%
\providecommand \Eprint [0]{\href }%
\providecommand \doibase [0]{http://dx.doi.org/}%
\providecommand \selectlanguage [0]{\@gobble}%
\providecommand \bibinfo  [0]{\@secondoftwo}%
\providecommand \bibfield  [0]{\@secondoftwo}%
\providecommand \translation [1]{[#1]}%
\providecommand \BibitemOpen [0]{}%
\providecommand \bibitemStop [0]{}%
\providecommand \bibitemNoStop [0]{.\EOS\space}%
\providecommand \EOS [0]{\spacefactor3000\relax}%
\providecommand \BibitemShut  [1]{\csname bibitem#1\endcsname}%
\let\auto@bib@innerbib\@empty
\bibitem [{\citenamefont {Benic}\ \emph {et~al.}(2019)\citenamefont {Benic},
  \citenamefont {Hatta}, \citenamefont {Li},\ and\ \citenamefont
  {Yang}}]{Benic:2019zvg}%
  \BibitemOpen
  \bibfield  {author} {\bibinfo {author} {\bibfnamefont {S.}~\bibnamefont
  {Benic}}, \bibinfo {author} {\bibfnamefont {Y.}~\bibnamefont {Hatta}},
  \bibinfo {author} {\bibfnamefont {H.-n.}\ \bibnamefont {Li}}, \ and\ \bibinfo
  {author} {\bibfnamefont {D.-J.}\ \bibnamefont {Yang}},\ }\href {\doibase
  10.1103/PhysRevD.100.094027} {\bibfield  {journal} {\bibinfo  {journal}
  {Phys. Rev. D}\ }\textbf {\bibinfo {volume} {100}},\ \bibinfo {pages}
  {094027} (\bibinfo {year} {2019})},\ \Eprint
  {http://arxiv.org/abs/1909.10684} {arXiv:1909.10684 [hep-ph]} \BibitemShut
  {NoStop}%
\bibitem [{\citenamefont {Beni\'c}\ \emph {et~al.}(2021)\citenamefont
  {Beni\'c}, \citenamefont {Hatta}, \citenamefont {Kaushik},\ and\
  \citenamefont {Li}}]{Benic:2021gya}%
  \BibitemOpen
  \bibfield  {author} {\bibinfo {author} {\bibfnamefont {S.}~\bibnamefont
  {Beni\'c}}, \bibinfo {author} {\bibfnamefont {Y.}~\bibnamefont {Hatta}},
  \bibinfo {author} {\bibfnamefont {A.}~\bibnamefont {Kaushik}}, \ and\
  \bibinfo {author} {\bibfnamefont {H.-n.}\ \bibnamefont {Li}},\ }\href
  {\doibase 10.1103/PhysRevD.104.094027} {\bibfield  {journal} {\bibinfo
  {journal} {Phys. Rev. D}\ }\textbf {\bibinfo {volume} {104}},\ \bibinfo
  {pages} {094027} (\bibinfo {year} {2021})},\ \Eprint
  {http://arxiv.org/abs/2109.05440} {arXiv:2109.05440 [hep-ph]} \BibitemShut
  {NoStop}%
\bibitem [{\citenamefont {Abele}\ \emph {et~al.}(2022)\citenamefont {Abele},
  \citenamefont {Aicher}, \citenamefont {Piacenza}, \citenamefont {Sch\"afer},\
  and\ \citenamefont {Vogelsang}}]{Abele:2022spu}%
  \BibitemOpen
  \bibfield  {author} {\bibinfo {author} {\bibfnamefont {M.}~\bibnamefont
  {Abele}}, \bibinfo {author} {\bibfnamefont {M.}~\bibnamefont {Aicher}},
  \bibinfo {author} {\bibfnamefont {F.}~\bibnamefont {Piacenza}}, \bibinfo
  {author} {\bibfnamefont {A.}~\bibnamefont {Sch\"afer}}, \ and\ \bibinfo
  {author} {\bibfnamefont {W.}~\bibnamefont {Vogelsang}},\ }\href {\doibase
  10.1103/PhysRevD.106.014020} {\bibfield  {journal} {\bibinfo  {journal}
  {Phys. Rev. D}\ }\textbf {\bibinfo {volume} {106}},\ \bibinfo {pages}
  {014020} (\bibinfo {year} {2022})},\ \Eprint
  {http://arxiv.org/abs/2204.13967} {arXiv:2204.13967 [hep-ph]} \BibitemShut
  {NoStop}%
\bibitem [{\citenamefont {Boughezal}\ \emph {et~al.}(2023)\citenamefont
  {Boughezal}, \citenamefont {de~Florian}, \citenamefont {Petriello},\ and\
  \citenamefont {Vogelsang}}]{Boughezal:2023ooo}%
  \BibitemOpen
  \bibfield  {author} {\bibinfo {author} {\bibfnamefont {R.}~\bibnamefont
  {Boughezal}}, \bibinfo {author} {\bibfnamefont {D.}~\bibnamefont
  {de~Florian}}, \bibinfo {author} {\bibfnamefont {F.}~\bibnamefont
  {Petriello}}, \ and\ \bibinfo {author} {\bibfnamefont {W.}~\bibnamefont
  {Vogelsang}},\ }\href {\doibase 10.1103/PhysRevD.107.075028} {\bibfield
  {journal} {\bibinfo  {journal} {Phys. Rev. D}\ }\textbf {\bibinfo {volume}
  {107}},\ \bibinfo {pages} {075028} (\bibinfo {year} {2023})},\ \Eprint
  {http://arxiv.org/abs/2301.02304} {arXiv:2301.02304 [hep-ph]} \BibitemShut
  {NoStop}%
\bibitem [{\citenamefont {Kane}\ \emph {et~al.}(1978)\citenamefont {Kane},
  \citenamefont {Pumplin},\ and\ \citenamefont {Repko}}]{Kane:1978nd}%
  \BibitemOpen
  \bibfield  {author} {\bibinfo {author} {\bibfnamefont {G.~L.}\ \bibnamefont
  {Kane}}, \bibinfo {author} {\bibfnamefont {J.}~\bibnamefont {Pumplin}}, \
  and\ \bibinfo {author} {\bibfnamefont {W.}~\bibnamefont {Repko}},\ }\href
  {\doibase 10.1103/PhysRevLett.41.1689} {\bibfield  {journal} {\bibinfo
  {journal} {Phys. Rev. Lett.}\ }\textbf {\bibinfo {volume} {41}},\ \bibinfo
  {pages} {1689} (\bibinfo {year} {1978})}\BibitemShut {NoStop}%
\bibitem [{\citenamefont {Dharmaratna}\ and\ \citenamefont
  {Goldstein}(1996)}]{Dharmaratna:1996xd}%
  \BibitemOpen
  \bibfield  {author} {\bibinfo {author} {\bibfnamefont {W.~G.~D.}\
  \bibnamefont {Dharmaratna}}\ and\ \bibinfo {author} {\bibfnamefont {G.~R.}\
  \bibnamefont {Goldstein}},\ }\href {\doibase 10.1103/PhysRevD.53.1073}
  {\bibfield  {journal} {\bibinfo  {journal} {Phys. Rev. D}\ }\textbf {\bibinfo
  {volume} {53}},\ \bibinfo {pages} {1073} (\bibinfo {year}
  {1996})}\BibitemShut {NoStop}%
\bibitem [{\citenamefont {Carlitz}\ and\ \citenamefont
  {Willey}(1992)}]{Carlitz:1992fv}%
  \BibitemOpen
  \bibfield  {author} {\bibinfo {author} {\bibfnamefont {R.~D.}\ \bibnamefont
  {Carlitz}}\ and\ \bibinfo {author} {\bibfnamefont {R.~S.}\ \bibnamefont
  {Willey}},\ }\href {\doibase 10.1103/PhysRevD.45.2323} {\bibfield  {journal}
  {\bibinfo  {journal} {Phys. Rev. D}\ }\textbf {\bibinfo {volume} {45}},\
  \bibinfo {pages} {2323} (\bibinfo {year} {1992})}\BibitemShut {NoStop}%
\bibitem [{\citenamefont {Yokoya}(2007)}]{Yokoya:2007xe}%
  \BibitemOpen
  \bibfield  {author} {\bibinfo {author} {\bibfnamefont {H.}~\bibnamefont
  {Yokoya}},\ }\href {\doibase 10.1143/PTP.118.371} {\bibfield  {journal}
  {\bibinfo  {journal} {Prog. Theor. Phys.}\ }\textbf {\bibinfo {volume}
  {118}},\ \bibinfo {pages} {371} (\bibinfo {year} {2007})},\ \Eprint
  {http://arxiv.org/abs/0705.2481} {arXiv:0705.2481 [hep-ph]} \BibitemShut
  {NoStop}%
\bibitem [{\citenamefont {Bhattacharya}\ \emph {et~al.}(2022)\citenamefont
  {Bhattacharya}, \citenamefont {Kang}, \citenamefont {Metz}, \citenamefont
  {Penn},\ and\ \citenamefont {Pitonyak}}]{Bhattacharya:2021twu}%
  \BibitemOpen
  \bibfield  {author} {\bibinfo {author} {\bibfnamefont {S.}~\bibnamefont
  {Bhattacharya}}, \bibinfo {author} {\bibfnamefont {Z.-B.}\ \bibnamefont
  {Kang}}, \bibinfo {author} {\bibfnamefont {A.}~\bibnamefont {Metz}}, \bibinfo
  {author} {\bibfnamefont {G.}~\bibnamefont {Penn}}, \ and\ \bibinfo {author}
  {\bibfnamefont {D.}~\bibnamefont {Pitonyak}},\ }\href {\doibase
  10.1103/PhysRevD.105.034007} {\bibfield  {journal} {\bibinfo  {journal}
  {Phys. Rev. D}\ }\textbf {\bibinfo {volume} {105}},\ \bibinfo {pages}
  {034007} (\bibinfo {year} {2022})},\ \Eprint
  {http://arxiv.org/abs/2110.10253} {arXiv:2110.10253 [hep-ph]} \BibitemShut
  {NoStop}%
\bibitem [{\citenamefont {Bauer}\ \emph {et~al.}(2023)\citenamefont {Bauer},
  \citenamefont {Pitonyak},\ and\ \citenamefont {Shay}}]{Bauer:2022mvl}%
  \BibitemOpen
  \bibfield  {author} {\bibinfo {author} {\bibfnamefont {B.}~\bibnamefont
  {Bauer}}, \bibinfo {author} {\bibfnamefont {D.}~\bibnamefont {Pitonyak}}, \
  and\ \bibinfo {author} {\bibfnamefont {C.}~\bibnamefont {Shay}},\ }\href
  {\doibase 10.1103/PhysRevD.107.014013} {\bibfield  {journal} {\bibinfo
  {journal} {Phys. Rev. D}\ }\textbf {\bibinfo {volume} {107}},\ \bibinfo
  {pages} {014013} (\bibinfo {year} {2023})},\ \Eprint
  {http://arxiv.org/abs/2210.14334} {arXiv:2210.14334 [hep-ph]} \BibitemShut
  {NoStop}%
\bibitem [{\citenamefont {Meng}\ \emph {et~al.}(1992)\citenamefont {Meng},
  \citenamefont {Olness},\ and\ \citenamefont {Soper}}]{Meng:1991da}%
  \BibitemOpen
  \bibfield  {author} {\bibinfo {author} {\bibfnamefont {R.-b.}\ \bibnamefont
  {Meng}}, \bibinfo {author} {\bibfnamefont {F.~I.}\ \bibnamefont {Olness}}, \
  and\ \bibinfo {author} {\bibfnamefont {D.~E.}\ \bibnamefont {Soper}},\ }\href
  {\doibase 10.1016/0550-3213(92)90230-9} {\bibfield  {journal} {\bibinfo
  {journal} {Nucl. Phys. B}\ }\textbf {\bibinfo {volume} {371}},\ \bibinfo
  {pages} {79} (\bibinfo {year} {1992})}\BibitemShut {NoStop}%
\bibitem [{\citenamefont {Ratcliffe}(1986)}]{Ratcliffe:1985mp}%
  \BibitemOpen
  \bibfield  {author} {\bibinfo {author} {\bibfnamefont {P.~G.}\ \bibnamefont
  {Ratcliffe}},\ }\href {\doibase 10.1016/0550-3213(86)90495-5} {\bibfield
  {journal} {\bibinfo  {journal} {Nucl. Phys. B}\ }\textbf {\bibinfo {volume}
  {264}},\ \bibinfo {pages} {493} (\bibinfo {year} {1986})}\BibitemShut
  {NoStop}%
\bibitem [{\citenamefont {Efremov}\ and\ \citenamefont
  {Teryaev}(1985)}]{Efremov:1984ip}%
  \BibitemOpen
  \bibfield  {author} {\bibinfo {author} {\bibfnamefont {A.~V.}\ \bibnamefont
  {Efremov}}\ and\ \bibinfo {author} {\bibfnamefont {O.~V.}\ \bibnamefont
  {Teryaev}},\ }\href {\doibase 10.1016/0370-2693(85)90999-2} {\bibfield
  {journal} {\bibinfo  {journal} {Phys. Lett. B}\ }\textbf {\bibinfo {volume}
  {150}},\ \bibinfo {pages} {383} (\bibinfo {year} {1985})}\BibitemShut
  {NoStop}%
\bibitem [{\citenamefont {Qiu}\ and\ \citenamefont
  {Sterman}(1991)}]{Qiu:1991pp}%
  \BibitemOpen
  \bibfield  {author} {\bibinfo {author} {\bibfnamefont {J.-w.}\ \bibnamefont
  {Qiu}}\ and\ \bibinfo {author} {\bibfnamefont {G.~F.}\ \bibnamefont
  {Sterman}},\ }\href {\doibase 10.1103/PhysRevLett.67.2264} {\bibfield
  {journal} {\bibinfo  {journal} {Phys. Rev. Lett.}\ }\textbf {\bibinfo
  {volume} {67}},\ \bibinfo {pages} {2264} (\bibinfo {year}
  {1991})}\BibitemShut {NoStop}%
\bibitem [{\citenamefont {Eguchi}\ \emph {et~al.}(2007)\citenamefont {Eguchi},
  \citenamefont {Koike},\ and\ \citenamefont {Tanaka}}]{Eguchi:2006mc}%
  \BibitemOpen
  \bibfield  {author} {\bibinfo {author} {\bibfnamefont {H.}~\bibnamefont
  {Eguchi}}, \bibinfo {author} {\bibfnamefont {Y.}~\bibnamefont {Koike}}, \
  and\ \bibinfo {author} {\bibfnamefont {K.}~\bibnamefont {Tanaka}},\ }\href
  {\doibase 10.1016/j.nuclphysb.2006.11.016} {\bibfield  {journal} {\bibinfo
  {journal} {Nucl. Phys. B}\ }\textbf {\bibinfo {volume} {763}},\ \bibinfo
  {pages} {198} (\bibinfo {year} {2007})},\ \Eprint
  {http://arxiv.org/abs/hep-ph/0610314} {arXiv:hep-ph/0610314} \BibitemShut
  {NoStop}%
\bibitem [{\citenamefont {Koike}\ and\ \citenamefont
  {Yoshida}(2012)}]{Koike:2011nx}%
  \BibitemOpen
  \bibfield  {author} {\bibinfo {author} {\bibfnamefont {Y.}~\bibnamefont
  {Koike}}\ and\ \bibinfo {author} {\bibfnamefont {S.}~\bibnamefont
  {Yoshida}},\ }\href {\doibase 10.1103/PhysRevD.85.034030} {\bibfield
  {journal} {\bibinfo  {journal} {Phys. Rev. D}\ }\textbf {\bibinfo {volume}
  {85}},\ \bibinfo {pages} {034030} (\bibinfo {year} {2012})},\ \Eprint
  {http://arxiv.org/abs/1112.1161} {arXiv:1112.1161 [hep-ph]} \BibitemShut
  {NoStop}%
\bibitem [{\citenamefont {Yoshida}(2016)}]{Yoshida:2016tfh}%
  \BibitemOpen
  \bibfield  {author} {\bibinfo {author} {\bibfnamefont {S.}~\bibnamefont
  {Yoshida}},\ }\href {\doibase 10.1103/PhysRevD.93.054048} {\bibfield
  {journal} {\bibinfo  {journal} {Phys. Rev. D}\ }\textbf {\bibinfo {volume}
  {93}},\ \bibinfo {pages} {054048} (\bibinfo {year} {2016})},\ \Eprint
  {http://arxiv.org/abs/1601.07737} {arXiv:1601.07737 [hep-ph]} \BibitemShut
  {NoStop}%
\bibitem [{\citenamefont {Metz}\ \emph {et~al.}(2006)\citenamefont {Metz},
  \citenamefont {Schlegel},\ and\ \citenamefont {Goeke}}]{Metz:2006pe}%
  \BibitemOpen
  \bibfield  {author} {\bibinfo {author} {\bibfnamefont {A.}~\bibnamefont
  {Metz}}, \bibinfo {author} {\bibfnamefont {M.}~\bibnamefont {Schlegel}}, \
  and\ \bibinfo {author} {\bibfnamefont {K.}~\bibnamefont {Goeke}},\ }\href
  {\doibase 10.1016/j.physletb.2006.11.009} {\bibfield  {journal} {\bibinfo
  {journal} {Phys. Lett. B}\ }\textbf {\bibinfo {volume} {643}},\ \bibinfo
  {pages} {319} (\bibinfo {year} {2006})},\ \Eprint
  {http://arxiv.org/abs/hep-ph/0610112} {arXiv:hep-ph/0610112} \BibitemShut
  {NoStop}%
\bibitem [{\citenamefont {Kovchegov}\ and\ \citenamefont
  {Sievert}(2012)}]{Kovchegov:2012ga}%
  \BibitemOpen
  \bibfield  {author} {\bibinfo {author} {\bibfnamefont {Y.~V.}\ \bibnamefont
  {Kovchegov}}\ and\ \bibinfo {author} {\bibfnamefont {M.~D.}\ \bibnamefont
  {Sievert}},\ }\href {\doibase 10.1103/PhysRevD.86.034028} {\bibfield
  {journal} {\bibinfo  {journal} {Phys. Rev. D}\ }\textbf {\bibinfo {volume}
  {86}},\ \bibinfo {pages} {034028} (\bibinfo {year} {2012})},\ \bibinfo {note}
  {[Erratum: Phys.Rev.D 86, 079906 (2012)]},\ \Eprint
  {http://arxiv.org/abs/1201.5890} {arXiv:1201.5890 [hep-ph]} \BibitemShut
  {NoStop}%
\bibitem [{\citenamefont {Schlegel}(2013)}]{Schlegel:2012ve}%
  \BibitemOpen
  \bibfield  {author} {\bibinfo {author} {\bibfnamefont {M.}~\bibnamefont
  {Schlegel}},\ }\href {\doibase 10.1103/PhysRevD.87.034006} {\bibfield
  {journal} {\bibinfo  {journal} {Phys. Rev. D}\ }\textbf {\bibinfo {volume}
  {87}},\ \bibinfo {pages} {034006} (\bibinfo {year} {2013})},\ \Eprint
  {http://arxiv.org/abs/1211.3579} {arXiv:1211.3579 [hep-ph]} \BibitemShut
  {NoStop}%
\bibitem [{\citenamefont {Beni\'c}\ \emph {et~al.}(2022)\citenamefont
  {Beni\'c}, \citenamefont {Horvati\'c}, \citenamefont {Kaushik},\ and\
  \citenamefont {Vivoda}}]{Benic:2022qzv}%
  \BibitemOpen
  \bibfield  {author} {\bibinfo {author} {\bibfnamefont {S.}~\bibnamefont
  {Beni\'c}}, \bibinfo {author} {\bibfnamefont {D.}~\bibnamefont {Horvati\'c}},
  \bibinfo {author} {\bibfnamefont {A.}~\bibnamefont {Kaushik}}, \ and\
  \bibinfo {author} {\bibfnamefont {E.~A.}\ \bibnamefont {Vivoda}},\ }\href
  {\doibase 10.1103/PhysRevD.106.114025} {\bibfield  {journal} {\bibinfo
  {journal} {Phys. Rev. D}\ }\textbf {\bibinfo {volume} {106}},\ \bibinfo
  {pages} {114025} (\bibinfo {year} {2022})},\ \Eprint
  {http://arxiv.org/abs/2210.10353} {arXiv:2210.10353 [hep-ph]} \BibitemShut
  {NoStop}%
\bibitem [{\citenamefont {Xing}\ and\ \citenamefont
  {Yoshida}(2019)}]{Xing:2019ovj}%
  \BibitemOpen
  \bibfield  {author} {\bibinfo {author} {\bibfnamefont {H.}~\bibnamefont
  {Xing}}\ and\ \bibinfo {author} {\bibfnamefont {S.}~\bibnamefont {Yoshida}},\
  }\href {\doibase 10.1103/PhysRevD.100.054024} {\bibfield  {journal} {\bibinfo
   {journal} {Phys. Rev. D}\ }\textbf {\bibinfo {volume} {100}},\ \bibinfo
  {pages} {054024} (\bibinfo {year} {2019})},\ \Eprint
  {http://arxiv.org/abs/1904.02287} {arXiv:1904.02287 [hep-ph]} \BibitemShut
  {NoStop}%
\bibitem [{\citenamefont {Ji}(1992)}]{Ji:1992eu}%
  \BibitemOpen
  \bibfield  {author} {\bibinfo {author} {\bibfnamefont {X.-D.}\ \bibnamefont
  {Ji}},\ }\href {\doibase 10.1016/0370-2693(92)91375-J} {\bibfield  {journal}
  {\bibinfo  {journal} {Phys. Lett. B}\ }\textbf {\bibinfo {volume} {289}},\
  \bibinfo {pages} {137} (\bibinfo {year} {1992})}\BibitemShut {NoStop}%
\bibitem [{\citenamefont {Hatta}\ \emph {et~al.}(2013)\citenamefont {Hatta},
  \citenamefont {Tanaka},\ and\ \citenamefont {Yoshida}}]{Hatta:2012jm}%
  \BibitemOpen
  \bibfield  {author} {\bibinfo {author} {\bibfnamefont {Y.}~\bibnamefont
  {Hatta}}, \bibinfo {author} {\bibfnamefont {K.}~\bibnamefont {Tanaka}}, \
  and\ \bibinfo {author} {\bibfnamefont {S.}~\bibnamefont {Yoshida}},\ }\href
  {\doibase 10.1007/JHEP02(2013)003} {\bibfield  {journal} {\bibinfo  {journal}
  {JHEP}\ }\textbf {\bibinfo {volume} {02}},\ \bibinfo {pages} {003} (\bibinfo
  {year} {2013})},\ \Eprint {http://arxiv.org/abs/1211.2918} {arXiv:1211.2918
  [hep-ph]} \BibitemShut {NoStop}%
\bibitem [{\citenamefont {Collins}\ and\ \citenamefont
  {Soper}(1977)}]{Collins:1977iv}%
  \BibitemOpen
  \bibfield  {author} {\bibinfo {author} {\bibfnamefont {J.~C.}\ \bibnamefont
  {Collins}}\ and\ \bibinfo {author} {\bibfnamefont {D.~E.}\ \bibnamefont
  {Soper}},\ }\href {\doibase 10.1103/PhysRevD.16.2219} {\bibfield  {journal}
  {\bibinfo  {journal} {Phys. Rev. D}\ }\textbf {\bibinfo {volume} {16}},\
  \bibinfo {pages} {2219} (\bibinfo {year} {1977})}\BibitemShut {NoStop}%
\bibitem [{\citenamefont {Tangerman}\ and\ \citenamefont
  {Mulders}(1995)}]{Tangerman:1994eh}%
  \BibitemOpen
  \bibfield  {author} {\bibinfo {author} {\bibfnamefont {R.~D.}\ \bibnamefont
  {Tangerman}}\ and\ \bibinfo {author} {\bibfnamefont {P.~J.}\ \bibnamefont
  {Mulders}},\ }\href {\doibase 10.1103/PhysRevD.51.3357} {\bibfield  {journal}
  {\bibinfo  {journal} {Phys. Rev. D}\ }\textbf {\bibinfo {volume} {51}},\
  \bibinfo {pages} {3357} (\bibinfo {year} {1995})},\ \Eprint
  {http://arxiv.org/abs/hep-ph/9403227} {arXiv:hep-ph/9403227} \BibitemShut
  {NoStop}%
\bibitem [{\citenamefont {Arnold}\ \emph {et~al.}(2009)\citenamefont {Arnold},
  \citenamefont {Metz},\ and\ \citenamefont {Schlegel}}]{Arnold:2008kf}%
  \BibitemOpen
  \bibfield  {author} {\bibinfo {author} {\bibfnamefont {S.}~\bibnamefont
  {Arnold}}, \bibinfo {author} {\bibfnamefont {A.}~\bibnamefont {Metz}}, \ and\
  \bibinfo {author} {\bibfnamefont {M.}~\bibnamefont {Schlegel}},\ }\href
  {\doibase 10.1103/PhysRevD.79.034005} {\bibfield  {journal} {\bibinfo
  {journal} {Phys. Rev. D}\ }\textbf {\bibinfo {volume} {79}},\ \bibinfo
  {pages} {034005} (\bibinfo {year} {2009})},\ \Eprint
  {http://arxiv.org/abs/0809.2262} {arXiv:0809.2262 [hep-ph]} \BibitemShut
  {NoStop}%
\bibitem [{\citenamefont {Boglione}\ and\ \citenamefont
  {Melis}(2011)}]{Boglione:2011zw}%
  \BibitemOpen
  \bibfield  {author} {\bibinfo {author} {\bibfnamefont {M.}~\bibnamefont
  {Boglione}}\ and\ \bibinfo {author} {\bibfnamefont {S.}~\bibnamefont
  {Melis}},\ }\href {\doibase 10.1103/PhysRevD.84.034038} {\bibfield  {journal}
  {\bibinfo  {journal} {Phys. Rev. D}\ }\textbf {\bibinfo {volume} {84}},\
  \bibinfo {pages} {034038} (\bibinfo {year} {2011})},\ \Eprint
  {http://arxiv.org/abs/1103.2084} {arXiv:1103.2084 [hep-ph]} \BibitemShut
  {NoStop}%
\bibitem [{\citenamefont {Ji}\ \emph {et~al.}(2006)\citenamefont {Ji},
  \citenamefont {Qiu}, \citenamefont {Vogelsang},\ and\ \citenamefont
  {Yuan}}]{Ji:2006vf}%
  \BibitemOpen
  \bibfield  {author} {\bibinfo {author} {\bibfnamefont {X.}~\bibnamefont
  {Ji}}, \bibinfo {author} {\bibfnamefont {J.-w.}\ \bibnamefont {Qiu}},
  \bibinfo {author} {\bibfnamefont {W.}~\bibnamefont {Vogelsang}}, \ and\
  \bibinfo {author} {\bibfnamefont {F.}~\bibnamefont {Yuan}},\ }\href {\doibase
  10.1103/PhysRevD.73.094017} {\bibfield  {journal} {\bibinfo  {journal} {Phys.
  Rev. D}\ }\textbf {\bibinfo {volume} {73}},\ \bibinfo {pages} {094017}
  (\bibinfo {year} {2006})},\ \Eprint {http://arxiv.org/abs/hep-ph/0604023}
  {arXiv:hep-ph/0604023} \BibitemShut {NoStop}%
\bibitem [{\citenamefont {Zhou}\ and\ \citenamefont
  {Metz}(2012)}]{Zhou:2010ui}%
  \BibitemOpen
  \bibfield  {author} {\bibinfo {author} {\bibfnamefont {J.}~\bibnamefont
  {Zhou}}\ and\ \bibinfo {author} {\bibfnamefont {A.}~\bibnamefont {Metz}},\
  }\href {\doibase 10.1103/PhysRevD.86.014001} {\bibfield  {journal} {\bibinfo
  {journal} {Phys. Rev. D}\ }\textbf {\bibinfo {volume} {86}},\ \bibinfo
  {pages} {014001} (\bibinfo {year} {2012})},\ \Eprint
  {http://arxiv.org/abs/1011.5871} {arXiv:1011.5871 [hep-ph]} \BibitemShut
  {NoStop}%
\bibitem [{\citenamefont {Pire}\ and\ \citenamefont
  {Ralston}(1983)}]{Pire:1983tv}%
  \BibitemOpen
  \bibfield  {author} {\bibinfo {author} {\bibfnamefont {B.}~\bibnamefont
  {Pire}}\ and\ \bibinfo {author} {\bibfnamefont {J.~P.}\ \bibnamefont
  {Ralston}},\ }\href {\doibase 10.1103/PhysRevD.28.260} {\bibfield  {journal}
  {\bibinfo  {journal} {Phys. Rev. D}\ }\textbf {\bibinfo {volume} {28}},\
  \bibinfo {pages} {260} (\bibinfo {year} {1983})}\BibitemShut {NoStop}%
\bibitem [{\citenamefont {Patel}(2015)}]{Patel:2015tea}%
  \BibitemOpen
  \bibfield  {author} {\bibinfo {author} {\bibfnamefont {H.~H.}\ \bibnamefont
  {Patel}},\ }\href {\doibase 10.1016/j.cpc.2015.08.017} {\bibfield  {journal}
  {\bibinfo  {journal} {Comput. Phys. Commun.}\ }\textbf {\bibinfo {volume}
  {197}},\ \bibinfo {pages} {276} (\bibinfo {year} {2015})},\ \Eprint
  {http://arxiv.org/abs/1503.01469} {arXiv:1503.01469 [hep-ph]} \BibitemShut
  {NoStop}%
\bibitem [{\citenamefont {Korner}\ \emph {et~al.}(2000)\citenamefont {Korner},
  \citenamefont {Melic},\ and\ \citenamefont {Merebashvili}}]{Korner:2000zr}%
  \BibitemOpen
  \bibfield  {author} {\bibinfo {author} {\bibfnamefont {J.~G.}\ \bibnamefont
  {Korner}}, \bibinfo {author} {\bibfnamefont {B.}~\bibnamefont {Melic}}, \
  and\ \bibinfo {author} {\bibfnamefont {Z.}~\bibnamefont {Merebashvili}},\
  }\href {\doibase 10.1103/PhysRevD.62.096011} {\bibfield  {journal} {\bibinfo
  {journal} {Phys. Rev. D}\ }\textbf {\bibinfo {volume} {62}},\ \bibinfo
  {pages} {096011} (\bibinfo {year} {2000})},\ \Eprint
  {http://arxiv.org/abs/hep-ph/0002302} {arXiv:hep-ph/0002302} \BibitemShut
  {NoStop}%
\bibitem [{\citenamefont {Korner}\ and\ \citenamefont
  {Schuler}(1985)}]{Korner:1984xd}%
  \BibitemOpen
  \bibfield  {author} {\bibinfo {author} {\bibfnamefont {J.~G.}\ \bibnamefont
  {Korner}}\ and\ \bibinfo {author} {\bibfnamefont {G.}~\bibnamefont
  {Schuler}},\ }\href {\doibase 10.1007/BF01551799} {\bibfield  {journal}
  {\bibinfo  {journal} {Z. Phys. C}\ }\textbf {\bibinfo {volume} {26}},\
  \bibinfo {pages} {559} (\bibinfo {year} {1985})}\BibitemShut {NoStop}%
\bibitem [{\citenamefont {Hagiwara}\ \emph {et~al.}(1984)\citenamefont
  {Hagiwara}, \citenamefont {Hikasa},\ and\ \citenamefont
  {Kai}}]{Hagiwara:1984hi}%
  \BibitemOpen
  \bibfield  {author} {\bibinfo {author} {\bibfnamefont {K.}~\bibnamefont
  {Hagiwara}}, \bibinfo {author} {\bibfnamefont {K.-i.}\ \bibnamefont
  {Hikasa}}, \ and\ \bibinfo {author} {\bibfnamefont {N.}~\bibnamefont {Kai}},\
  }\href {\doibase 10.1103/PhysRevLett.52.1076} {\bibfield  {journal} {\bibinfo
   {journal} {Phys. Rev. Lett.}\ }\textbf {\bibinfo {volume} {52}},\ \bibinfo
  {pages} {1076} (\bibinfo {year} {1984})}\BibitemShut {NoStop}%
\bibitem [{\citenamefont {Hagiwara}\ \emph {et~al.}(1983)\citenamefont
  {Hagiwara}, \citenamefont {Hikasa},\ and\ \citenamefont
  {Kai}}]{Hagiwara:1982cq}%
  \BibitemOpen
  \bibfield  {author} {\bibinfo {author} {\bibfnamefont {K.}~\bibnamefont
  {Hagiwara}}, \bibinfo {author} {\bibfnamefont {K.-i.}\ \bibnamefont
  {Hikasa}}, \ and\ \bibinfo {author} {\bibfnamefont {N.}~\bibnamefont {Kai}},\
  }\href {\doibase 10.1103/PhysRevD.27.84} {\bibfield  {journal} {\bibinfo
  {journal} {Phys. Rev. D}\ }\textbf {\bibinfo {volume} {27}},\ \bibinfo
  {pages} {84} (\bibinfo {year} {1983})}\BibitemShut {NoStop}%
\bibitem [{\citenamefont {Santiago}(2023)}]{Santiago:2023rfl}%
  \BibitemOpen
  \bibfield  {author} {\bibinfo {author} {\bibfnamefont {M.~G.}\ \bibnamefont
  {Santiago}},\ }\href@noop {} {\  (\bibinfo {year} {2023})},\ \Eprint
  {http://arxiv.org/abs/2310.02231} {arXiv:2310.02231 [hep-ph]} \BibitemShut
  {NoStop}%
\bibitem [{\citenamefont {Ethier}\ \emph {et~al.}(2017)\citenamefont {Ethier},
  \citenamefont {Sato},\ and\ \citenamefont {Melnitchouk}}]{Ethier:2017zbq}%
  \BibitemOpen
  \bibfield  {author} {\bibinfo {author} {\bibfnamefont {J.~J.}\ \bibnamefont
  {Ethier}}, \bibinfo {author} {\bibfnamefont {N.}~\bibnamefont {Sato}}, \ and\
  \bibinfo {author} {\bibfnamefont {W.}~\bibnamefont {Melnitchouk}},\ }\href
  {\doibase 10.1103/PhysRevLett.119.132001} {\bibfield  {journal} {\bibinfo
  {journal} {Phys. Rev. Lett.}\ }\textbf {\bibinfo {volume} {119}},\ \bibinfo
  {pages} {132001} (\bibinfo {year} {2017})},\ \Eprint
  {http://arxiv.org/abs/1705.05889} {arXiv:1705.05889 [hep-ph]} \BibitemShut
  {NoStop}%
\bibitem [{\citenamefont {Owens}\ \emph {et~al.}(2013)\citenamefont {Owens},
  \citenamefont {Accardi},\ and\ \citenamefont {Melnitchouk}}]{Owens:2012bv}%
  \BibitemOpen
  \bibfield  {author} {\bibinfo {author} {\bibfnamefont {J.~F.}\ \bibnamefont
  {Owens}}, \bibinfo {author} {\bibfnamefont {A.}~\bibnamefont {Accardi}}, \
  and\ \bibinfo {author} {\bibfnamefont {W.}~\bibnamefont {Melnitchouk}},\
  }\href {\doibase 10.1103/PhysRevD.87.094012} {\bibfield  {journal} {\bibinfo
  {journal} {Phys. Rev. D}\ }\textbf {\bibinfo {volume} {87}},\ \bibinfo
  {pages} {094012} (\bibinfo {year} {2013})},\ \Eprint
  {http://arxiv.org/abs/1212.1702} {arXiv:1212.1702 [hep-ph]} \BibitemShut
  {NoStop}%
\bibitem [{\citenamefont {Lai}\ \emph {et~al.}(2010)\citenamefont {Lai},
  \citenamefont {Guzzi}, \citenamefont {Huston}, \citenamefont {Li},
  \citenamefont {Nadolsky}, \citenamefont {Pumplin},\ and\ \citenamefont
  {Yuan}}]{Lai:2010vv}%
  \BibitemOpen
  \bibfield  {author} {\bibinfo {author} {\bibfnamefont {H.-L.}\ \bibnamefont
  {Lai}}, \bibinfo {author} {\bibfnamefont {M.}~\bibnamefont {Guzzi}}, \bibinfo
  {author} {\bibfnamefont {J.}~\bibnamefont {Huston}}, \bibinfo {author}
  {\bibfnamefont {Z.}~\bibnamefont {Li}}, \bibinfo {author} {\bibfnamefont
  {P.~M.}\ \bibnamefont {Nadolsky}}, \bibinfo {author} {\bibfnamefont
  {J.}~\bibnamefont {Pumplin}}, \ and\ \bibinfo {author} {\bibfnamefont
  {C.~P.}\ \bibnamefont {Yuan}},\ }\href {\doibase 10.1103/PhysRevD.82.074024}
  {\bibfield  {journal} {\bibinfo  {journal} {Phys. Rev. D}\ }\textbf {\bibinfo
  {volume} {82}},\ \bibinfo {pages} {074024} (\bibinfo {year} {2010})},\
  \Eprint {http://arxiv.org/abs/1007.2241} {arXiv:1007.2241 [hep-ph]}
  \BibitemShut {NoStop}%
\bibitem [{\citenamefont {Cammarota}\ \emph {et~al.}(2020)\citenamefont
  {Cammarota}, \citenamefont {Gamberg}, \citenamefont {Kang}, \citenamefont
  {Miller}, \citenamefont {Pitonyak}, \citenamefont {Prokudin}, \citenamefont
  {Rogers},\ and\ \citenamefont {Sato}}]{Cammarota:2020qcw}%
  \BibitemOpen
  \bibfield  {author} {\bibinfo {author} {\bibfnamefont {J.}~\bibnamefont
  {Cammarota}}, \bibinfo {author} {\bibfnamefont {L.}~\bibnamefont {Gamberg}},
  \bibinfo {author} {\bibfnamefont {Z.-B.}\ \bibnamefont {Kang}}, \bibinfo
  {author} {\bibfnamefont {J.~A.}\ \bibnamefont {Miller}}, \bibinfo {author}
  {\bibfnamefont {D.}~\bibnamefont {Pitonyak}}, \bibinfo {author}
  {\bibfnamefont {A.}~\bibnamefont {Prokudin}}, \bibinfo {author}
  {\bibfnamefont {T.~C.}\ \bibnamefont {Rogers}}, \ and\ \bibinfo {author}
  {\bibfnamefont {N.}~\bibnamefont {Sato}} (\bibinfo {collaboration} {Jefferson
  Lab Angular Momentum}),\ }\href {\doibase 10.1103/PhysRevD.102.054002}
  {\bibfield  {journal} {\bibinfo  {journal} {Phys. Rev. D}\ }\textbf {\bibinfo
  {volume} {102}},\ \bibinfo {pages} {054002} (\bibinfo {year} {2020})},\
  \Eprint {http://arxiv.org/abs/2002.08384} {arXiv:2002.08384 [hep-ph]}
  \BibitemShut {NoStop}%
\bibitem [{\citenamefont {Barry}\ \emph {et~al.}(2021)\citenamefont {Barry},
  \citenamefont {Ji}, \citenamefont {Sato},\ and\ \citenamefont
  {Melnitchouk}}]{Barry:2021osv}%
  \BibitemOpen
  \bibfield  {author} {\bibinfo {author} {\bibfnamefont {P.~C.}\ \bibnamefont
  {Barry}}, \bibinfo {author} {\bibfnamefont {C.-R.}\ \bibnamefont {Ji}},
  \bibinfo {author} {\bibfnamefont {N.}~\bibnamefont {Sato}}, \ and\ \bibinfo
  {author} {\bibfnamefont {W.}~\bibnamefont {Melnitchouk}} (\bibinfo
  {collaboration} {Jefferson Lab Angular Momentum (JAM)}),\ }\href {\doibase
  10.1103/PhysRevLett.127.232001} {\bibfield  {journal} {\bibinfo  {journal}
  {Phys. Rev. Lett.}\ }\textbf {\bibinfo {volume} {127}},\ \bibinfo {pages}
  {232001} (\bibinfo {year} {2021})},\ \Eprint
  {http://arxiv.org/abs/2108.05822} {arXiv:2108.05822 [hep-ph]} \BibitemShut
  {NoStop}%
\bibitem [{\citenamefont {Aghasyan}\ \emph {et~al.}(2017)\citenamefont
  {Aghasyan} \emph {et~al.}}]{COMPASS:2017jbv}%
  \BibitemOpen
  \bibfield  {author} {\bibinfo {author} {\bibfnamefont {M.}~\bibnamefont
  {Aghasyan}} \emph {et~al.} (\bibinfo {collaboration} {COMPASS}),\ }\href
  {\doibase 10.1103/PhysRevLett.119.112002} {\bibfield  {journal} {\bibinfo
  {journal} {Phys. Rev. Lett.}\ }\textbf {\bibinfo {volume} {119}},\ \bibinfo
  {pages} {112002} (\bibinfo {year} {2017})},\ \Eprint
  {http://arxiv.org/abs/1704.00488} {arXiv:1704.00488 [hep-ex]} \BibitemShut
  {NoStop}%
\bibitem [{\citenamefont {Alexeev}\ \emph {et~al.}(2023)\citenamefont {Alexeev}
  \emph {et~al.}}]{COMPASS:2023vqt}%
  \BibitemOpen
  \bibfield  {author} {\bibinfo {author} {\bibfnamefont {G.~D.}\ \bibnamefont
  {Alexeev}} \emph {et~al.} (\bibinfo {collaboration} {COMPASS}),\ }\href@noop
  {} {\  (\bibinfo {year} {2023})},\ \Eprint {http://arxiv.org/abs/2312.17379}
  {arXiv:2312.17379 [hep-ex]} \BibitemShut {NoStop}%
\bibitem [{\citenamefont {Echevarria}\ \emph {et~al.}(2021)\citenamefont
  {Echevarria}, \citenamefont {Kang},\ and\ \citenamefont
  {Terry}}]{Echevarria:2020hpy}%
  \BibitemOpen
  \bibfield  {author} {\bibinfo {author} {\bibfnamefont {M.~G.}\ \bibnamefont
  {Echevarria}}, \bibinfo {author} {\bibfnamefont {Z.-B.}\ \bibnamefont
  {Kang}}, \ and\ \bibinfo {author} {\bibfnamefont {J.}~\bibnamefont {Terry}},\
  }\href {\doibase 10.1007/JHEP01(2021)126} {\bibfield  {journal} {\bibinfo
  {journal} {JHEP}\ }\textbf {\bibinfo {volume} {01}},\ \bibinfo {pages} {126}
  (\bibinfo {year} {2021})},\ \Eprint {http://arxiv.org/abs/2009.10710}
  {arXiv:2009.10710 [hep-ph]} \BibitemShut {NoStop}%
\bibitem [{\citenamefont {Bury}\ \emph {et~al.}(2021)\citenamefont {Bury},
  \citenamefont {Prokudin},\ and\ \citenamefont {Vladimirov}}]{Bury:2021sue}%
  \BibitemOpen
  \bibfield  {author} {\bibinfo {author} {\bibfnamefont {M.}~\bibnamefont
  {Bury}}, \bibinfo {author} {\bibfnamefont {A.}~\bibnamefont {Prokudin}}, \
  and\ \bibinfo {author} {\bibfnamefont {A.}~\bibnamefont {Vladimirov}},\
  }\href {\doibase 10.1007/JHEP05(2021)151} {\bibfield  {journal} {\bibinfo
  {journal} {JHEP}\ }\textbf {\bibinfo {volume} {05}},\ \bibinfo {pages} {151}
  (\bibinfo {year} {2021})},\ \Eprint {http://arxiv.org/abs/2103.03270}
  {arXiv:2103.03270 [hep-ph]} \BibitemShut {NoStop}%
\bibitem [{\citenamefont {Gurjar}\ and\ \citenamefont
  {Mondal}(2024)}]{Gurjar:2023uho}%
  \BibitemOpen
  \bibfield  {author} {\bibinfo {author} {\bibfnamefont {B.}~\bibnamefont
  {Gurjar}}\ and\ \bibinfo {author} {\bibfnamefont {C.}~\bibnamefont
  {Mondal}},\ }\href {\doibase 10.1103/PhysRevD.109.014038} {\bibfield
  {journal} {\bibinfo  {journal} {Phys. Rev. D}\ }\textbf {\bibinfo {volume}
  {109}},\ \bibinfo {pages} {014038} (\bibinfo {year} {2024})},\ \Eprint
  {http://arxiv.org/abs/2308.14528} {arXiv:2308.14528 [hep-ph]} \BibitemShut
  {NoStop}%
\bibitem [{\citenamefont {Brown}\ \emph {et~al.}(2014)\citenamefont {Brown}
  \emph {et~al.}}]{Brown:2014sea}%
  \BibitemOpen
  \bibfield  {author} {\bibinfo {author} {\bibfnamefont {C.}~\bibnamefont
  {Brown}} \emph {et~al.},\ }\href {\doibase 10.2172/1296770} {\  (\bibinfo
  {year} {2014}),\ 10.2172/1296770}\BibitemShut {NoStop}%
\bibitem [{\citenamefont {Chen}\ \emph {et~al.}(2019)\citenamefont {Chen} \emph
  {et~al.}}]{SeaQuest:2019hsx}%
  \BibitemOpen
  \bibfield  {author} {\bibinfo {author} {\bibfnamefont {A.}~\bibnamefont
  {Chen}} \emph {et~al.} (\bibinfo {collaboration} {SeaQuest}),\ }\href
  {\doibase 10.22323/1.346.0164} {\bibfield  {journal} {\bibinfo  {journal}
  {PoS}\ }\textbf {\bibinfo {volume} {SPIN2018}},\ \bibinfo {pages} {164}
  (\bibinfo {year} {2019})},\ \Eprint {http://arxiv.org/abs/1901.09994}
  {arXiv:1901.09994 [nucl-ex]} \BibitemShut {NoStop}%
\end{thebibliography}%

\end{document}